\documentclass[11pt, a4paper, oneside]{Thesis} % Paper size, default font size and one-sided paper
\usepackage{wrapfig}
\usepackage{lscape}
\usepackage{rotating}
\usepackage{graphicx}
\usepackage{caption}
\usepackage{amsmath}
\usepackage{amsmath}
\usepackage{cite}
\usepackage{amsfonts}
\usepackage{amssymb}
\usepackage{wasysym}
\usepackage{graphicx}
\usepackage{mathrsfs}
\usepackage{adjustbox}
\usepackage{caption}
\usepackage{IEEEtrantools}
\newcommand\myeq{\mathrel{\stackrel{\makebox[0pt]{\mbox{\normalfont\tiny (a)}}}{=}}}
\graphicspath{{Pictures/}} % Specifies the directory where pictures are stored
\usepackage[square, numbers]{natbib} % Use the natbib reference package - read up on this to edit the reference style; if you want text (e.g. Smith et al., 2012) for the in-text references (instead of numbers), remove 'numbers' v

\hypersetup{urlcolor=black, colorlinks=true} % Colors hyperlinks in blue - change to black if annoyingv`	
\title{\ttitle} % Defines the thesis title - don't touch this

\begin{document}
\makeatletter
\renewcommand*{\NAT@nmfmt}[1]{\textsc{#1}}
\makeatother

% prints author names as small caps

\frontmatter % Use roman page numbering style (i, ii, iii, iv...) for the pre-content pages

\setstretch{1.6} % Line spacing of 1.6 (double line spacing)

% Define the page headers using the FancyHdr package and set up for one-sided printing
\fancyhead{} % Clears all page headers and footers
\rhead{\thepage} % Sets the right side header to show the page number
\lhead{} % Clears the left side page header

\pagestyle{fancy} % Finally, use the "fancy" page style to implement the FancyHdr headers

\newcommand{\HRule}{\rule{\linewidth}{0.5mm}} % New command to make the lines in the title page

% PDF meta-data
\hypersetup{pdftitle={\ttitle}}
\hypersetup{pdfsubject=\subjectname}
\hypersetup{pdfauthor=\authornames}
\hypersetup{pdfkeywords=\keywordnames}

%----------------------------------------------------------------------------------------
%	TITLE PAGE
%----------------------------------------------------------------------------------------

\begin{titlepage}
\begin{center}

\graphicspath{ {./Figures/} }
\begin{figure}[hb]
  \centering
  \includegraphics[width=0.4\linewidth]{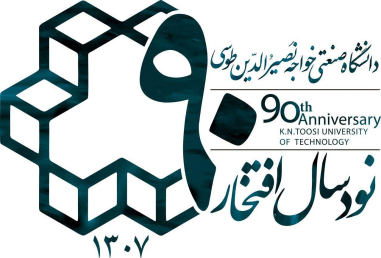}
\end{figure}

Faculty of Electrical Engineering\\ % Research group name and department name
\textsc{ K. N. Toosi University of Technology}\\[1.5cm] % University name

\HRule \\[0.4cm] % Horizontal line
{\huge \bfseries Improving Privacy-Preserving Techniques for Smart Grid using Lattice-based Cryptography}\\[0.4cm] % Thesis title
\HRule \\[1.5cm] % Horizontal line
 
\large \textit{A thesis submitted in fulfilment of the requirements\\ for the Master's of Science degree}\\[0.3cm] % University requirement text
\textit{by}\\[0.4cm]

\href{http://home.iitk.ac.in/~saiwal}{\textbf{Saleh Darzi}}

\vfill
\large March 2021\\[2cm] % Date

\end{center}

\end{titlepage}

%----------------------------------------------------------------------------------------
%	DECLARATION PAGE
%	Your institution may give you a different text to place here
%----------------------------------------------------------------------------------------

%%\Declaration{\addtocontents{toc}{\vspace{1em}} % Add a gap in the Contents, for aesthetics

%%It is certified that the work contained in this thesis entitled ''\ttitle'' by ''\authornames'' has been carried out under my supervision and that it has not been submitted elsewhere for a degree.
%%\\[2cm]

%%\begin{minipage}{0.4\textwidth}
%%	\begin{flushleft} \large
	%%	\emph{\large \today}\\[2cm] % Date
%%	\end{flushleft}
%%\end{minipage}
%%\begin{minipage}{0.65\textwidth}
%%	{\begin{center} \large
	%%	\supname
%%		\begin{center}
	%%		{Professor\\ 
		%%	\normalsize{\deptname\\
			%%\univname}}
%%		\end{center}
%%	\end{center}}
%%\end{minipage}
%%\vfill{}}

\clearpage % Start a new page

%----------------------------------------------------------------------------------------
%	ABSTRACT PAGE
%----------------------------------------------------------------------------------------

\addtotoc{Abstract} % Add the "Abstract" page entry to the Contents

 % Add a gap in the Contents, for aesthetics
\begin{abstract}
{\footnotesize \vspace{-5mm}With the advancement of communication structures and information technologies, a reliable distributed\vspace{-2mm} network, named Smart Grid was introduced. Although the smart grid provides efficient \vspace{-2mm}transmission of\vspace{-2mm} energy and data, the frequent gathering of users’ consumption data discloses users’ privacy. Plenty of data \vspace{-2mm}aggregation schemes have been introduced to preserve the privacy of users’ private information. Unfortunately,\vspace{-2mm} with the advent of quantum machines, most of these schemes will be rendered vulnerable and insecure.\vspace{-2mm} Hence, to preserve users' privacy in the smart grid, in the first part of this thesis, we attempt to introduce a secure lattice-based privacy-preserving \vspace{-2mm}multi-functional and multi-dimensional data aggregation scheme namely LPM2DA. Specifically, our proposed\vspace{-2mm} LPM2DA scheme utilizes a lattice-based homomorphic encryption for data aggregation and an efficient\vspace{-2mm} lattice-based signature to ensure integrity and authentication and it is resistant against various attacks.\vspace{-2mm} The LPM2DA enables the control center to acquire temporal and spatial aggregation of multi-dimensional\vspace{-2mm} data via our proposed polynomial Chinese remainder theoremin technique in a privacy-preserving\vspace{-2mm} manner. Also, it empowers the control center to calculate different statistical\vspace{-2mm} functions such as mean, variance, and skewness on users’ multi-dimensional data. Another\vspace{-2mm} drawbacks of the traditional data aggregation schemes are their weak and unrealistic architecture of their\vspace{-2mm} network model. Besides, these schemes are vulnerable to centralization problems such as single point of\vspace{-2mm} failure, mistrust, and they are prone to intimidation and pressure. Hence, to protect the privacy of users’\vspace{-2mm} sensitive data in a more pragmatic architecture and realistic network model, in the second part of this\vspace{-2mm} thesis, we have proposed a secure privacy-preserving data aggregation scheme for blockchain-based smart grid\vspace{-2mm} (SPDBlock). The SPDBlock not only preserves the privacy of users’ individual and aggregated data, but\vspace{-2mm} also can guarantee other security criteria like integrity, authentication, and resistance to smart grid-based\vspace{-2mm} and blockchain-based attacks. Based on our novel construction, the SPDBlock scheme enables the detection and prosecution of malicious\vspace{-2mm} entities in the network. Note that relying on the employment of our proposed Chinese remainder theorem\vspace{-2mm} technique, the SPDBlock scheme is also capable of transmitting multi-dimensional data efficiently. As a result of employing distributed decryption and secret\vspace{-2mm} sharing technique, the control center is only able to decrypt the approved aggregation of valid ciphertexts\vspace{-2mm} with a few smart meters’ assistance. Finally, in the performance evaluation, we have\vspace{-2mm} demonstrated that the SPDBlock scheme accomplishes great achievements in comparison with traditional\vspace{-2mm} schemes and achieves the lowest communication and computational costs among recent blockchain-based\vspace{-2mm} data aggregation schemes.

}
\end{abstract}

\clearpage % Start a new page

%----------------------------------------------------------------------------------------
%	ACKNOWLEDGEMENTS
%----------------------------------------------------------------------------------------

\setstretch{1.3} % Reset the line-spacing to 1.3 for body text (if it has changed)

\acknowledgements{\addtocontents{toc}{\vspace{1em}} % Add a gap in the Contents, for aesthetics

First and foremost, I am truly grateful to my supervisor, Prof. Bahareh Akhbari. Thank you for your constant guidance, support, and encouragement throughout my Master’s study. As a person who devotes his life to science, your guidance formed the basis of my academic life. Being your student was the best decision of my entire academic life and I hope to continue our cooperation for good. Without your guidance, patience, and valuable inputs on my research ideas and writings, this work would have not been possible.

I would like to express my deep gratitude to my advisor, Prof. Hassan Khodaiemehr. Relying on your pragmatic and specific view on mathematical and engineering problems, not only I learned to be a better researcher, but your precious inputs, remarks, and assistance helped increase the quality of this thesis. 

I wish to extend my special thanks to some of my course professors. I sincerely thank Prof. Mahmoud Ahmadian Attari because he was my role model in academic-discipline as well as tutor in science. Moreover, I would like to thank Prof. Taraneh Eghlidos since it was not possible to expand my education without her assistance and valuable comments. And, I am gratefully acknowledging Prof. Lotfollah Beygi for his various consult through my studying and his help for my academic future; and other K. N. Toosi University of Technology professors who helped me through my Master’s.

Finally, my deepest appreciation and grateful thanks go to my family, and especially MOTHER for her great sacrifices and encouragement during my education. And, I wish to show my sincere gratitude to my sister for all the reinforcement and to her partner-in-life who showed me the path.

}
\clearpage % Start a new page

%----------------------------------------------------------------------------------------
%	DEDICATION
%----------------------------------------------------------------------------------------
%
\setstretch{1.3} % Return the line spacing back to 1.3
\pagestyle{empty} % Page style needs to be empty for this page
\dedicatory{Dedicated to my MOTHER} % Dedication text
\addtocontents{toc}{\vspace{2em}} % Add a gap in the Contents, for aesthetics

%---------------------------------------------------------------------------------------
%Tozihat
%--------------------------------------------------------------------------------------
  \HRule \\[0.4cm] % Horizontal line
 Name of the student:	 \textbf{\authornames}		\\
		Thesis title: \textbf{Improving Privacy-Preserving Techniques for Smart Grid using Lattice-based Cryptography}\\
		Thesis supervisor:  \textbf{Bahareh Akhbari Ph.D.} \\ Thesis advisor:  \textbf{Hassan Khodaiemehr Ph.D.}\\
 Degree for which submitted:  \textbf{Master of Science} \\ Department:  \textbf{Electrical and Computer Engineering} \\
		Month and year of thesis submission: \textbf{{\large March 2021}\\[0.4cm] }
		\HRule \\[1.5cm] % Horizontal line	

%----------------------------------------------------------------------------------------
%	LIST OF CONTENTS/FIGURES/TABLES PAGES
%----------------------------------------------------------------------------------------

\pagestyle{fancy} % The page style headers have been "empty" all this time, now use the "fancy" headers as defined before to bring them back

\lhead{\emph{Contents}} % Set the left side page header to "Contents"
\tableofcontents % Write out the Table of Contents

\lhead{\emph{List of Figures}} % Set the left side page header to "List of Figures"
\listoffigures % Write out the List of Figures

\lhead{\emph{List of Tables}} % Set the left side page header to "List of Tables"
\listoftables % Write out the List of Tables

\mainmatter % Begin numeric (1,2,3...) page numbering

\pagestyle{fancy} % Return the page headers back to the "fancy" style

% Include the chapters of the thesis as separate files from the Chapters folder
% Uncomment the lines as you write the chapters

%----------------------------------------------------------------------------------------
% Introduction
%----------------------------------------------------------------------------------------
\chapter{Introduction}
\lhead{\emph{Introduction}} % Set the left side page header to "Introduction"

With the improvement of communication structures and information technologies, the smart grid as the new generation of power grid was officially defined by Energy Independence and Security Act of 2007 (EISA-2007)\cite{EISA}. In respect to this definition, Smart Grid (SG) is an electrical grid that incorporates smart communication and connections among electricity consumers, suppliers, and other grid components to achieve a reliable, economical, and secure electricity supply \cite{SGdef}. As opposed to the traditional grid that was comprised of dumb components with no data connections between them, the smart grid allows for two-way communication of data and energy between its components \cite{Survey}. 

Moreover, since the traditional grid is comprised of dispensing generated electricity to the consumers, two additional concerns of this grid are a massive power storage system and shortage of an appropriate framework for collecting electricity demands. However, the smart grid is specifically designed to address these concerns and challenges. In fact, the smart grid is a distributed network that utilizes additional renewable sources of energy, therefore it is reliable and makes it easy to monitor and control the energy distribution. Specifically, the smart grid provides fault detection and self-healing. As a result, it is dependable and efficient for the transmission of energy. Thus, the term ``smart" is expressed to define all these new grid’s abilities and facilities. 
Concretely, the traditional grid’s failures and blackouts like Northeastern Blackout of 2003 which affected more than fifty-five million people, stands in total contrast to the new energy grid \cite{Jokar}.

Among seven domains of smart grid defined by the National Institute of Standards and Technology (NIST), consider the customer-side network in which all the appliances in the residential area (or in a house) are reporting their data to the smart meter (SM) installed in each home area network (HAN). Subsequently, SM reports the collection of consumption data of all these appliances to the control center (CC) whose obligation is to scrutinize these data and monitor the energy distribution and the grid’s stability \cite{Vahedi}.

Although two-way communication gave advantages to the smart grid over the traditional grid, it also could bring some vulnerabilities, privacy disclosure, and security menaces. Energy theft, fraud, impersonation, and reducing network reliability and learning personal patterns are the most common privacy and security threats of the smart grid. Specifically, the privacy of consumers’ information is of paramount significance, because some private information can be deduced from users’ reports, such as their habits, lifestyle, the number of people residing in a household, and in some cases the type of appliances can be recognized \cite{ErkinZ}.

To tackle these threats and maintaining privacy, diverse approaches have been adopted: 
$1)$ Hardware equipment to mask the transmitting data. This approach is costly and not suitable for smart grid networks considering the numerous number of smart meters.
$2)$ Concealing each smart meter’s data with noise. Specifically, in this approach, the addition of private data with a particular noise will be transmitted to the control center. The main drawback to this approach is the low accuracy of data recovered by CC.
$3)$ Utilization of cryptographic techniques to ensure the privacy of users. This approach comprises of three different types: anonymization, authentication, and data aggregation (DA) schemes \cite{Book}.

By means of anonymization techniques, each entity disguises its true identification by a pseudo-ID. However, for the purpose of traceability of errant entities, these lightweight techniques need an online trusted authority (TA) who knows the connection between the real and pseudo ID. 
Authentication techniques are deployed for authenticity confirmation of each entity in each phase. Therefore, by the growth of users in the network, these authentication techniques add a certain amount of delay to each phase\cite{Book}.
To preserve privacy and to hinder the managing unit CC, or even a fraudulent employee in the operation center from acquiring personal information of each individual, data aggregation (DA) schemes have been suggested. The structure of a DA scheme requires a powerful entity called gateway (GW) between CC and the residential area’s smart meters, to combine the consumption information of that residential area. In order to preserve private information of individual HANs, data aggregation schemes deploy homomorphic encryption systems like Paillier, BGN, El-Gamal, RSA, \textit{etc.} \cite{Book}.

In recent years, there are two types of pivotal improvement to the data aggregation schemes: achieving post-quantum security and utilizing blockchain technology. More specifically, since the alteration of smart grid infrastructure is extremely expensive and considering the imminent quantum attacks, one novel approach towards preserving the privacy of customer-side network is to employ post-quantum cryptography for post-quantum security. 
Security of traditional privacy-preserving schemes relies on the hardness of integer factorization and/or discrete logarithm problems, which have been broken by Shor’s quantum algorithm \cite{Shor}. This obviously indicates the need for using secure and effective post-quantum cryptography like lattice-based cryptography, which is resistant to quantum attacks and entails straightforward procedures like polynomial addition and multiplication. However, only a few research works exploit lattice-based cryptography to preserve privacy. Therefore, to achieve security against high-level attacks, we have proposed a secure Lattice-based Privacy-preserving Multi-functional and Multi-dimensional Data Aggregation scheme (LPM2DA) in smart grid.

Furthermore, the typical architecture of traditional DA schemes is centralized, and thereby they are susceptible to “single point of failure”, “single point of trust”, and prone to intimidation and mistrust. Moreover, the typical smart grid network architecture requires some semi-honest authorities in their model, and therefore they are not practical or similar to real-world situations \cite{S6}. Also, the advent of blockchain technology which is now integrated into various areas such as cryptocurrencies, financial services, energy trading, and communication systems, resolves these challenges. Besides, in recent and current years, the utilization of blockchain-based architecture in SG privacy-preserving schemes has risen drastically. Thus, similar to LPM2DA, to render the SG network a safer place even against some high-level insider attacks, we have proposed a secure privacy-preserving data aggregation scheme for blockchain-based smart grid (SPDBlock).

%%%%%%%%%GOALS and CONTRIBUTIONS
\section{Goals and Contributions}
Regarding the privacy problems which are the most significant threat for the SG implementation in the perspective of the customer-side network and residential area users, and considering other security concerns, we have presented two novel privacy-preserving schemes to address these challenges. To be precise, counting on the post-quantum security and blockchain-based network architecture, we have proposed one lattice-based and one blockchain-based privacy-preserving DA scheme. In the following paragraphs, we provide a comprehensive illustration of each scheme’s construction.

\begin{itemize}
\item In the first part of this thesis, to preserve privacy, ensure integrity and provide authentication, we propose a secure Lattice-based Privacy-preserving Multi-functional and Multidimensional Data Aggregation scheme (LPM2DA) in smart grid, which enables CC to acquire temporal and spatial aggregation of multi-dimensional data. Specifically, by means of acquiring the spatial aggregation of users’ data, CC could realize electricity theft or power leakage and make better conscious pricing choices; and by computing temporal aggregation, CC could attain the bill. The practical calculation of various aggregations and multiple functions on the users’ data in a privacy-preserving manner, is the consequence of employing an efficient R-LWE based homomorphic encryption along with an appropriate R-LWE based signature scheme. 

Besides, since the lattice-based encryption scheme has a large plaintext space and the smart grid communications are usually very small, it would not be efficient to send a single piece of data via each transmission. One way to make a scheme more efficient is to transmit multi-dimensional data in each transmission instead of only sending one data. Moreover, transmitting multi-dimensional data is considerably practical in the smart grid network because the consumption data is composed of various information about multiple appliances of each house and the heating and lighting system’s data. In this scheme, we propose a novel Chinese remainder theorem-based technique to transmit multi-dimensional data. In addition, in order to find out about the balance and uniformity of usage data in the smart grid network, our scheme empowers CC to calculate different statistical functions such as the mean, variance, and skewness on users’ multi-dimensional data. Hence, the LPM2DA scheme can compute various aggregations, average, variance, and skewness of each dimensional data. 

Concretely, we show that our proposed scheme is resistant against various attacks like modification, impersonation, replay, and man-in-the-middle attack. Due to the system model and employed encryption in our scheme, the decryption of aggregated data does not depend on the presence of all the SMs in the residential area, which makes our scheme fault-tolerant. Especially, not only our DA scheme sustains intact by decreasing the number of users, but it also can support an acceptable increase in the number of users. Therefore, our proposed scheme allows dynamic users.

\item The second part of this thesis presents the SPDBlock scheme in-depth. The SPDBlock is a blockchain-based DATA scheme with a network that is very similar to real-world situations. Besides the key management center (KMC) who is an outsider authority and only distributes the keys at the beginning of the system configuration, no SG entity is assumed to be honest or honest-but-curious. This strong assumption in the network model is the direct result of our blockchain-based construction wherein two blockchains (the Sidechain and Mainchain) work in parallel to one another. More specifically, in this construction, each SG entity is being observed and the observers can gain advantages or even money from their observing. 

The Sidechain blocks are generated by some powerful and pre-approved smart meters in the customer-side of the SG network. Also, the Mainchain blocks on which the whole SG network operates is created by the GW. The GW is merely a relay that copies the general information of the Sidechain blocks in its own created blocks. Despite a new blockchain-based architecture, we also propose a novel consensus mechanism namely, hash onion, so that the miner selection is performed in a really efficient way. This scheme not only centers its attention on preserving the privacy of users’ individual, aggregated, and the bill data in a real-world network model, but also can ensure other security criteria such as data integrity, source authentication, detection of various attacks, and prosecution of the insider attackers.

Regardless of the spatial aggregation which the control center needs for grid analysis, to calculate the bill in a privacy-preserving manner, our scheme should be able to compute temporal aggregation of each users’ consumption. Also, since our proposed CRT-based technique in the LPM2DA is independent of the deployed encryption scheme, we specialize this technique to utilize it with our employed Elliptic curve-based homomorphic encryption in the SPDBlock. As a result, the SPDBlock allows for the transmission of multi-dimensional data which is significantly useful for grid monitoring and it is practical in the SG network because each smart meter’s data is truly multi-dimensional. Besides, it enables the CC to obtain various aggregations of each dimensional data. Notwithstanding the security criteria such as privacy, integrity, and authentication, our proposed scheme is resistant to numerous SG attacks like replay, modification, impersonation, and MITM attacks, and some blockchain-based attacks such as fork, collusion, and $51\%$ attack.

Due to the utilization of distributed decryption and secret sharing technique in our scheme construction, at least a specific number of smart meters should assist the CC so that he could decrypt the aggregated consumption data and the bills. Furthermore, akin to the real-world situations, the SPDBlock scheme also considers smart meters’ crashing, being hacked, getting manipulated by the user, and the absence of old smart meters or the addition of new smart meters. Thus, it accomplishes dynamic user and fault-tolerance properties.
\end{itemize}

\section{Outline of the Thesis}
The outline of this thesis is structured as follows. In chapter 2, we first delineate the smart grid network more thoroughly concerning the SG architecture, infrastructure, and communication models. Successively, we expound on smart grid security concerns and threats. With respect to the SG problems, we focus on security concerns, describing various approaches towards solving these problems and centering our attention on privacy-preserving schemes. In the next two chapters, we propose two novel privacy-preserving schemes. The first proposed scheme namely, LPM2DA, is presented in chapter 3. Regarding this scheme, we initially depict the system model which is comprised of the network model, the adversarial model, and the security requirements, and then introduce the preliminary notations and assumptions. Subsequently, we present the proposed LPM2DA construction and provide a detailed security analysis based on different parameters such as privacy, integrity, authentication, and resistance against various attacks. Lastly, the performance evaluation of the LPM2DA is illustrated relying on the computation costs and communication efficiency. For chapter 4, we have presented the SPDBlock scheme. In exactly the same way as the previous chapter, we first represent the system model that contains the network model, the adversarial model, and security requirements. Successively, we define some notations and approaches in blockchain technology and Elliptic curve cryptography as a preliminary. Afterward, we explicate the SPDBlock construction and provide a thorough secure analysis with respect to preserving privacy, ensuring integrity and authentication, and resistance to SG and blockchain-based attacks. Finally, to show the practicality and efficiency of the SPDBlock scheme, we meticulously evaluate its performance relying upon computational costs, communication overheads, and its achievements comparing to other related works. In the end, we conclude the thesis on the grounds of our lattice-based and blockchain-based approaches and present the future works within chapter 5.

%----------------------------------------------------------------------------------------------
% Lit Rev
%----------------------------------------------------------------------------------------------

\chapter{Literature Review}
\lhead{\emph{Literature Review}} % Set the left side page header to "Literature Review"
In this chapter, we plan to study the fundamental background, network models, security and privacy concerns of the SG, then explicate some similar research works. Concretely, after introducing some important background information about the SG, we exhaustively expound on the SG domains, frameworks, and communication network models. Subsequently, a comprehensive inspection of SG implementation problems, security and privacy challenges, and attack models is presented. Finally, relying on traditional privacy-preserving schemes, achieving post-quantum security, and utilizing blockchain architecture, we demonstrate several similar privacy-preserving DA schemes.

%Beginning of Lit Rev
The SG network that incorporates the recent improvement of communication structures and information technologies could be viewed as an application in the state-of-the-art Internet of Things (IoT). Regarding the improvement of technologies and the fact that every “thing” is going to be connected to the internet, the advent of a smart electrical grid was inevitable. However, relying on some blackouts, especially the American blackout in August 2003, the idea of a reliable and more secure electrical grid was introduced. Subsequently, in 2007, the Energy Independence and Security Act (EISA-2007) formally defined the smart grid and established some distinguishing characteristics of the SG\cite{EISA}. Generally, these characteristics were: a more reliable, secure, and efficient communication; dynamic optimization; incorporation of distributed resources; resilience against various attacks and disruptions; inclusion of smart devices; and \textit{etc.} Since then, the improvement of smart grid has been considered in almost all the countries of the world and it plays a significant part regarding many aspects like economic, social, and industrial growth. Nevertheless, the cybersecurity coordination task group (CSCTG) was organized in 2009 to address the cybersecurity issues of the SG network \cite{NIST668}.

%%%%%%%%%%%%%%%SG NETWORK
\section{Smart Grid Network}
The National Institute of Standards and Technology (NIST) presents nine priority areas in the smart grid that requires a great deal of attention and the establishment of standards \cite{NIST246}. To better understand the necessary requirements for research work in the SG application, we enumerate all of these areas: Demand response and consumer energy efficiency; Wide-area situational awareness; Distributed Energy Resources (DER); Energy Storage; Electric transportation; Network communications; Advanced metering infrastructure (AMI); Distribution grid management; Cybersecurity. In this thesis, we concentrate on the cybersecurity area which includes privacy, integrity, authentication, availability, and \textit{etc} \cite{NIST246}.

\begin{figure}
\includegraphics[width=15cm, height=10cm]{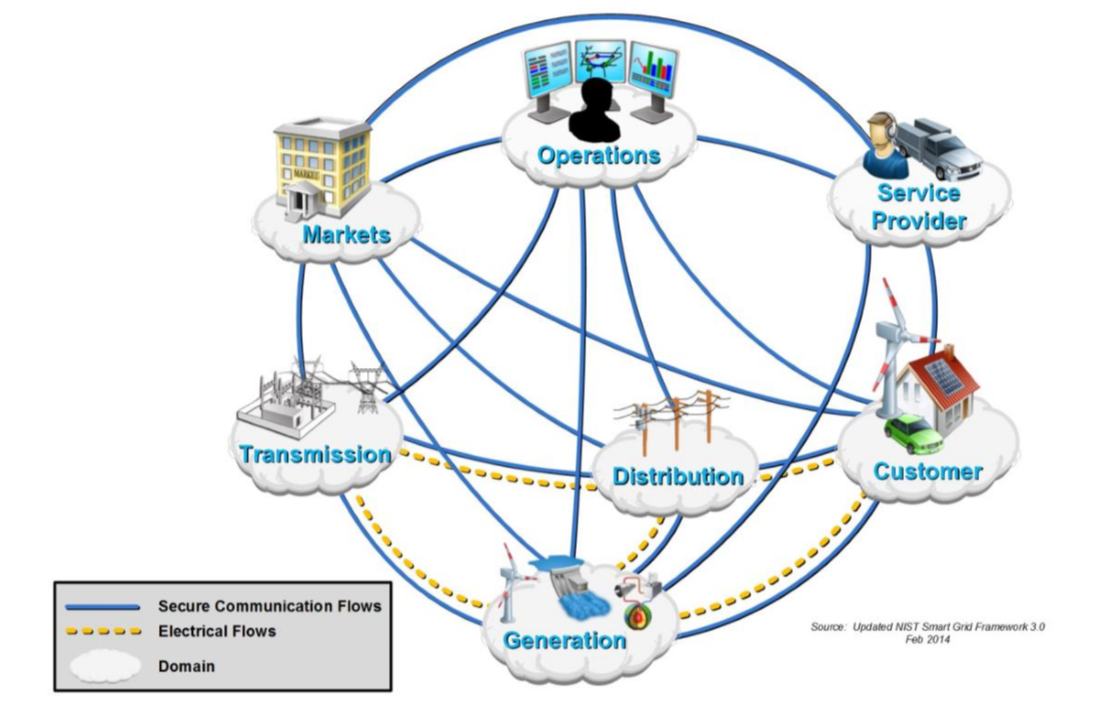}
\caption{Interaction of Actors in Different Smart Grid Domains through Secure Communication Flows \cite{NIST668}}
\label{Fig-copy}
\end{figure}

The \textit{NIST Framework and Roadmap} document establishes the smart grid based on seven major domains: Transmission, Distribution, Operations, Generation, Markets, Customer, and Service Provider \cite{NIST668}. The specific interaction of each domain including communication and electricity is depicted in FIGURE \ref{Fig-copy}. 
\begin{itemize}
\item \textbf{Operations}: This operation domain is truly the administrator of the electricity movement. Specifically, this domain handles the administrating operations such as monitoring, control, fault management, analysis, reporting and statistics, network calculations, and \textit{etc.}
\item \textbf{Generation}: This domain is responsible for the generation of electricity from various sources of energy like water, wind, solar, or chemical combustion. The generation domain might be able to store some energy for later distribution.
\item \textbf{Transmission}: As demonstrated in FIGURE \ref{Fig-copy}, the transmission domain is generally accountable for transferring electricity from the generation to the distribution domain.
\item \textbf{Distribution}: The direct connection between the transmission domain and the customers is the distribution domain. This domain might as well generate, store, or control electricity. 
\item \textbf{Markets}: The business operations and trading are performed in this domain. Besides, the Market domain is to all other domains for a reliable exchange price and stabilizing the supply and demand in the grid.
\item \textbf{Service Provider}: As its name indicates, the service provider domain carries out procedures such as customer management, billing, home-related services, installation, and maintenances.
\item \textbf{Customer}: The whole SG is designed to realize the customer needs. Apparently, the most electricity consumption happens in this domain which is comprised of three main areas: Home area, commercial/building area, and industrial area. Apart from the consumption of energy, some generation and storage of electricity also achieved in this domain. Most cybersecurity attacks and privacy disclosures are executed about this domain. 
\end{itemize}
A great reference for each domain’s function is thoroughly illustrated in the \cite{NIST668} document.

\begin{figure}
\includegraphics[width=15cm, height=10cm]{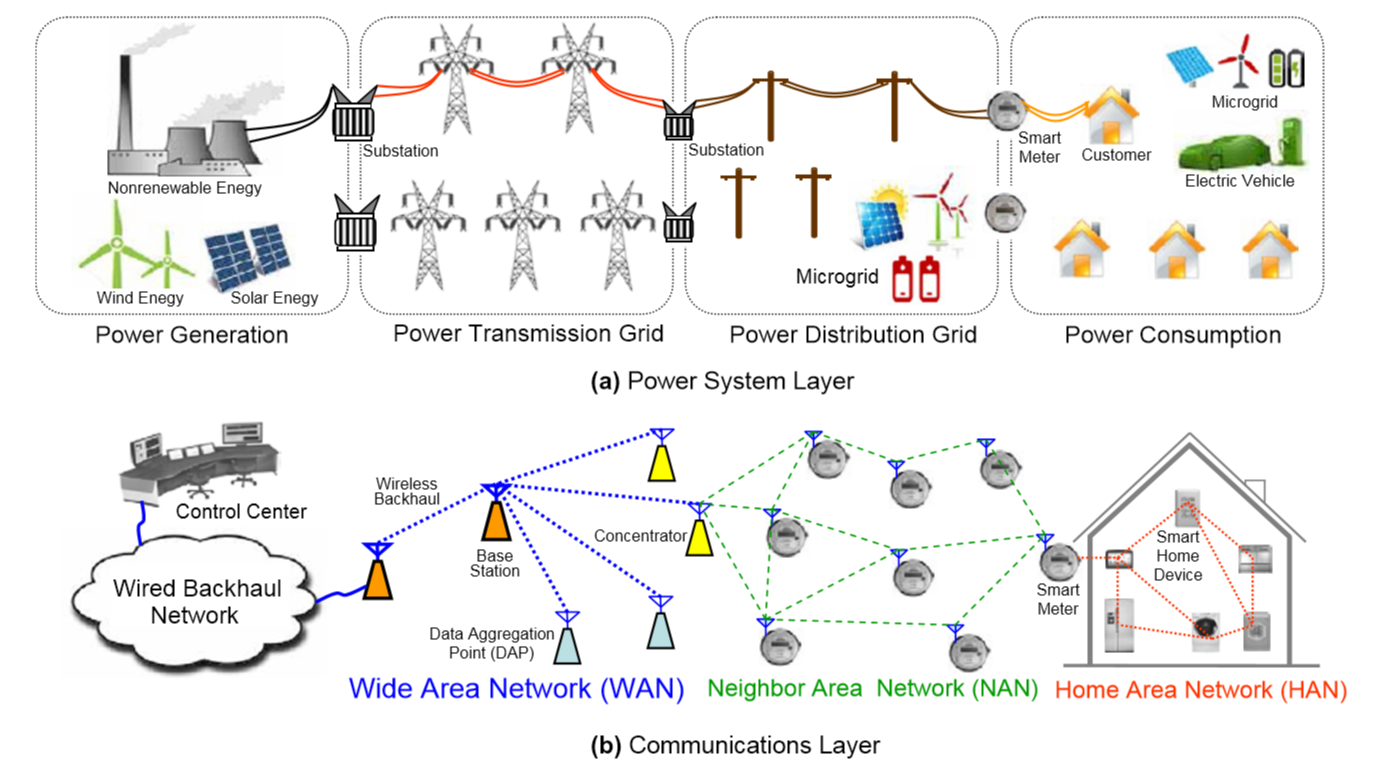}
\caption{The overall layered architecture of Smart Grid \cite{ho2013}}
\label{Fig-copy2ho}
\end{figure}

%%%%%SG Communication Network 
To illustrate a more detailed description of the SG network with the focus on communication layers, the SG typical communications network is demonstrated in FIGURE \ref{Fig-copy2ho}. Apart from the domains mentioned above, we expound on the communication structure of each network area presented in FIGURE \ref{Fig-copy2ho}. The following network areas are accounted for in almost every SG privacy-preserving scheme \cite{ho2013}.
\begin{enumerate}
\item \textbf{Customer Premise Area Network (HAN/BAN)}: The customer premise area network is comprised of building area network (BAN) and home area network (HAN). The BAN is a vast geographical area or a huge building comprised of numerous units or apartments. Each BAN has a gateway (GW) to relay information between users and the control center. Basically, the HAN is those units or apartments mentioned in the BAN subsection. Each HAN has a smart meter to garner the real-time consumption data. Specifically, due to the existence of different types of appliances in each house and the fact that HANs could generate electricity besides their consumption, smart meters collect those multi-dimensional data in specified periods like every 15 or 30 minutes and transmit them to the network authorities. Usually, the communication among BAN and HAN entities is through economical communication technologies for instance Wi-Fi, Bluetooth, Zig-Bee, or power line carrier (PLC) \cite{Sur-wiley}. 
\item \textbf{NAN}: The neighbor area network is accountable for relaying numerous information and heavy data from the aggregators and substations to the wide-area network (WAN) entities. Thus, as opposed to HAN and BAN, NAN requires a higher and stronger communication link with data rates ranging between $100$ kbps to $10$ Mbps for broad spaces \cite{Sur-wiley}. 
\item \textbf{WAN}: The wide area network covers an extensive area of about hundreds and thousands of square kilometers. This network area is the final place of connection with the data aggregators and it receives data at high rates ranging from $10$ Mbps to $1$ Gbps. Hence, it demands a higher communication vessel like cellular or Wi-MAX technologies \cite{Sur-wiley}.
\end{enumerate}

\begin{figure}
\includegraphics[width=15cm, height=10cm]{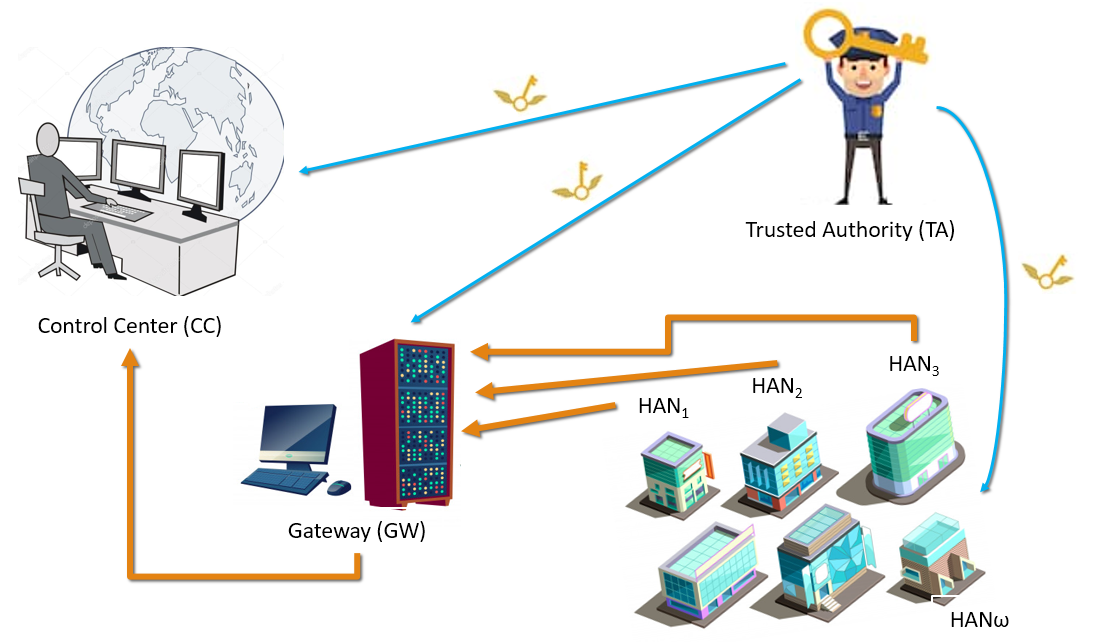}
\caption{A typical model of the SG network}
\label{MyFigure}
\end{figure}

%My Typical figure
However, for the proposition of a privacy-preserving scheme, customarily, only a few components of the above communication network are considered. Although the network model hinges on the construction of each privacy-preserving scheme and its assumptions, we have depicted a typical network model in FIGURE \ref{MyFigure}.
In FIGURE \ref{MyFigure}, besides the smart meter of each HAN and the gateway (GW) of each BAN, we have taken into account a trusted authority for the system configurations and the CC as a representative of the network authority/administer.

Another aspect of SG network classification is particularly applied in DA schemes. Owing to the development of communication technologies in recent years, DA schemes could be divided into three different architectures. The first architecture is the traditional network architecture shown above. For the next one, to increase the communication efficiency and lowering the delay, the fog/edge computing architecture is presented as a novel prospect. And lastly, the blockchain architecture is introduced to resolve the centralization problems in the SG network \cite{S6}. 

%%%%%%%%%%%%%%%%%%%%

%%%%%%%%%%%%%%%%%%

%%%%%%%%%%%%%%%%%%

%%%%%%%%%SG Sec and Priv
\section{Smart Grid’s Security and Privacy}
In any system or application, from the perspective of cryptography, the core security requirements are confidentiality, integrity, and availability. In the smart grid network, akin to other applications, there exist some additional application-specific security criteria such as non-repudiation, access control, secure storage, and scalability that should be achieved. We expound on these criteria for the SG network as follows \cite{Sur-wiley, Sur-manu}:
\begin{itemize}
\item \textbf{Confidentiality and Privacy}: The privacy and confidentiality terms that are used interchangeably are defined as the restriction on the private information accessible to other unauthorized entities in the network. Since the smart meters’ data are highly sensitive and might reveal some information about the users such as their lifestyle, their activities at home, and even their religion, the privacy of their data which are transmitted in open channels should be protected. 

The NIST delineates the privacy challenges based on four types \cite{NIST33}: $1)$~Personal information: this aspect of privacy which considers a user’s lifestyle and habits is the most serious threat for the SG network. $2)$~Personal privacy: it is connected to the physical characteristics of a user such as its body-related information which is critical for health-care systems. $3)$~Behavioral privacy: this aspect considers each user’s right for preserving the privacy of their everyday life. $4)$~Personal communication privacy: Another aspect of importance for the SG network is the privacy of each entity’s communication in the network. To be exact, the adversary should not be able to link an entity’s communicated data. 

By and large, privacy is of significant importance because the revelation of individual data not only discloses the privacy, but also it enables the adversary to forecast the subsequent action of each entity. Notwithstanding the above classification, there exist some extra attacks associated with privacy such as the renowned differential privacy.
\item \textbf{Integrity}: The integrity criterion is defined as the ability to detect and prevent the modification or alteration of the transmitted data. Since each entity in the SG network requires to transmits its data through public channels where an adversary may reside, integrity is chiefly important for the SG network and it is, in fact, a complementary criterion to privacy. Besides, some classic attacks like modification, replay, and MITM attack which we will explain in the next subsection, are correlated with invalidating the integrity. Therefore, by certifying the integrity, SG can resist these attacks.
\item \textbf{Authentication}: This criterion is akin to the integrity and deals with the transmitted data. More particularly, owing to the communication in public channels in SG, the authentication asserts that the identity of each entity should be established so that no adversary or an unauthorized entity could impersonate other individuals or apply an active or passive MITM attack.
\item \textbf{Availability}: The availability criterion states that each entity should have access to the required data without any delay or disruption. In the SG network that enables a reliable and efficient electricity supply, and considering the number of communication and computations each entity needs to perform to be certain of privacy, integrity, and authentication, accomplishing availability is a prime factor. Aside from the availability itself, the SG network also requires to detect and resist some availability-related attacks, for instance, denial of service (DoS), distributed DoS (DDoS), and even malware attacks.
\item \textbf{Other criteria}: Since the SG network necessitates to store sensitive data in its system in multiple locations, all the above-cited criteria should also be achieved for these storages. Furthermore, due to the numerous types of communication links and devices in the SG infrastructure, the scalability of an SG-designed scheme should be accomplished. Finally, no SG entity should be capable of accessing unauthorized data or denying their participation in the network, thereby the SG network should realize the access control and non-repudiation properties, respectively.
\end{itemize}
In spite of the above security requirements, we expound on the threat models and different kinds of attackers in the SG network. Regarding the employed factor, we can classify the attackers in 2 ways: the first classification concentrates on the attacks themselves, the passive and active; the second classification is relying upon the types of the attacker, insider, and outsider. The definition of each attack hinges on the system model and the construction of a scheme. However, in the following paragraphs, we generally demonstrate some typical attacks on the SG network \cite{Sur-manu, Sur-kumar, Sur-sanjab}.
\begin{itemize}
\item Replay attack: Replay attack is the transmission of a message that has been sent before. It could be applied by an insider attacker for electricity theft and by an outsider adversary to invalidate the network reliability.
\item MITM attack: This type of attack includes a passive style wherein the adversary passively eavesdrops on the communicated data between two entities without being detected; and an active style in which the adversary alters the communicated data between two entities without being detected or caught by any of them.
\item Impersonation and Modification: These attacks are correlated with the integrity and authentication criteria. In fact, few schemes could resist impersonation and modification attacks without ensuring integrity and authentication. 
\item DoS attack: The well-known denial-of-service attack aims to discredit the system availability. As a result of the DoS attack, the system cannot provide services for legitimate customers. For instance, in a DDoS attack, the powerful adversary who has access to some entities in the network tries to request the grid operator from different locations so that the operator halts and cannot provide any regular and legal services. It is worth noting that this attack is normally performed by a powerful outsider attacker.
\item Collusion attacks: The collusion attacks are normally considered as an internal attack wherein the network entities attempt to cooperate for malicious reasons such as disclosure of privacy and invalidation of integrity and authentication. Nonetheless, the adversary could also collude with one or some network entities for a similar reason. Despite the dependence of the collusion attacks on the system assumptions, a typical collusion attack happens among the network authorities with greater access control.
\end{itemize}
Other than the above attacks, there are several threats and special attacks on the SG network like: key-based, malware, human-factor-aware differentially aggregated, and bad data injection attacks. For more information, a good reference is \cite{Sur-manu}.

%%%%%%%%%%%Pr-Pr schemes
\section{Privacy-Preserving Schemes}
As mentioned in the introduction chapter, there are several approaches towards addressing the security and privacy concerns of the SG network. Besides the utilization of hardware equipment, concealing each smart meter’s data with noise, and physical layer approaches, in this section, we exhaustively concentrate on the cryptographic approaches and more particularly the privacy-preserving DA schemes. Regarding the above evaluation of security requirements, adversarial model, and the network architecture, each privacy-preserving scheme focuses its attention on achieving a part of the aforementioned criteria and properties. To convey a clearer sense about our proposed scheme, the explication of a handful of privacy-preserving schemes are presented based on three different approaches: traditional DA schemes, post-quantum secure DA schemes, and finally DA schemes for the blockchain-based smart grid.

%%%%%%%%%%
\begin{itemize}
\item \textbf{Traditional privacy-preserving DA schemes:}\\
Lu \textit{et al.}'s scheme, EPPA \cite{EPPA}, preserves the privacy of individual reports by employing the Paillier encryption scheme. By relying on the super-increasing sequence, each user could transmit its multi-dimensional data in a way that the control center could calculate the aggregation of each dimension via a recursive algorithm. It also utilizes batch verification to decrease the authentication overhead.

Zhang \textit{et al.} \cite{SMing} has also proposed a DA scheme based on the Paillier encryption scheme, that could aggregate users’ information in both temporal and spatial form. In this scheme’s system model, to achieve temporal and spatial aggregation, a network is formed in which users could communicate and transmit shares of their data to each other without using encryption.

Chen \textit{et al.} \cite{MuDA} adopted an additive homomorphic encryption scheme namely, BGN cryptosystem, to present its DA scheme MuDA. By means of computing and transmitting one, two, and three kinds of aggregation by GW, the control center could obtain the average, variance, and one-way analysis of variance (ANOVA) of users’ data, respectively. Since the GW adds a particular noise to the aggregated data to make the scheme resistant to differential attack, it would render the aggregated data obtained by CC less accurate.

Ge \textit{et al.} \cite{FGDA} proposed another DA scheme called FGDA, which is capable of calculating the average, variance, and skewness via transmitting a single aggregated data by GW. Furthermore, it supports users' fault-tolerance by trusted authority’s participation in the aggregation phase. Based on a session key between users and GW, the authentication of the message would be assured.

Ni \textit{et al.} \cite{Range-based} proposed a DA scheme relying on the lifted El-Gamal encryption and BBS signature scheme. Based on GW’s role in noise addition and aggregation of honest smart meters’ data, the features like fault-tolerance and differential privacy are guaranteed. Moreover, it utilizes zero-knowledge-proof techniques to screen the unusual reports.

In the DA scheme proposed by Ming \textit{et al.} in \cite{Ming}, to efficiently preserve privacy, heavy operations such as bilinear pairings are eliminated, and the elliptic curve and El-Gamal encryption schemes are employed. In addition, their scheme is also resistant to various attacks and provides the transmission of multi-dimensional data. 

Finally, the Boudia \textit{et. al.}’s scheme \cite{S8} employs an Elliptic Curve El-Gamal encryption and ECDSA signature to preserve privacy and ensure integrity and authentication in a DA construction, respectively. Furthermore, owing to the deployment of multi-recipient encryption, this scheme is capable of transmitting multi-dimensional data. And lastly, it is worth noting that the Boudia \textit{et. al.}’s scheme transmits the CC’s feedback analysis by deploying secure symmetric encryption.

\item \textbf{Lattice-based privacy-preserving DA schemes:}\\
Only a few researches exploit lattice-based cryptography to preserve privacy \cite{LRSPPP, AbdallahNTRU, Abdallah2018, PDA}. LRSPPP \cite{LRSPPP} utilizes a ring learning with error (R-LWE) based cryptosystem and an R-LWE based signature scheme. Abdallah et al.’s scheme \cite{AbdallahNTRU} uses the revised version of the NTRU cryptosystem and the new NTRU signature scheme (NSS). Both schemes \cite{LRSPPP, AbdallahNTRU} estimate the energy demand for a constellation of households by means of a forecasting function. Technically, load forecasting provides prudent decision-making in the network and could be achieved through different algorithms \cite{Load1, Load2}.
On one hand, forecasting lessens the communication overhead. On the other hand, it adds a database to each building area network (BAN) for storing the demands that BAN has access to it.

Abdallah et al. \cite{Abdallah2018} proposed another scheme, in which a lattice-based homomorphic scheme is used to encrypt and sign the individual and aggregated data. It categorizes the smart household appliances into four groups and lets them aggregate their consumption without SM’s participation. 

Concentrating on privacy assurance, Li et al. \cite{PDA} introduced a DA scheme namely PDA, relying on an R-LWE based somewhat homomorphic encryption scheme. It accomplishes multi-dimensional data and empowers CC to calculate functions like the mean and variance on these data.

\item \textbf{Blockchain-based privacy-preserving DA schemes:}\\
One of the first blockchain-based privacy-preserving schemes for SG was proposed by Guan \textit{et. al.} \cite{S2}. This scheme operates based on private blockchains for different groups of people with similar consumption data. The protection of privacy is founded on creating multiple pseudonyms for each user and the authentication is guaranteed via the bloom filter. Guan \textit{et. al.} propose a novel proof-of-work mechanism in which whoever has a clear consumption to the average consumption data gets to be the miner of that round. In their scheme, the communications are in plaintext and the bill is calculated by transmitting the whole private blockchain to the billing center at the end of each month.

The Fan \textit{et. al.} introduced a DA scheme that functions based upon Consortium blockchains and delegated proof-of-stake mechanism \cite{S3}. The transmission of multi-dimensional data to multiple receivers (the CC, the grid operator, and the equipment supplier) is achieved through a hybrid signcryption scheme and the integrity and authentication are certified via a signature scheme. Owing to the introduction of the grid operator’s smart contract, this scheme is capable of sending the feedback of the network analysis and some grid warnings to the users.

Wang \textit{et. al.} \cite{S4} proposed a DA scheme based on utilizing blockchain technology along with homomorphic encryption. Their scheme is constructed in a two-tier hierarchical blockchain including several regional cluster blockchains (RC-BCs) and one wide-area blockchain (WA-BC). The privacy of users’ consumption data is protected by employing the Paillier encryption and the integrity and authentication are confirmed via a double-hash and a signature scheme. Their consensus mechanism is a simple PBFT mechanism wherein only $f$ node out of $n$ network node could be malicious where $n\geq 3f+1$.

Fan \textit{et. al.} proposed a privacy-preserving DA scheme called DPPDA \cite{S5}. The DPPDA scheme is in fact a decentralized DA scheme that is constructed to work without the assistance of any trusted third party (TTP). The employment of Paillier encryption and the BLS signature scheme protects the privacy, integrity, and authentication of the DPPDA. Furthermore, they also proposed a miner election algorithm to select a miner out of several candidates relying on the vote of all the nodes.

The DA-SADA scheme \cite{S6} is introduced by Chen \textit{et. al.} not only operates based on blockchain architecture, but also allows for fog computing architecture. Technically, this scheme is structured in a three-tier architecture and enables the aggregation of consumption data by deploying the Paillier encryption. It should be noted that besides double-blockchain creation, the two-layer aggregations are performed anonymously. Moreover, the DA-SADA utilizes the bloom filter for a fast authentication of pseudonyms, deploys signature for integrity, and uses batch verification for lower overheads.

In the DA scheme proposed by Zhang \textit{et. al.} \cite{S7}, to protect the privacy of users’ consumption data efficiently, heavy operations such as bilinear pairing and exponentiation operations are eliminated, and instead, the Elliptic Curve-based operations are performed. Also, the Consortium blockchain is constructed relying upon the PBFT mechanism. And for the protection of privacy and data integrity, it employs a ring signcryption algorithm. 

The above blockchain-based schemes are chosen specifically based on their purpose which is preserving the privacy of users’ consumption data. Although there exist various types of blockchain-based schemes such as authentication, anonymization, and data aggregation for the smart grid network, the cited schemes cover all the data aggregation privacy-preserving schemes and they are the best and most efficient of each type. 
\end{itemize}

%---------------------------------------------------------------------------------------
% LPM2DA
%---------------------------------------------------------------------------------------
\chapter{LPM2DA: A Lattice based Privacy preserving Multi Functional and Multi dimensional Data Aggregation Scheme for Smart Grid}
\lhead{\emph{LPM2DA}} % Set the left side page header to "LPM2DA"
The outline of this chapter is structured as follows. In section 3.1, we demonstrate the network model, attack scenarios and establish security requirements and the scheme’s objective. Section 3.2 introduces the fundamental notation and assumptions of our homomorphic encryption scheme. The explication of our proposed LPM2DA scheme is in section 3.3. We assess the security and performance of our scheme separately in section 3.4 and 3.5, and then draw the scheme’s conclusion and summary within section 3.6.

%system model
\section{SYSTEM MODEL}
In this section, we describe the network model and its components thoroughly, then establish our aim and the security requirements.
%network model
\subsection{Network Model}
Typically, the smart grid communication network consists of one managing unit who administers the communication and information flow; a large number of users which mainly consume energy; and an authority who monitors all entities. 
In this scheme, we name the managing unit, control center (CC). In addition, inside each building area network (BAN), we categorize users in collection of home area networks (HAN) which could be units or apartments equipped with smart meters to gather energy consumption data. And lastly, TA signifies the trusted authority. 
As shown in Fig.1, our system model is comprised of one control center, one trusted authority, and a building area network composed of $\omega$ home area networks.
\begin{figure}
\includegraphics[width=15cm, height=10cm]{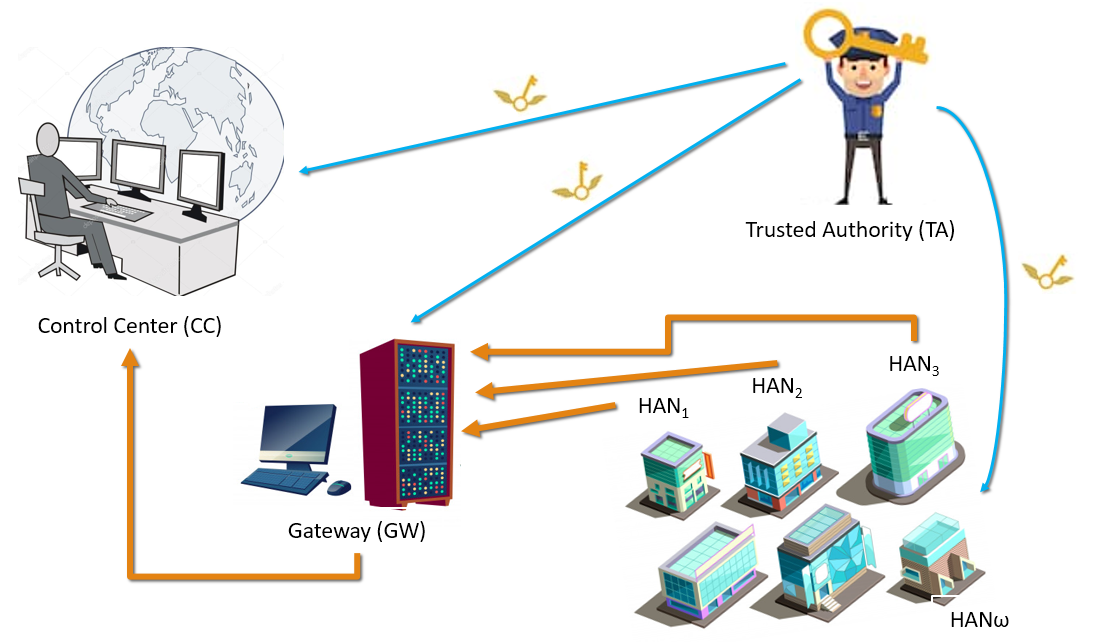}
\caption{LPM2DA's System Model}
\label{Fig1}
\end{figure}

\textbf{CC}: The CC's role could be played by an organization or government. Its responsibility is mainly administering the load balance, collecting consumption information and evaluating different functions on these data. It also settles down any wrongdoings, and even produces bills.

\textbf{BAN}: The BAN is a powerful entity whose gateway (GW) acts as a communication relay between CC and smart meters. Precisely, the GW mostly executes statistical functions on the legitimate data in order that CC obtains various aggregations. The communication between GW and smart meters is by means of somewhat inexpensive WiFi technology, whereas, the GW communicates with the CC via high bandwidth or wired links.

\textbf{HAN}: Each HAN which could be units or apartments equipped with smart meters to gather real-time information of the energy consumed by the appliances in the apartment. At some intervals, like every 15 or 30 minutes, each smart meter has a unique ID and records the activity and energy usage of all the appliances. Then, it sends these consumption data to CC through BAN gateway.

\textbf{TA}: If we take CC as the brain of the smart grid network, TA is representative of this network's eyes. The TA is a trustworthy entity who is responsible for the whole system setup and key distributions.

%adversary model
\subsection{ADVERSARY MODEL and SECURITY REQUIREMENTS}
To consider the real world status, we identify CC and GW as honest-but-curious \textit{i.e.}, they play their role in the system honestly, but they are inquisitive entities and will take any chances they have got to find out about comprehensive consumption data of each customer. As opposed to traditional meters, it is really harder for users to tamper with smart meters, so we consider the SM to be an honest entity.

Concretely, there exists a strong adversary $\mathcal{A}$ residing in the smart grid network which can damage the smart meters and interfere with databases of CC and GW. The adversary mostly eavesdrops and alters the communication flow or even sets up some undetectable malware in GW to get a hold of consumption data, and exploits it to its advantage.
More seriously, to disclose privacy and violate integrity and authentication of the network, not only adversary could be passive and just eavesdrop the communication, but also could carry out some active attacks such as, impersonation, modification, man-in-the-middle (MITM), and replay attack.

Therefore, to transmit data in a privacy-preserving manner, ensuring integrity of the information, authenticating the source and impeding malicious deeds, we should gratify the security requirements established beneath:

\textbf{Privacy preservation}: In the smart grid communication, neither smart grid components (GW, CC) nor the adversary needs to know the comprehensive consumption data. Therefore, the eavesdropping adversary who deployed malicious malware, or even an unhappy employee in the control center  should not be  capable of unveiling the consumption information. Accordingly, knowing only the residential area's data will suffice CC to balance loads and monitor the grid.

\textbf{Integrity and authentication}: Since the transmission of data is through public links, each entity in the smart grid network should be uniquely identified in order to participate in the communication.
Therefore, in order to hinder adversaries from forging or altering the transmitted data, every entity's authenticity and validity of its data should be checked after each transaction.

\textbf{Resistance against attacks}: Typically, like any other open network, smart grid is susceptible to numerous attacks, for instance: replay attack, impersonation, modification, differential, MITM and other internal attacks.

%design goal
\subsection{DESIGN GOAL}
In this scheme, we aim to preserve privacy and guarantee data integrity and source authentication meanwhile empowering CC to scrutinize residential area's data thoroughly by means of computing different kinds of aggregation. Furthermore, our data aggregation scheme is not susceptible to imminent quantum attacks. Concretely, the addressed security requirements and calculation of the statistical functions should be accomplished efficiently. Albeit, considering the computation overhead is indispensable to ensure the availability of the smart grid network (providing CC with real-time data every 15 minutes), the communication costs should also be minimized.

%priliminaries
\section{PRELIMINARIES}
\subsection{Notation}
  For a sufficiently large prime modulus $q$, we define the ring $\mathbb{Z}/q\mathbb{Z}$ (denoted as $\mathbb{Z}_q$) in $(-q/2, q/2) \cap \mathbb{Z} $. We consider the quotient ring $R_q=R/qR$ based on the prime modulus $q$ and a cyclotomic  ring $R$ which is defined next.  The ring $R$ is a polynomial ring of the form $R=\mathbb{Z}[x]/<f_m(x)>$ where $f_m(x)=x^n+1$ is the irreducible $m$th cyclotomic polynomial; $n$ is a power of 2; and $m=2n$. We identify the error distribution $\chi$ as the discrete Gaussian distribution $D_{\mathbb{Z}^n,r}$, in which every vector drawn from $D_{\mathbb{Z}^n,r}$ with standard deviation $r>0$ is of length $r\sqrt{n}$ \cite{signature}. 

The symbol $``\cdot"$ specifies all types of multiplications such as matrix and polynomial multiplications. The notation like $A_{M\times N}$ are used to represent matrices of rings with dimensions $M$ and $N$, where all elements are from $R_q$; and the identity matrix of size $N\times N$ is denoted by $I_{N\times N}$. For simplicity, we denote the row vectors of length $n$ as $[a_1,\ldots,a_n]$, and the column vectors of length $n$ as $[a_1;\ldots;a_n]$. 

In the encryption system \cite{SHIELD} deployed in our scheme, there exists one function named ``bit decomposition" denoted as \textrm{BD(.)}, which can take integers, polynomials, or matrix of polynomials as input and yields the extended version of them. Specifically, BD(.) function outputs a size-$\ell$ vector consisting of the bit decomposition of the input elements. Therefore, it outputs an $\ell \times n$ matrix for the degree-$n$ polynomial with $\ell$-bit integer coefficients as input. Subsequently, by substituting the bit representation of each integer coefficient, \textrm{BD(.)} function produces a matrix of size-$x\times y\ell$ with depth $n$ for the input matrix of polynomials with dimension $x\times y$ (wherein each polynomial is of degree $n$ with $\ell$-bit integer coefficients).
The inverse of the BD(.) function, which is a function denoted by BDI(.), is also used in the encryption part of our scheme.
Technically, BDI(.) function collects $\ell$ successive coefficients and yields the $\ell$-bit  integer of the BD(.)'s input. Precisely, BDI(.) function can be identified as the multiplication of its input matrix and a matrix named $\Upsilon_{y\ell \times y}$ as described below\cite{SHIELD}:
\begin{IEEEeqnarray}{ll}
B_{x\times y} = \mbox{BDI}(\tilde{B}_{x\times y\ell})= \tilde{B}_{x\times y\ell}\cdot \Upsilon_{y\ell \times y}.\\\nonumber
\Upsilon_{y\ell \times y}=\begin{bmatrix}
1&0&0&\cdots&0\\ 2&0&0&\cdots&0\\ \vdots&\vdots &\vdots&\ddots&\vdots \\2^\ell&0&0&\cdots&0\\
0&1&0&\cdots&0\\ 0&2&0&\cdots&0\\ \vdots&\vdots &\vdots&\ddots&\vdots \\0&2^\ell&0&\cdots&0\\
0&0&1&\cdots&0\\ \vdots&\vdots &\vdots&\ddots&\vdots \\0&0&0&\cdots&1\\
0&0&0&\cdots&2\\ \vdots&\vdots &\vdots&\ddots&\vdots \\0&0&0&\cdots&2^\ell
\end{bmatrix}.
\end{IEEEeqnarray}

%assumtion
\subsection{Assumption}
The Ajtai's breakthrough paper \cite{Ajtai} which showed a connection between the worst-case and the average-case problems for lattices, affords confidence in adopting lattice-based schemes in cryptography which holds a great promise for post quantum cryptography.
In 2005, with the seminal work of Regev \cite{Onlattices}, the average-case problem of learning with errors was announced.
Since then, the LWE and its variants seemed to be a versatile problem in the most encryption schemes. In this paper, we utilize a homomorphic scheme named, SHIELD \cite{SHIELD} which is based on the ring variant of LWE problem introduced by Lyubashevsky \textit{et al.}\cite{Onideallattices}.
The hardness of R-LWE problem is linked to the worst-case problems on ideal lattices.
For a uniformly random $s\in R_q$ (called the secret key), there are two distributions on $R_q \times R_q$: the first distribution is $(a,b=a\cdot s+e)\in R_q\times R_q$ where $a$ is uniform, randomly chosen from $R_q$ and $e$ is an independent error chosen from the error distribution $\chi \subset R_q$; and the second distribution is formed by choosing $(a,b)\in R_q\times R_q$ uniformly. The decision R-LWE problem is distinguishing between these two distributions with non-negligible advantage.

%LPM2DA
\section{The Proposed Scheme: LPM2DA}
Our proposed LPM2DA scheme is formed by five phases, named: system setup, user report generation, report aggregation, secure report reading, and temporal aggregation.
%overview
First, we present an overview of our proposed DA scheme. The LPM2DA scheme mainly focuses on the computational competence such as computing multiple statistical functions, different aggregations, and calculation proficiency. 

At the beginning, TA initializes the system via settling both system and Chinese remainder theorem's (CRT) parameters\cite{CRT}. Then, it generates and issues different key pairs of all entities. One of the efficient points of this scheme is its ability to handle multi-dimensional data by exploiting polynomial CRT. Technically, based on polynomial CRT, each smart meter gathers all of its multi-dimensional data into a single appropriate data and then encrypts that data.
Subsequently, each smart meter signs its encrypted data in order to achieve integrity, authentication, and securing the scheme against various attacks.

Despite the fact that the GW is merely a relay between residential area's smart meters and the CC, it verifies the signatures, computes different kinds of aggregations on the valid data, and also computes the bill at the end of each bill period. Finally, after verifying the aggregated data which is signed by GW, CC decrypts the aggregated data. Due to the deployed homomorphic encryption scheme and GW's collaboration, CC could obtain any kind of aggregation it prefers. Specifically, relying on the CRT's features, CC could acquire various aggregations of each dimension's data without disclosing privacy. Now, the comprehensive description of each phase of our scheme is presented below.

%system setup
\subsection{System Setup}
In this phase, TA plays the main role for the configuration of the system parameters and generates the keys for signing and encryption.
First, TA yields system parameters including: the maximum degree of polynomials $n$; the prime modulus $q = 1 \pmod {2n}$; the standard deviation of discrete Gaussian distributions of key space and ciphertext space from which the key and the ciphertext error are taken from and denoted by $\sigma_k$ and $\sigma_c$, respectively; $\ell=\lceil$log$q\rceil$, and $N=2\ell$ is a parameter used in the ciphertext matrix dimensions and  manages the ciphertext size.
The entire system parameters and the complexity of the operations evaluated on ciphertexts are managed by $\lambda$, which is the security parameter and typically is set to be $80$ or $120$ bits.

%CRT
Then, to utilize the CRT technique\cite{CRT} for transmitting $k$-dimensional data, TA chooses $k$ pairwise coprime polynomials in the ring $R$, denoted by $p_j(x)$ for $j=1,2,...,k$. We denote the   degree of each $p_j(x)$ with $\mbox{deg}(p_j)$ and  set $D=\sum_{j=1}^{k}{\mbox{deg}(p_j)}$. These parameters are chosen such that for each smart meters' data polynomial like $p_{data}(x)$, $\mbox{deg}(p_{data})<\mbox{deg}(p_j)$.
TA denotes the multiplication of all of these $k$ polynomials with $Q(x)=\prod_{j=1}^{k}{p_j(x)}$ and defines $Q_j(x)=Q(x)/p_j(x)$ for $j=1,2,...,k$. Then, it finds $k$ polynomials $T_j(x)$ with $\mbox{deg}(T_j(x))<\mbox{deg}(p_j(x))$ by the partial fraction decomposition of $1/Q(x)=\sum_{j=1}^{k}{T_j(x)/p_j(x)}$.

%CC KEY GEN
Based on the notations and the system parameters, to build CC's public and private keys, TA chooses the polynomial $t_{cc}$ from the discrete Gaussian error distribution $D_{R_q,\sigma_k}$. Subsequently, the CC's secret key is set to be the column vector of polynomials $SK_{cc}=[1;-t_{cc}]$. 
Then, it samples $a_{cc}\leftarrow R_q$, $e_{cc}\leftarrow D_{R_q,\sigma_k}$ uniformly and yields $b_{cc}=a_{cc}\cdot t_{cc}+e_{cc}$. Hence, CC's public key is the row vector of polynomials $PK_{cc}=[b_{cc}~a_{cc}]$. From the definition of $b_{cc}$, we can see that the inner product of public and private keys over the ring $R_q$ is:
\vspace{-2mm}\begin{equation}
PK_{cc}\cdot SK_{cc}=[b_{cc}~~a_{cc}]\cdot[1;-t_{cc}]=b_{cc}-a_{cc}\cdot t_{cc}=e_{cc} .
\label{eq2}
\end{equation}

%GW KEY GEN
Next, TA samples GW's secret key $s_{GW}$ from $R_q$. Given a prime integer $t_{GW}\in {\mathbb{Z}_q}^*$ and drawing $a_{GW}\leftarrow R_q$ and $e_{GW}\leftarrow \chi$, where $\chi=D_{\mathbb{Z}^n,r}$ is the error distribution, TA sets $b_{GW}=a_{GW}\cdot s_{GW}+t_{GW}\cdot e_{GW}$ and lets the GW's public key to be $PK_{GW}=(a_{GW},b_{GW})\in R_q\times R_q$. 

%SM KEY GEN
Similarly, to generate keys for each smart meter, TA draws a ring element $s_i\leftarrow R_q$ for $i=1,2,...,\omega$ and $SK_i=s_i$. By computing $b_i= a_i\cdot s_i+t_i\cdot e_i\in R_q$ where $t_i\in {\mathbb{Z}_q}^*$, $a_i$ is drawn from $R_q$ and $e_i$ is chosen from the error distribution $\chi=D_{\mathbb{Z}^n,r}$, TA assigns the $SM_i$'s public key as $PK_i=(a_i,b_i)\in R_q\times R_q$ for $i=1,2,...,\omega$.

%Key Distr.
Finally, TA chooses a secure collision resistant hash function $H:~\{ 0,1\}^*~\rightarrow~R_q$ which will be utilized in signing procedure. At the end, TA issues the key pairs to each entity through a secure channel, and publishes the public parameters.

%User Report Gen
\subsection{User Report Generation}
At every time instants $T_\gamma$, for example every $15$ or $30$ minutes, each smart meter $SM_i$ for $i=1,2,...,\omega$ uses the polynomial CRT to pack its $k$-dimensional data ($d_{i1},d_{i2},...,d_{ik}$ for $i=1,2,...,\omega$) into one message and encrypts it with CC's public key.
Specifically, each $SM_i$ in residential area performs the following procedure:
\begin{enumerate}
\item $SM_i$ multiplies each one of its data by the polynomials $Q_j(x)$ and $T_j(x)$, and computes the summation of them to attain $m_i(x)\in R_q$:
\begin{IEEEeqnarray}{ll}
m_i(x)=d_{i1}(x)\cdot T_1(x) Q_1(x)+d_{i2}(x)\cdot T_2(x) Q_2(x) +\cdots +d_{ik}(x)\cdot T_k(x) Q_k(x) \nonumber\\
~~~~~~~~ =\sum_{j=1}^{k}d_{ij}\cdot T_j(x) Q_j(x) \in R_q .
\label{eq3}
\end{IEEEeqnarray}
\item At the interval $T_\gamma$, to encrypt the polynomial message $m_i\in R_q$ with CC's public key, $SM_i$ utilizes an error matrix $E_{N\times 2}\leftarrow D_{R_q^{N\times 2}, \sigma_c}$ and an $N\times 1$ matrix of polynomials in which every random coefficient is in $\{ 0,1\}$. Then, it can obtain the $N\times 2$ matrix of ciphertext, as shown below:
\begin{IEEEeqnarray}{l}
C_{i\gamma}=m_i\cdot \mbox{BDI}(I_{N\times N})+ r_{N\times 1}\cdot PK_{cc}+E_{N\times 2} .\hspace{5mm}
\label{eq4}
\end{IEEEeqnarray}
\end{enumerate}

%Sign the Ctxt
To achieve data integrity and source authentication, every smart meter $SM_i$ needs to sign the ciphertext with its own secret key. Specially, $SM_i$ for $i=1,2,...,\omega$ hashes the ciphertext $C_i$ (for simplicity we use the notation $C_i$ instead of $C_{i\gamma}$) and the timestamp $T_\gamma$, and then creates the signature as presented below:
\begin{IEEEeqnarray}{l}
u_i=\big(v_i+H(C_i,T_\gamma)\big)\cdot s_i+ t_i\cdot e'_i ~, \sigma_i=(u_i,v_i) ,
\label{eq5}
\end{IEEEeqnarray}
where $v_i$ is drawn from a uniform distribution over $R_q$ and the error term $e'_i\leftarrow \chi$ is different from $e_i$ which was sampled in the system setup phase. Finally, $SM_i$ sends the ciphertext $C_i$, timestamp $T_\gamma$, and the signature $\sigma_i=(u_i,v_i)$ to the GW.

%Report Aggr.
\subsection{Report Aggregation}
The GW checks the integrity and authenticity of the data received by verifying the signature and the timestamp.
Precisely, GW verifies the signature by assessing the following conditions for every $SM_i, i=1,2,...,\omega$:
\vspace{-2mm}\begin{IEEEeqnarray}{ll}
1)  ~\sigma_i \in R_q\times R_q,\nonumber\\ 
2) ~ [-a_i\cdot u_i+b_i\cdot v_i] \bmod t_i=-b_i\cdot H(C_i,T_\gamma) \bmod t_i. \nonumber
\end{IEEEeqnarray}
If both of the mentioned conditions hold, the GW aggregates ciphertexts in a privacy preserving manner. The verification correctness is demonstrated at the end of this section. In our proposed scheme, the utilization of the homomorphic encryption scheme lets CC to compute multiple functions with various circuit depth\footnote{The circuit depth of a function is the number of multiplication levels required for the implementation of that function.}, such as mean, variance, skewness, and one-way analysis of variance (ANOVA), etc.
Nevertheless, here we just demonstrate the aggregation needed for the computation of mean, variance, and skewness. Surely, GW accomplishes different aggregations as steps shown below:
\begin{enumerate}
\item GW computes the homomorphic addition of all the verified ciphertexts: 
\vspace{-3mm}\begin{IEEEeqnarray}{l}
C_{Add}=\sum_{i=1}^{\omega}C_i .
\label{eq6}
\end{IEEEeqnarray}
\item Then, GW calculates the homomorphic multiplication of each ciphertext with itself. Moreover, it aggregates the result:
\vspace{-2mm}\begin{IEEEeqnarray}{ll}
A_i=\mbox{BD}(C_i)\cdot C_i  ~ \mathrm{for} ~ i=1,2,\ldots,\omega , \nonumber \\
C_{Mult}=\sum_{i=1}^{\omega}A_i .
\label{eq7}
\end{IEEEeqnarray}
\item For the last aggregation, GW calculates the homomorphic multiplication of each $C_i$ with $A_i$ obtained from the equation \eqref{eq7}, and then computes the summation of them:
\vspace{-2mm}\begin{IEEEeqnarray}{ll}
B_i=\mbox{BD}(A_i)\cdot C_i ~ \mathrm{for} ~ i=1,2,\ldots,\omega , \nonumber \\
C_{Skew}=\sum_{i=1}^{\omega}B_i .
\label{eq8}
\end{IEEEeqnarray}
\end{enumerate}

%Sign Aggr. result
Now, to provide integrity and authentication, GW signs the timestamp and the aggregations. Specifically, after calculating the hash value of $T_\gamma, C_{Add}, C_{Mult}$, and $C_{Skew}$, GW achieves $\sigma_{GW}=(u_{GW},v_{GW})$ by evaluating $u_{GW}=(v_{GW}+H(C_{Add}, C_{Mult}, C_{Skew}, T_\gamma ))\cdot s_{GW}+t_{GW}\cdot e'_{GW}$ where $v_{GW}\in R_q$ and the error term $e'_{GW}$ is different from $e_{GW}$ in the system setup phase, and drawn from the error distribution $\chi$.
Finally, GW transmits different aggregations $C_{Add}, C_{Mult}, C_{Skew}$, timestamp $T_\gamma$, and $\sigma_{GW}=(u_{GW},v_{GW})$ to the CC.

%Verif correctness
\textbf{Verification Correctness}: The first condition's correctness is apparent; and the second condition's correctness is presented below. We first analyze the left-hand side of the second condition's equation:
\begin{IEEEeqnarray}{lll}
[-a_i\cdot u_i+b_i\cdot v_i] \bmod t_i \nonumber\\\hspace{-2mm}=\hspace{-1mm}[-a_i \Big(\big(v_i+\hspace{-1mm}H(C_i,T_\gamma )\big) s_i+t_i e'_i\Big)\hspace{-1mm}+\hspace{-1mm}(a_i s_i+t_i e_i)v_i] \bmod t_i \nonumber\\
\hspace{-2mm}=\hspace{-1mm}[-a_i v_i s_i-a_i H(C_i,T_\gamma) s_i\hspace{-1mm}-a_i t_i e'_i\hspace{-1mm}+\hspace{-1mm}a_i s_i v_i\hspace{-1mm}+\hspace{-1mm}t_i e_i v_i] \bmod t_i \nonumber\\
\hspace{-2mm}=\hspace{-1mm}-a_i\cdot H(C_i,T_\gamma)\cdot s_i .
\end{IEEEeqnarray}
Since the right-hand side of the second condition's equation is simplified as below, it is obvious that both sides of the equation are equal, and the verification is correct.
\begin{IEEEeqnarray}{lll}
[-b_i\cdot H(C_i,T_\gamma)\cdot s_i] \bmod t_i \nonumber\\
=[-(a_i\cdot s_i+t_i\cdot e_i)\cdot H(C_i,T_\gamma)] \bmod t_i \nonumber\\
= -a_i\cdot s_i\cdot H(C_i,T_\gamma) .
\end{IEEEeqnarray}

%Report Reading
\subsection{Secure Report Reading}
When CC receives data from GW, it first checks the validity of the timestamp and signatures. Indeed, CC examines the equations $\sigma_{GW}\in R_q\times R_q$ and $[-a_{GW}\cdot u_{GW}+b_{GW}\cdot v_{GW}] \bmod t_{GW}=-b_{GW}\cdot H(C_{Add}, C_{Mult}$, $ C_{Skew}, T_\gamma) \bmod t_{GW}$, and if they hold, CC could decrypt the legitimate data to obtain different aggregations of user's data (in each dimension). 
Surely, by decrypting each of the ciphertexts $C_{Add}, C_{Mult}, \mbox{and} ~C_{Skew}$, CC will acquire $\sum_{i=1}^{\omega}m_i$, $ \sum_{i=1}^{\omega}m^2_i, \mbox{and}~ \sum_{i=1}^{\omega}m^3_i$, respectively. For decryption, CC just needs to multiply the ciphertext by its private key $SK_{cc}$. The decryption and homomorphic correctness are demonstrated at the end of this section.

Next, CC acquires the $k$-dimensional data, by applying CRT techniques. These aggregations are exhibited in the form:
\begin{IEEEeqnarray}{lll}
\sum_{i=1}^{\omega}m_i=\sum_{i=1}^{\omega}\Big(\sum_{j=1}^{k}d_{ij}(x) T_j(x) Q_j(x)\Big)  ,\\
\sum_{i=1}^{\omega}m^2_i=\sum_{i=1}^{\omega}\Big(\sum_{j=1}^{k}d_{ij}(x) T_j(x) Q_j(x)\Big)^2 ,\\
\sum_{i=1}^{\omega}m^3_i=\sum_{i=1}^{\omega}\Big(\sum_{j=1}^{k}d_{ij}(x) T_j(x) Q_j(x)\Big)^3  .
\end{IEEEeqnarray}
Based on CRT's features, the remainder of the Euclidean division of $\sum_{i=1}^{\omega}m_i, \sum_{i=1}^{\omega}m^2_i$, and $\sum_{i=1}^{\omega}m^3_i$, by $p_j(x)$ is $\sum_{i=1}^{\omega}d_{ij}(x), \sum_{i=1}^{\omega}d_{ij}^2(x)$, and $\sum_{i=1}^{\omega}d_{ij}^3(x)$, respectively. The correctness of CRT features used in above statement is demonstrated at the end of this section.
Eventually, CC could compute various functions of the aggregated $k$-dimensional data for $j=1,2,...,k$, as follows:
\begin{IEEEeqnarray}{lll}
\mbox{Mean} = \frac{1}{\omega}\sum_{i=1}^{\omega}d_{ij} ,\\
\mbox{Variance} =\frac{1}{\omega}(\sum_{i=1}^{\omega}d^2_{ij})-\mbox{Mean}^2 ,\\
\mbox{Skewness} = \frac{\frac{1}{\omega}(\sum_{i=1}^{\omega}d^3_{ij})-3 \mbox{Mean}\cdot\mbox{Variance}-\mbox{Mean}^3}{\mbox{Variance}^{3/2}}.
\end{IEEEeqnarray}

%Dec Correctness
\textbf{Decryption Correctness}: Consider the ciphertext $C_i$ as a representative of $m_i$'s encryption, then we have:
\begin{IEEEeqnarray}{lll}
C_i\cdot SK_{cc}=\big(m_i\cdot \mbox{BDI}(I_{N\times N}) + r_{N\times 1}\cdot PK_{cc}
+E_{N\times 2}\big) \cdot SK_{cc}\nonumber \\
=m_i\cdot \mbox{BDI}(I_{N\times N})\cdot SK_{cc} +  r_{N\times 1}\cdot PK_{cc}\cdot SK_{cc} +E_{N\times 2}\cdot SK_{cc}\nonumber\\
%=m_i\cdot BDI(I_{N\times N})\cdot SK_{cc} +  r_{N\times 1}\cdot [b_{cc}~a_{cc}]\cdot [1~-t_{cc}]\nonumber \\+E_{N\times 2}\cdot SK_{cc}
=m_i\cdot \mbox{BDI}(I_{N\times N})\cdot SK_{cc} +  r_{N\times 1}\cdot e_{cc}+E_{N\times 2}\cdot SK_{cc}\nonumber\\
=m_i\cdot \mbox{BDI}(I_{N\times N})\cdot SK_{cc} + error .
\label{eq15}
\end{IEEEeqnarray}
Due to errors $e_{cc}$ and $E_{N\times 2}$, we have $error=r_{N\times 1}\cdot e_{cc}+E_{N\times 2}\cdot SK_{cc}$. Now, the first $\ell$ coefficients in \eqref{eq15} are of the form $m_i, 2m_i,..., 2^{\ell -1}m_i$ in addition to $error$, as shown below:
\vspace{-2mm}\begin{IEEEeqnarray}{lll}
m_i\cdot \mbox{BDI}(I_{N\times N})\cdot [1~;-t_{cc}]
=m_i\cdot\begin{bmatrix}
1 & 2 &...& 2^{\ell -1} & -t_{cc} & -2t_{cc} &...& -2^{\ell -1}t_{cc}\end{bmatrix}^T.
\hspace{+3mm}\end{IEEEeqnarray}
So, considering $n$-degree polynomial, for each coefficient of this polynomial and with the assumption that $error<q/2$, the most significant bit of each part of the above matrix has one bit of $m_i$. 

%Homo correctness
\textbf{Homomorphic correctness}: Considering the above correctness, legitimacy of homomorphic addition is evident. Nevertheless, for the homomorphic multiplication, the correctness is slightly tricky. Assuming $C_i$ is the encryption of $m_i$, the homomorphic multiplication is asymmetric based on the input; and as a result, the noise growth is lower. Matrix dimensions are omitted for simplicity.
\begin{IEEEeqnarray}{lll}
\mbox{BD}(C_i)\cdot C_i \cdot SK_{cc}=\mbox{BD}(C_i)\cdot \big(m_i\cdot \mbox{BDI}(I) +  r\cdot PK_{cc}+E\big) \cdot SK_{cc} \nonumber \\
=\mbox{BD}(C_i)\cdot \big(m_i\cdot \mbox{BDI}(I)\cdot SK_{cc} +r\cdot e_{cc}+E\cdot SK_{cc}\big)\nonumber\\
=m_i\cdot \hspace{-1mm}\mbox{BD}(C_i)\cdot\hspace{-1mm} \mbox{BDI}(I)\cdot \hspace{-1mm}SK_{cc} \hspace{-1mm}+\hspace{-1mm}\mbox{BD}(C_i)\big(r\cdot e_{cc}\hspace{-1mm}+\hspace{-1mm}E\cdot\hspace{-1mm} SK_{cc}\big) \hspace{-6mm}\nonumber\\
=m_i\cdot \hspace{-1mm}\mbox{BD}(C_i)\cdot \mbox{BDI}(I)\cdot SK_{cc} + error_1\nonumber\\
\myeq m_i\cdot C_i \cdot SK_{cc}+ error_1\nonumber\\
%%\end{IEEEeqnarray}
%%\begin{IEEEeqnarray}{lll}
%=BD(C_i)\cdot (m_i\cdot BDI(I)\cdot SK_{cc} +error_1)\nonumber\\
=m_i\cdot \big(m_i\cdot \mbox{BDI}(I) + r\cdot PK_{cc}+E\big)\cdot SK_{cc}+ error_1\nonumber\\
=m^2_i\cdot \mbox{BDI}(I)\cdot  SK_{cc}+m_i\cdot r\cdot PK_{cc}\cdot SK_{cc} + m_i\cdot E\cdot SK_{cc}+error_1\nonumber\\
=m^2_i\cdot \mbox{BDI}(I)\cdot  SK_{cc}\hspace{-1mm}+\hspace{-1mm}m_i\cdot\big( r\cdot e_{cc}\hspace{-1mm}+\hspace{-1mm}E\cdot SK_{cc}\big)\hspace{-1mm}+\hspace{-1mm}error_1\nonumber\\
=m^2_i\cdot \mbox{BDI}(I)\cdot  SK_{cc}+error_2
\end{IEEEeqnarray}
where (a) is due to the fact that $\textrm{BD}(C_i)\cdot \mbox{BDI}(I_{N\times N})=I_{N\times N}~\cdot~C_{N\times 2}=C_i$.
Based on the errors $e_{cc}$ and $E$, we consider $error_1=\mbox{BD}(C_i)\big(r\cdot e_{cc}+E\cdot SK_{cc}\big)$ and $error_2=m_i\cdot\big( r\cdot e_{cc}+E\cdot SK_{cc}\big)+error_1$.
It should be observed that the decryption of $\textrm{BD}(C_i)\cdot C_i$ is of the form $m^2_i$.
It can be noted that by means of an accumulator-like procedure, we can multiply ciphertexts efficiently\cite{SHIELD}. The homomorphic correctness of depth-$3$ multiplication can be done similarly.

%CRT correctness
\textbf{CRT Correctness}: The Chinese remainder theorem for polynomials states that for the modulo $p_j(x)$, for $j=1,2,...,k$, with degree $deg(p_j)$, $D=\sum_{j=1}^{k}deg(p_j)$ and polynomials $d_{ij}(x)$ where $d_{ij}(x)=0$ or $deg(d_{ij}(x))\hspace{-1mm}<deg(p_j(x))$, there is a unique polynomials $P(x)$ with $deg(P(x))<D$ for which the following equation holds\cite{CRT}:
\begin{IEEEeqnarray}{l}
P(x)=d_{ij}(x)~\big(\bmod p_j(x)\big) ~\mbox{for}~j=1,2,...,k.
\end{IEEEeqnarray}
Using the parameters determined in the system setup phase like $Q(x)=\prod_{j=1}^{k}p_j(x), Q_j(x)=Q(x)/p_j(x)$, and by substituting $p_j(x)=Q(x)/Q_j(x)$ in $1/Q(x)=\sum_{j=1}^{k}T_j(x)/p_j(x)$, we have:
\begin{IEEEeqnarray}{lll}
\sum_{j=1}^{k}T_j(x) Q_j(x)=1.
\end{IEEEeqnarray}
Now, by considering $d_1, d_2, ..., d_k$ as the $k$-dimensional data, and based on the above analysis, we have:
\begin{IEEEeqnarray}{lll}
\sum_{j=1}^{k}d_j(x)T_j(x)Q_j(x)=
d_j(x) + \sum_{J=1}^{k}\big(d_J(x)-d_j(x)\big)T_J(x)Q_J(x)\nonumber\\
\equiv d_j(x)  ~~\big(\bmod p_j(x)\big) , ~~\mbox{for}~ j=1,2,...,k.
\end{IEEEeqnarray}
Therefore, to compute each dimension of $SM_i$'s aggregated data, CC could perform as follows:
\begin{IEEEeqnarray}{ll}
\sum_{i=1}^{\omega}m_i=\sum_{i=1}^{\omega}\bigg(\sum_{j=1}^{k}d_{ij}(x)~T_j(x)~Q_j(x)\bigg) \nonumber\\
~~~~~~~=\sum_{j=1}^{k}\bigg(\big(\sum_{i=1}^{\omega}d_{ij}(x)\big)~T_j(x)~Q_j(x)\bigg) \nonumber\\
~~~~~~~=\Big(\sum_{i=1}^{\omega}d_{i1}(x)\Big)T_1(x)Q_1(x)+\Big(\sum_{i=1}^{\omega}d_{i2}(x)\Big)T_2(x)\nonumber\\~~~~~~~~~~~Q_2(x)
+\cdots+\Big(\sum_{i=1}^{\omega}d_{ik}(x)\Big)T_k(x)Q_k(x) .
\end{IEEEeqnarray}
Then, CC could acquire the aggregation of each dimension with the following congruences:
\begin{IEEEeqnarray}{ll}
\sum_{i=1}^{\omega}m_i \equiv \sum_{i=1}^{\omega}d_{ij}(x) ~ \big(\hspace{-1mm}\bmod p_j(x)\big),~\mbox{for}~j=1,2,...,k.~~\hspace{+5mm}
\end{IEEEeqnarray}
Similarly, the CC could compute the congruences for the $\sum_{i=1}^{\omega}m^2_i,~\sum_{i=1}^{\omega}m^3_i$.
%temporal aggr.
\subsection{Temporal Aggregation}
The purpose of this phase is to compute an imperative aggregation named, temporal aggregation. As opposed to spatial aggregation computed in the prior phases to accomplish privacy of users' real-time data, temporal aggregation will be calculated regularly (every bill period)  to obtain the bill.
Given the timestamp $T_\gamma$ for $\gamma=1,2,...,b,b+1,...,2b,...,3b,...$; at the end of each bill period, 
%CC request GW to compute the temporal aggregation of smart meters' data. 
GW computes the temporal aggregation of each smart meter's data. Lets assume that we are at the end of the first bill period $T_b$.
%CC signs the $request_i$, $ID_i$, and $T_b$ with part of his  private key ($t_{cc}$) and sends it to the GW. Specifically, CC samples $v_{cc}\in R_q$ and computes $u_{cc}=(v_{cc}+H(request_i, ID_i, T_b))\cdot t_{cc}+t_2\cdot e'_{cc}$ where $t_2\in \mathbb{Z}^*_q$ and $e'_{cc}\leftarrow \chi$. Finally, CC sends the $request_i, ID_i, T_b$, and $\sigma_{cc}=(u_{cc},v_{cc})$ to the GW.
%After receiving these data, GW inspects two conditions $\sigma_{cc}\in R_q\times R_q$ and $[-b_{cc}\cdot+a_{cc}\cdot v_{cc}] \bmod t_2=-a_{cc}\cdot H(request_i, ID_i, T_b) \bmod t_2$, to find out whether signature is valid or not. If both conditions hold, 
 The GW aggregates the $SM_i$'s data relying on the data stored in its database connected to the $SM_i$'s identity ($ID_i$), and then signs the aggregated data as shown below:
\begin{IEEEeqnarray}{clll}
C_{Bi}=\sum_{\gamma=1}^{b}C_{i\gamma} ,\\
v_{GW}\leftarrow R_q ~, ~~u_{Gw}=\big(v_{GW}+H(C_{Bi},ID_i, T_b)\big)\cdot s_{GW}
+t_{GW}\cdot e'_{GW}.
\end{IEEEeqnarray}
Lastly, GW sends $C_{Bi}, ID_i, T_b$, and $\sigma_{GW}=(u_{GW},v_{GW})$ to CC. 
After verifying the timestamp and signature by checking the validity of two conditions:
\begin{IEEEeqnarray}{ll}
1)~\sigma_{GW}\in R_q\times R_q \\ \nonumber
2)~[-a_{GW}\cdot u_{GW}+b_{GW}\cdot v_{GW}] \bmod t_{GW}\nonumber\\~~~=-b_{GW}\cdot H(C_{Bi}, ID_i, T_b) \bmod t_{GW} 
\end{IEEEeqnarray}
then, CC decrypts the temporal aggregation in a similar way and obtains the bill.

%security analysis
\section{Security Analysis}
This section comprises of the analysis of three parts: privacy preservation, integrity and authentication, and resistance against other attacks. Concretely, as mentioned in the adversary model, assume there exists a strong adversary $\mathcal{A}$ who aims to obtain the consumption data by means of eavesdropping, using malware and attacking entities. Therefore, we focus on the privacy preserving assessment because it is of paramount importance for the smart grid network. In addition, we demonstrate that our proposed scheme can guarantee integrity of data, authenticity of sources and resist attacks carried out by the adversary.

%privacy
\textbf{Privacy Preservation}: There exists three components in the smart grid network who communicate the consumption reports through two links: SM-to-GW and GW-to-CC. Since, every reports transmitted through public channels, the strong adversary $\mathcal{A}$ inhabits in the customer-side of the network in order to eavesdrop the energy consumption data. However, the energy consumption data is a really small value and the adversary might try brute-force attack. Since our utilized homomorphic encryption scheme \cite{SHIELD} is IND-CPA secure (\textit{i.e.}, indistinguishability under chosen plaintext attack) and every message is encrypted, no adversary is capable of unveiling smart meters or aggregated data without the knowledge of CC's private key.

Let's imagine that the honest-but-curious entity, GW, desires to acquire the energy consumption data. Since, GW is simply a relay and its main obligation is to gather encrypted data and aggregate them homomorphically, GW cannot get any individual's data. Thus, the role of the GW as a relay will not disclose any private information about the transmitted data. Hence, any computationally strong entity can play GW's role.

Envision the brain of the smart grid, CC, wants to obtain the individual reports. The CC cannot attain any private information about the households, because we have designed the system model to be a data aggregation scheme and only aggregated data is received by the CC. Therefore, our system structure permits CC to achieve the bill and the aggregated data as it needs.

Now, consider that the strong adversary could utilize malicious malware in the CC or GW, or even he/she is potent enough to interfere with GW or CC's database. Evidently, we can infer from the above assessments, this adversary could not get any individual information, because the GW and CC themselves could not acquire the individual's energy consumption information. As a result of deploying a homomorphic encryption scheme which is IND-CPA secure and the fact that the CC's private key is distributed by TA over a secure channel, the adversary's attempts to achieve private information about individual users will fail or at best gives him/her aggregated reports' statistics. Hence, our proposed LPM2DA scheme preserves the confidentiality of users' information.

%integrity
\textbf{Integrity and Authentication}:
In our proposed scheme, in order to achieve integrity of communicated data, each entity must sign its reporting data. Specially, by relying on a collision resistance hash function and the unforgeability of the R-LWE based signature scheme \cite{signature} under chosen message attack, the integrity of each transmitted data is assured.
Recall that only TA allocates private keys via a secure channel and there is a straightforward equivalence between discovering private keys from transmitted data and solving the R-LWE problem.
Thus, since the users report the signatures alongside their data ($C_i, \sigma_i$), send the signature of aggregated data by GW ($C_{GW}, \sigma_{GW}$) and report the IDs in the bill period, it is impossible for any adversary to modify the reports without the private keys. 
Since the first step in every phase is to authorize each entity and its data and the invalid data would be discarded, the adversary cannot tamper the aggregated data and interrupt the system. Consequently, there is no way in which the adversary could forge the signature or modify the data without detection.

Since each entity has its own public and private keys assigned by TA over a secure channel, it would be fairly hard for the adversary to impersonate an entity. As a result of sending every reports with its signature, each entity will be authenticated before any further steps. Accordingly, data integrity and source authentication of users' data are provided.

%attacks
\textbf{Resistance against other Attacks}:
In addition to the criteria addressed above, our proposed scheme is resistant to various attacks and supports some practical properties. As mentioned earlier, no adversary could yield a valid signature of his/her reports to engage in fraud or corrupt the residential area's data. Consequently, the modification and impersonation attacks can be detected. Similarly, if the adversary resides in the links between SM-to-GW or GW-to-CC, the execution of an MITM attack will be detected by checking the legitimacy of each entity. The timestamp is used in every messages and signatures; hence the adversary could not carry out replay attack. Furthermore, whether the adversary comprises some smart meters or even occasionally smart meters do not send any data, the GW still can aggregate the received ciphertexts due to the employed homomorphic encryption scheme. Therefore, our proposed scheme achieves fault-tolerance.
In our proposed scheme, each GW is generally connected with $\omega$ number of users on average. We determined $\omega$ in a way that with a few increase or decrease in the number of users, the CC could decrypt the aggregated data appropriately. As a consequence, if a new SM needs to connect to a near GW, it would suffice to register himself to the TA to get his ID and key pairs. Thus, our LPM2DA scheme supports dynamic users.

%performance evaluation
\section{PERFORMANCE EVALUATION}
In this section, we aim to assess the computational performance and communicational overhead of our LPM2DA scheme. In order to show the efficiency of our proposed scheme, we compare its performance with schemes like EPPA\cite{EPPA}, MuDA\cite{MuDA}, FGDA\cite{FGDA}, Shen’s scheme\cite{Cube-data} and lattice-based schemes like PDA\cite{PDA} and Abdallah’s scheme\cite{Abdallah2018}. It should be noted that traditional schemes are not secure against quantum attacks. The lattice-based schemes in \cite{LRSPPP, AbdallahNTRU} are relying on forecasting the electricity demand and in their system model a secure database is required, which makes their system model different from DA schemes’ model. 
Therefore, the performance of these schemes are not fairly comparable. The Abdallah’s scheme utilizes a homomorphic encryption based on lattices. However, its system model considers the appliances before the smart meters, which is different from our considered model that is a typical system model in DA schemes. Hence, we assess the efficiency of LPM2DA and Abdallah’s scheme in a separate subsection.

%Comp Cost
\textbf{Computation Costs}:
Due to the fact that the cost of system setup phase is truly a one-time expense, we only focus on the cost of user report generation, report aggregation and secure report reading phase. In order to analyze the computational efficiency of our proposed scheme, we compare LPM2DA with various schemes described subsequently. The lattice-based DA scheme namely, PDA\cite{PDA}, which uses a somewhat homomorphic encryption and could transmit multi-dimensional data via an algorithm and it is capable of computing mean and variance of users’ data. The available traditional schemes are as follows: EPPA\cite{EPPA}, which uses Paillier cryptosystem and bilinear pairing and could transmit multi-dimensional data via super-increasing sequence; DA schemes like MuDA\cite{MuDA} which uses BGN cryptosystem and bilinear pairing, and could calculate functions like average, variance and one-way ANOVA on users’ data; and FGDA\cite{FGDA} which is also capable of calculating average, variance and skewness of users’ data; and finally, Shen’s scheme\cite{Cube-data} which also utilizes Paillier cryptosystem and bilinear pairing, and could transmit multi-dimensional data via Horner’s rule. It should be noted that the computational costs are from \cite{EPPA, LRSPPP, PDA, SHIELD}.

Based on PBC \cite{PBC} and MIRACL \cite{MIRACL} libraries running on a 3.0-GHz Pentium IV processor with 512MB memory, and considering $|N|^2=2048$ and a 160-bit cyclic group $\mathbb{G}$, the computation costs of exponentiation operation in $\mathbb{Z}_{n^2}$ which is denoted by $C_e$, 
group-based multiplication $C_m$, group-based exponentiation $C_{et}$, bilinear pairing $C_{bp}$ and Pollard's lambda method $C_{pl}$ are demonstrated in TABLE \ref{TabCOMP}.
%exponentiation cost
\begin{table}[h]
\centering
\captionof{table}{Computational costs~1.}
%\begin{adjustbox}{width=0.6\columnwidth,center}
\begin{tabular}{l|l}
Operation & Time (ms) \\ \hline
$C_{ez}$ & 12.4 ms \\ \hline
$C_m$ & 6.4 ms \\ \hline
$C_{et}$ & 8.4 ms \\ \hline
$C_{bp}$ & 20 ms \\ \hline
$C_{pl}$ & 18.3 ms ($\omega$ =50), \\
& 25.8 ms ($\omega$ =100) \\
& 31.5 ms ($\omega$=150) \\
& 36.48 ms ($\omega$=200)\\\hline\hline
\end{tabular}
\label{TabCOMP}
%\end{adjustbox}
\end{table}

Based on a 1126 MHz GPU with 4 GB memory and a 3.5 GHz core i7 5930K with 15 MB cache size CPU; and using the abbreviations Enc, Dec, Add, Mult for encryption, decryption, addition and multiplication, respectively, and the notation SH for somewhat homomorphic, the computational costs are denoted in TABLE \ref{TabCOMP2}.

%Lattice-based cost
\vspace{-3mm}\begin{table}[h]
\centering
\captionof{table}{Computational costs~2}
\label{TabCOMP2}
\begin{tabular}{l|l}
LPM2DA Enc (CPU) & $<$ 100 ms\\
LPM2DA (GPU) & $<$ 10 ms\\\hline
LPM2DA Dec (CPU) & $<$ 100 ms\\ 
LPM2DA (GPU) & $<$ 3 ms\\\hline
LPM2DA Add (CPU) & $<$ 1 ms \\
LPM2DA Add (GPU) & $<$ 0.1 ms \\\hline
LPM2DA Mult (CPU) & $<$ 104 ms \\
LPM2DA Mult (GPU) & $<$ 1.4 ms \\ \hline
LPM2DA Sign & 0.36 ms \\\hline
LPM2DA Verify & 0.57 ms \\\hline\hline
PDA SHEnc & 348 ms \\\hline
PDA SHDec & 26 ms \\\hline
PDA SHAdd & 1 ms \\\hline
PDA SHMult & 41 ms \\\hline\hline
\end{tabular}
\end{table}
To lessen the computation overhead, the multiplication can be done using the number theoretic transform (NTT) which is similar to fast Fourier transform (FFT) in terms of computational cost in $O(n \log n)$. Our homomrphic encryption scheme can exploit the parallelism when implemented on a GPU platform, which achieves extremely huge speed up compared to the implementations based on CPU, IBM HLib and PDA’s somewhat homomorphic encryption and other traditional encryption schemes.
In the case of using NTT \cite{NTT1, NTT2}, since each dimensional data in smart grid communication is typically small, transmitting $k$-dimensional data via Chinese remainder theorem could cost nearly $k\times30\mu$sec which is practically insignificant in contrast to encryption part of our scheme.
According to the structure of each scheme and the costs presented above, each smart meter’s cost is presented in TABLE \ref{TabSM}.

%SM cost
\begin{table}[h]
\centering
\captionof{table}{SM's computational cost}
\label{TabSM}
\begin{tabular}{l|l}
& SM   \\\hline\hline
LPM2DA (CPU) & Enc + Sign $<100$ ms \\ \hline
LPM2DA (GPU) & Enc + Sign $<10$ ms \\ \hline
PDA\cite{PDA} & SHEnc = $348$ ms\\ \hline
EPPA\cite{EPPA} & ($k$ +1)$C_{ez}$+$C_m$+$4C_p$ \\ 
&= $12.4k$ + $98.8$ ms\\\hline
MuDA\cite{MuDA}& $3C_{et}$+$C_m$ = $31.4$ ms \\\hline
FGDA\cite{FGDA} & $C_{et}$+$2C_m$ = $21.2$ ms \\\hline
Shen's scheme\cite{Cube-data} & $2C_{ez}$+$C_m$ = $46.4$ ms \\\hline
\end{tabular}
\end{table}

\begin{figure}[t]
\includegraphics[width=15cm, height=9cm]{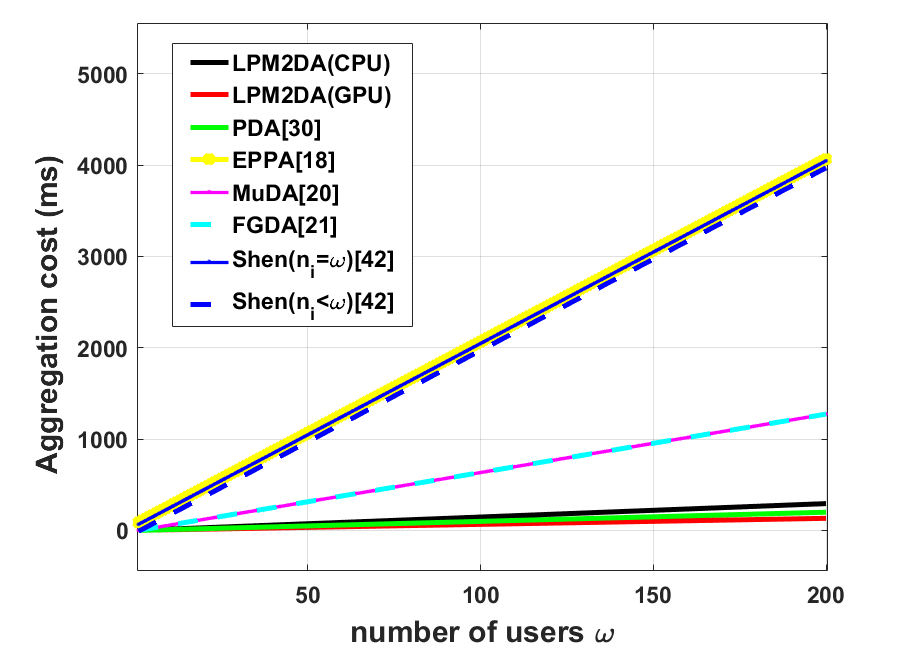}
\caption{Cost of aggregation computation.}
\label{Fig2}
\end{figure}

\begin{figure}[t]
\includegraphics[width=15cm, height=9cm]{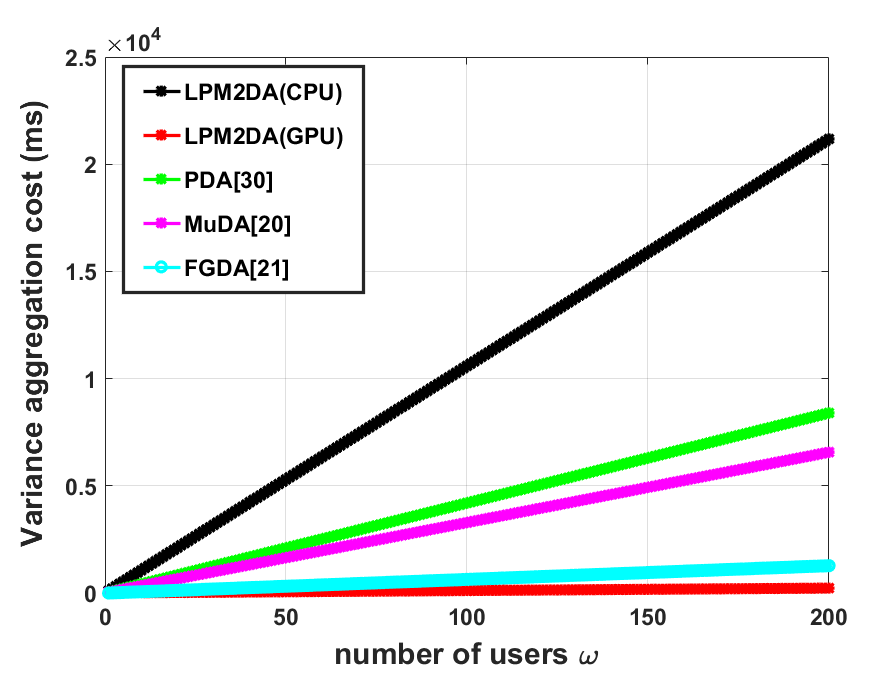}
\caption{Cost of variance aggregation computation.}
\label{Fig3}
\end{figure}

It should be noted that in Shen’s scheme\cite{Cube-data}, there exists two gateways: District gateway (DGW) and Residential area gateway (RAGW); and $n_i$ depicts the number of smart meters in the $i th$ residential area, where $n_i<\omega$. The GW’s computational cost is demonstrated in the table below \ref{TabGW} and the clear comparison of average and variance aggregation's costs are depicted in Fig.\ref{Fig2} and Fig.\ref{Fig3}. 

% GW cost
\begin{table}[h]
\centering
\captionof{table}{GW's computational cost}
\label{TabGW}
\begin{tabular}{l|l|l}
& GW (Aggr.) & GW (Variance Aggr.)   \\\hline\hline
LPM2DA & ($\omega$-$1$)Add+$1$Sign & ($\omega$-$1$)Add+$\omega$Mult \\ 
&+$\omega$ Verification&\\\hline
PDA\cite{PDA}& ($\omega$-$1$)SHAdd &($\omega$-$1$)SHAdd+$\omega$SHMult\\\hline
EPPA\cite{EPPA}& ($\omega$+$3$)$C_{bp}$+$C_m$ &---------\\ \hline
MuDA\cite{MuDA} & ($\omega$-$1$)$C_m$ & 2($\omega$-$1$)$C_m$+($\omega$+$1$)$C_{bp}$ \\\hline
FGDA\cite{FGDA} & ($\omega$-$1$)$C_m$ & ($\omega$-$1$)$C_m$ \\\hline
Shen's&RAGW:($n_i$+$2$)$C_{bp}$ &---------\\
scheme&$+$($\omega-n_i$)$C_{ez}$+$C_m$&\\
\cite{Cube-data}&DGW:($\omega_2$+$2$)$C_{bp}$&\\
&+$C_m$ &\\\hline
\end{tabular}
\end{table}

 %Fig2

%Fig3

For the homomorphic multiplication part, our encryption scheme uses the components of ciphertext $C_i$ and multiplies it by $C_i$’s bit-wise decompositions, which means it is asymmetric based on the input ciphertexts. Thus, due to the asymmetric error growth depending on two  subsequent ciphertexts, this multiplication method gives a slow noise growth, which is clearly better than PDA's symmetric noise growth. Hence, it is clear that our scheme not only has lower computation cost, but it also can support more users in a residential area.
Although the CC is a potent entity, we analyze CC’s computational costs in obtaining the aggregation and computing the average and variance of the users' data in TABLE \ref{TabCC}.

%CC cost
\begin{table}[h]
\centering
\captionof{table}{CC's computational cost}
\label{TabCC}
\begin{tabular}{l|l|l}
& CC (Aggregation) & CC(Variance)\\\hline\hline
LPM2DA & Dec $<100$ms & Dec $<100$ms \\
(CPU) & &\\\hline
LPM2DA & Dec $<3$ms &    Dec $<3$ms \\
(GPU) &&\\\hline
PDA\cite{PDA}& SHDec $=26$ms & SHDec $=26$ms\\ \hline
EPPA\cite{EPPA}& $2C_{bp}$+$C_{ez}$+$4C_m$&---------\\
&+$C_{et}=86.4$ms&\\\hline
MuDA\cite{MuDA} & $C_{et}$+$C_{pl}=34.3$ms & $2C_{et}$+$2C_{pl}=68.5$ms  \\\hline
FGDA\cite{FGDA} & $C_{et}$+$C_m=14.8$ms & $C_{et}$+$C_m=14.8$ms \\\hline
Shen's\cite{Cube-data} & $2C_{bp}=40$ms&---------\\\hline
\end{tabular}
\end{table}

Due to the evaluation performed earlier, we can safely conclude that in terms of computation cost and calculating average and variance aggregation, our proposed scheme is efficient in CPU and achieves the lowest computation cost with GPU. Specially, considering PDA’s algorithm and EPPA’s super-increasing sequence and the Horner’s rule utilized in Shen’s scheme, Chinese remainder theorem provides the handling of $k$-dimensional data more efficiently.

\textbf{Comparison with Abdallah's scheme \cite{Abdallah2018}}:
Since the scheme \cite{Abdallah2018} is a DA scheme with distinct system model, in this subsection we aim to compare LPM2DA with this lattice-based DA scheme based on computational costs. If we consider each household appliance's data in \cite{Abdallah2018} equivalent to each component of data in our scheme, we can practically compare the schemes. Considering $20$ appliances in each HAN, the smart meter and appliances in Abdallah's scheme need nearly $160$ ms for computations\cite{Abdallah2018}, while the smart meter in our scheme needs less than $100$ms (with CPU) and $10$ms (with GPU) to compute the 20-dimensional data. It should be noted that based on the data presented in \cite{Abdallah2018}, we can see that the key size of \cite{Abdallah2018} is $2.95$ Mb while our scheme's key is $63.4$ kb, which is significantly better. Therefore, our LPM2DA is much more efficient compared to Abdallah's scheme\cite{Abdallah2018} in terms of computational and communication costs.

\textbf{Communication efficiency}: In order to analyze the communication performance of each scheme in smart grid networks, we should consider two parts: communication between SM to GW and between GW to CC. Since GW and CC are powerful entities, the communicational efficiency of a DA scheme relies on the efficiency of smart meters’ performance. Specifically, to compare the efficiency of each scheme, we utilize the \textit{expansion metric} as formulated below:
\begin{IEEEeqnarray}{ll}
\mbox{Exp}=\frac{\mbox{bits transmitted}}{\mbox{data (bit)}}=\frac{1}{\mbox{Eff}}
\end{IEEEeqnarray}

To be accurate and fair, we compare the efficiency of LPM2DA with schemes which could transmit multi-dimensional data, such as PDA\cite{PDA}, EPPA\cite{EPPA} and Shen’s scheme\cite{Cube-data}. Since EPPA and Shen’s scheme are using Paillier cryptosystem and expand the $k$-dimensional data to $2\log(\mathcal{N})$, their expansions are better than lattice-based schemes like LPM2DA and PDA, but they are significantly slower and do not have quantum security. Moreover, the EPPA utilizes super-increasing sequence and Shen’s scheme utilizes Horner’s rule to transmit multi-dimensional data. Therefore, those schemes lose computational efficiency to gain $k$-dimensional data. For communication comparison with PDA, the expansions are presented in the TABLE \ref{TabComm}.
%Comm overhead
\begin{table}[h]
\centering
\captionof{table}{Communication overhead}
\label{TabComm}
\begin{tabular}{l|l}
& bits transmitted/data\\\hline\hline
LPM2DA & $\frac{(N\times 2)\times n\times \mbox{log}(q)}{k\times p_j\times \mbox{log}(q)}$\\\hline\hline
PDA\cite{PDA} &$\frac{2\times n\times \mbox{log}(q)}{k\times \mathcal{K}_1}$\\\hline\hline
%EPPA & $\frac{k\times \mathcal{K}_2}{2\times \mbox{log}(\mathcal{N})}$\\\hline\hline
%Shen & $\frac{k\times \mathcal{K}_3}{2\times \mbox{log}(\mathcal{N})}$\\\hline\hline
\end{tabular}
\end{table}

In TABLE \ref{TabComm}, $k$ denotes the number of dimensions of data. The PDA utilizes an algorithm to achieve $k$-dimensional data in which every component of data can be shown with $\mathcal{K}_1=10$ bits as mentioned in \cite{PDA} and for $128$-bit security, $n=1024$ and log($q$)$=58$. It should be noted that based on PDA's algorithm with $\mathcal{K}_1=10$, the $k$-dimensional data is practically limited to $k=4$. However, LPM2DA achieves $k$-dimensional data by deploying polynomial CRT. Technically, in our scheme, log($q$)$=31$, and each component of data is a polynomial with average degree  $p_j=100$ and the coefficients are of size log($q$) and there exists no limit on the dimension of data $k$.
Therefore, considering our proposed scheme’s quantum security and computational competence, LPM2DA is acceptable in communication overhead comparing to traditional schemes like EPPA\cite{EPPA} and Shen’s scheme\cite{Cube-data}, and it achieves better communication efficiency in contrast to lattice-based schemes like PDA and Abdallah’s scheme.

%conclusion
\section{CONCLUSION AND SUMMARY}
Due to the importance of smart grid network's privacy, in this scheme, we have introduced a secure lattice-based multi-functional and multi-dimensional data aggregation scheme (LPM2DA). In the situations where networks are in imminent danger of quantum attacks, our LPM2DA scheme not only preserves privacy of users’ consumption data, but also ensures data integrity and source authentication. Moreover, our LPM2DA scheme maintains fault-tolerant, allows dynamic users and is resistant against various attacks.
Based on homomorphic encryption scheme and Chinese remainder theorem, the control center could acquire temporal and spatial aggregations of users’ multi-dimensional data efficiently; and it is capable of calculating various statistical functions like mean, variance and skewness. Finally, we have demonstrated the computational and communication efficiency of our proposed scheme via comparing its performance with other data aggregation and lattice-based schemes.

%-------------------------------------------------------------------------------------
% SPDBlock
%------------------------------------------------------------------------------------
\chapter{SPDBlock: A Secure Privacy preserving Data aggregation scheme for Blockchain based smart grid}
\lhead{\emph{SPDBlock}} % Set the left side page header to "SPDBlock"
The sketch of this chapter is formed as follows. In section 4.1, we illustrate the network model, attack scenarios and delineate security requirements and the scheme’s goals. Section 4.2 introduces the basic notations and assumptions of blockchains and Elliptic Curve cryptography. The explanation of our proposed SPDBlock scheme is presented in section 4.3. We thoroughly analyze the security and evaluate the performance of our scheme separately in section 4.4 and 4.5, and then present  its summary within section 4.6.

%system model
\section{SYSTEM MODEL}
In this section, we depict the network model and describe all of its entities comprehensively. Afterward, we delineate the adversary model, the security requirements, and our goals.

%network model
\subsection{Network Model}
\begin{figure}
\includegraphics[width=15cm, height=10cm]{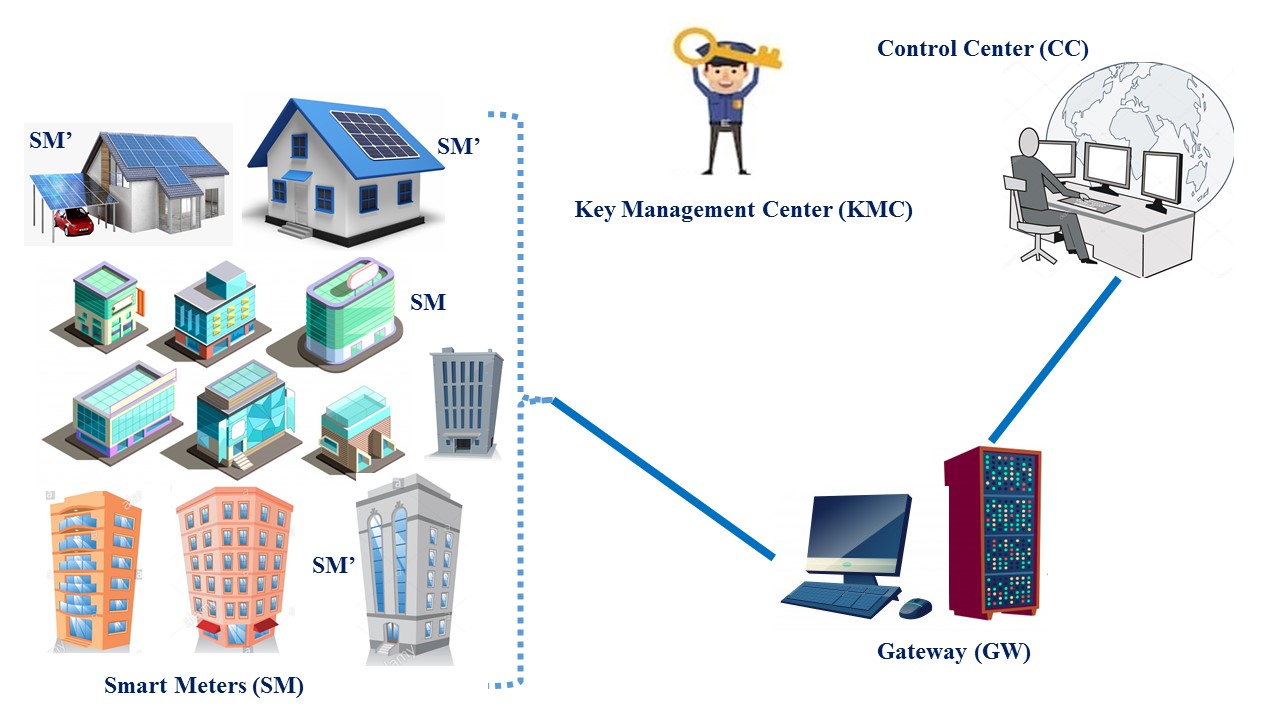}
\caption{SPDBlock's System Model}
\label{FigBC}
\end{figure}

Typically, to preserve privacy and gratify other security criteria in smart grid, the following entities are accounted in most network models: The control center (CC) who is in charge of administration and analysis of the grid; The users which primarily consumes energy and in some cases, they could generate electricity; An entity named, gateway (GW) which inhibits between the customer-side of the network and the authorities like CC, to play the role of a relay; And lastly, in some schemes there exists an authority accounted for security settings. As depicted in FIGURE \ref{FigBC}, our network model is also comprised of one control center (CC), one building area network gateway (GW), a trusted key management center (KMC), and home area networks (HAN) equipped with smart meters (SM). In addition, the $SM’$s denote computationally potent smart meters which will be introduced later. In our proposed scheme, we focus on preserving privacy and do not consider any communication challenges like data loss.

Fundamentally, our proposed scheme's network operates based on blockchain. In fact, we form two blockchains: the first blockchain namely, Sidechain is accounted for the customer-side network. The Sidechain is created by a number of approved smart meters ($SM’$) and contains individual energy consumption along with the aggregated data. The second blockchain which the whole smart grid network relies upon is constructed by the GW in parallel with the Sidechain based on the aggregated consumption data and general information of the network. It should be noted that only customer-side entities like users and the GW have access to sidechain’s blocks. However, every node (entity) in the network could observe Mainchain’s blocks.

\textbf{CC}: In most countries, the control center is normally administered by an organization affiliated with the government. Generally, controlling the grid balance, administering energy generation, and analyzing the consumption data are CC’s main responsibilities. To put it in a nutshell, settling any wrongdoings and every management decision is made by the CC.

\textbf{BAN}: The building area network is a vast geographical area or a huge building comprised of numerous units or apartments. Each BAN has a gateway (GW) to relay information between users and the CC. Usually, users and the GW communicate through economical WiFi technology; and high bandwidth or wired links are used to link the GW and CC. In our proposed scheme, the GW is the only generator of Mainchain’s blocks. In order to help the CC and to distribute the network’s computational overhead, the GW burdens itself with the calculation of each bill.

\textbf{HAN}: Basically, the home area network is those units or apartments mentioned in the BAN subsection. Each HAN has a smart meter to garner the real-time consumption data. Specifically, due to the existence of different types of appliances in each house and the fact that HANs could generate electricity besides their consumption, smart meters collect those multi-dimensional data in specified periods like every 15 or 30 minutes and transmit them to the network authorities. In our proposed scheme, a set of potent smart meters which are denoted by $SM’$ in Fig1, are in charge of data aggregation. Expressly, relying on communication and computational power, these $SM’$s are considered potent enough to play the role of the aggregator and generating Sidechain’s blocks.

\textbf{KMC}: The key management center is the only trustworthy entity in our proposed scheme which is in charge of network initiation, generation of key pairs, and establishing general parameters. It should be noted that the KMC only participates in the system setup phase and it will be offline for the rest.

%adversary model
\subsection{Adversary Model and Security Requirements} 
Typically, the privacy preservation schemes take entities like CC or GW as honest-but-curious  \textit{i.e.} they are inquisitive to figure out individual consumptions but they do their job and do not act maliciously. In the real world where each entity could get attacked or act maliciously, considering main authorities like the CC, GW, and $SM’$s to be honest-but-curious is not wise and realistic. 
In our proposed scheme, the GW and $SM’$s which construct the Mainchain and Sidechain’s blocks respectively, are considered to be curious and not necessarily honest. For this reason, in comparison with other privacy preservation schemes that have entities who could play their parts honestly, our proposed scheme is more pragmatic and practical. 

Although smart meters are not as easy to tamper with as traditional meters, they could get attacked or hacked. It should be noted that the smart meters in our proposed scheme runs based on smart contract codes wherein even the writer of the codes cannot be at an advantage. Due to the smart contract codes, smart meters could not act mischievously, otherwise, it will be detected and they’ll get caught and penalized accordingly. 

As expected, the control center which is the prime authority who analyzes the grid information should only receive the aggregated consumption data. To make our scheme more akin to the real-world situations, we consider the control center’s disgruntled employees as malicious parties who launch passive or active attacks such as differential attacks or collusion with other entities to gain access to the individual consumption data.

Concretely, there are strong adversaries $\mathcal{A}$ in the smart grid network trying to disclose privacy. Due to eavesdropping and accessing transmitted data or generated blocks, the adversaries primarily focus on disclosing individual consumption data and practical information. In addition to privacy, some adversaries may take advantage of the destruction of other security criteria like integrity and authentication. More seriously, the adversary $\mathcal{A}$ not only could eavesdrop and attack passively, but also could launch active attacks such as replay, modification, impersonation, denial-of-service (DoS), and man-in-the-middle (MITM) attacks. 

So as to preserve privacy in the customer-side network and satisfying other security standards, the following security requirements should be achieved thoroughly:

\textbf{Privacy preservation}: In the smart grid network, relying on the users’ prospect, the foremost prerequisite is privacy. Concretely, due to the frequent transmission of data, the customer-side users are deeply concerned about the individual consumption data against the adversary and other entities residing in the network. Additionally, even the aggregated consumption data should not be divulged to anyone besides the control center who needs these data for network analysis and grid balance.

\textbf{Integrity and authentication}: Since the data and blocks are transmitted via public connections, the transmitter ought to prove his identity and provide information to substantiate the integrity of its data. Besides, attacks like MITM, impersonation, and modification will be hampered by the ensuring of integrity and authentication. 

\textbf{Resistance against attacks}: Smart grid network is exposed to various attacks and different types of adversaries. To be specific, numerous outsider and insider attackers are carrying out attacks such as replay, differential, DoS, MITM attacks. Furthermore, owing to the utilization of blockchain for the smart grid infrastructure, blockchain-based attacks like double spending and $51\%$ attack will be added to the mentioned attacks.

%design goal
\subsection{Design Goal}
Considering the system model and security requirements, our proposed scheme chiefly concentrates on preserving privacy and guaranteeing integrity and authentication. The in-depth description of our design goals are as follows:

\textbf{Privacy-preserving}: Actually, as the adversary or even insider entities like CC could try to find out individual consumption data via differential attacks or collusion. Besides privacy preservation of users’ individual and aggregated consumption data, we consider other types of privacy like differential privacy and resistance against collusion. Moreover, despite data aggregation schemes that utilize homomorphic encryption systems that enable the CC to decrypt any individual data, our proposed scheme should only grant the decryption of aggregated data and provides individual data privacy. 

\textbf{Data integrity and source authentication}: The integrity of each data and every block should be verified to hinder the growth of faulty blockchain. Furthermore, in each phase, the transmitter of the data or the constructor of blocks should be authenticated. 

\textbf{Multi-dimensional data}: In the smart grid network which the users’ data are in fact diverse, transmitting several data with one proper multi-dimensional data would make the system works more efficiently and enables the CC to analyze the grid exhaustively.

\textbf{Detection and prosecution}: Since our system model is realistic and all the responsible authorities could act maliciously, our proposed blockchain-based scheme should enable detection of wrongdoings and defines a procedure to punish the malicious actions and encourage the prover of faulty behavior. 

\textbf{Temporal and spatial aggregation}: Regardless of spatial aggregation which the CC needs for grid analysis, in order to calculate the bill in a privacy-preserving manner, our scheme should be able to compute temporal aggregation of each users’ consumption.

\textbf{Resistance against attacks}: Since the communication is through open connections, our proposed SPDBlock scheme should be capable of resistance against attacks like replay, DoS, impersonation, modification, and MITM attacks.

\textbf{Fault-tolerance}: As in the real-world situations where smart meters could crash, be hacked, or get manipulated by the users, it would be crucial for a privacy-preserving scheme to remain fault-tolerant and support dynamic users.

\textbf{Efficiency}: An essential quality of any proposed scheme is to preserve privacy and ensure other criteria in an efficient manner with small computational costs and lower communication overhead.

%priliminaries
\section{PRELIMINARIES}
\subsection{Blockchain}
The advent of blockchain which is now integrated into various areas such as cryptocurrencies, financial services, energy trading, and communication systems was after Satoshi Nakamoto’s paper \cite{Nakamoto}. Concretely, the term ``blockchain” describes its true nature which is a chain of blocks containing information that could be transactions or any type of data. Technically, blockchain consists of two integral elements: distributed network and cryptography. In essence, blockchain aims to resolve centralized networks' problems such as single point of failure, prone to intimidation, strain, and mistrust. More seriously, blockchain technology empowers the majority of system users with the responsibility of network leadership and supervision\cite{Bitcoin}. 

Generally, blockchain falls within three main classifications: public blockchain, private blockchain and consortium blockchain. The public blockchain as used in the renowned cryptocurrency, Bitcoin, is a permissionless network in which every node could read and write anonymously. Conversely, the private blockchain is a permissioned network wherein every known identity (nodes) which has been screened by the owner of the network could participate in reading and writing in the blockchain. And eventually, the consortium blockchain is a chain with characteristics between private and public blockchains. In a consortium blockchain which is a permissioned network owned by multiple entities, all the pre-approved nodes could read the blocks and only a few nodes are allowed to write. What's more, the consortium blockchain is faster and lighter than the public blockchain; and in contrast to a private blockchain, allows multiple organizations to operate on the same network \cite{SoK}. 

Despite centralized networks which are governed by the rules of their owners, the blockchain’s rules and confirmation of block data are provided by the consensus mechanisms. The most well-known mechanisms are proof of work (PoW) which is utilized in Bitcoin and proof of stake (PoS) which is going to be employed in the new Etherium blockchain. Generally, by solving a puzzle in the PoW mechanism, participants prove that they have the required computational powers. To be exact, participants calculate different hashes based on a predefined target and difficulty, in which the only way to reach the target is the alteration of block’s data or nonce. However, in the PoS mechanism, the participants stake their money in the blockchain and the miner of each block gets selected by a decentralized random number generation (DRNG) scheme. Specifically, these DRNG schemes work relying on the participants’ reputation in the blockchain and the amount of money they have staked. Practically, blockchain schemes are mainly perform based on the prominent PoW and PoS mechanisms. Nevertheless, in some specific infrastructure or applications, they make use of other mechanisms like proof of activity (PoA), proof of space (PoSpace), and proof of efficiency (PoE) \cite{Bitcoin}.

One of the most important challenges in the blockchain is their scalability which means we have to alter the blockchain structure or devise a novice consensus mechanism to obtain an improved and efficient network for the real world applications. In order to solve the scalability challenge, some techniques like alteration of the block structure, reducing the intervals between blocks, changing nodes’ roles in the network, and utilization of payment channel networks have been presented. However, our proposed SPDBlock scheme employs a similar framework utilized in one of the scalability solutions for blockchain namely, the Plasma scheme \cite{Plasma}. Precisely, considering the fact that central networks have a lot more speed and efficiency comparing to decentralized networks, we implemented a central network that is administered via a decentralized network. In fact, we have two blockchains in which participants could monitor the blocks and prove any wrongdoings to the administering network and get rewarded \cite{Bitcoin}.
%ECC
\subsection{Elliptic Curve}
Elliptic curve cryptography is a significant concept introduced by Neal Kobiltz \cite{Kobiltz} and Victor S. Miller \cite{Miller} independently in 1985. The reason why Elliptic curve cryptography has gotten so much attention in recent years is because of its expediency and the fact that it provides an equal security level with shorter parameters and keys. For this reason, and because all nodes must store the whole blockchain in their database, the utilization of Elliptic curve cryptography is prevalent in blockchain-based schemes. 
The Elliptic curve equation deployed in our proposed SPDBlock scheme is: 
\begin{IEEEeqnarray}{lll}
E: y^2=x^3+ax+b \pmod{q} ~,\\
\mbox{wherein}~ a, b, x, y \in GF(q)~\mbox{and}~ 4a^3+27b^2\neq 0 \pmod {q}~. \nonumber
\end{IEEEeqnarray}
Furthermore, to map the values to points on the curve, the following multiplication will be computed:
\begin{IEEEeqnarray}{ll} 
d \in GF(q) \Longrightarrow D=d\cdot G ~\in E~,
\end{IEEEeqnarray}
where G is the generator point and $``\cdot"$ denotes point multiplication.

%SPDBlock
\section{The Proposed Scheme: SPDBlock}
In this part, we demonstrate our proposed SPDBlock scheme which is constructed of seven phases: system setup, miner selection, user report generation, Sidechain’s block generation, Mainchain’s block generation, secure report reading, and bill calculation. Succinctly, we delineate our proposed SPDBlock scheme.

%overview
\subsection{Overview}
At the beginning, the KMC initiates the system via generation and distribution of key pairs and public parameters, respectively. Since distributed decryption requires the cooperation of at least a predetermined number of users, the KMC computes each SM’s key pairs alongside the subsequent customer-side’s group-key (GK) based on different Elliptic curves and the secret sharing technique. Moreover, in order to transmit multi-dimensional data, KMC forms the Chinese remainder theorem’s parameters and publishes all the public parameters in the network. 

In spite of other blockchain-based schemes in which every SM broadcasts its data to all other SMs in the network and then the miner gets selected accordingly, to minimize the communication and computational overhead in our proposed scheme, the miner selection phase occurs before user report generation and transmission. Specifically, based on a truly random number formed by all the $SM’$s’ contributions, one $SM’$ gets selected. Afterward, all the smart meters in the customer-side network encrypt their multi-dimensional data and transmit them to the predetermined miner. Additionally, to help the last block's decryption process, acknowledgment of the last block's validity, and ensuring integrity and authentication, the smart meters compute an auxiliary ciphertext, calculate one hash and sign these data, respectively, and transmit them to the miner along with their consumption ciphertexts. Subsequently, the miner aggregates valid ciphertexts and constructs the customer-sidechain’s block based on each SM’s report. Then, the miner publishes the new block in the customer-side network. Now, each SM can verify the validity of each block based on the information it contains, and does one of the two procedures: If he approves the block, he will add this block to his customer-sidechain and sends an acknowledgment (ACK) the next round; but, whether his data is not included in the block or the block's data are erroneous, he will send proof of these wrongdoings to the GW directly and transmits NACK the next round.

We should note that our proposed scheme operates based on Mainchain’s blocks which are created by the GW. In fact, Mainchain is a blockchain to which every entity in the network has access and contains the general and practical information of the customer-side’s network. While Sidechain contains individual consumption data, the Mainchain blocks consist of the aggregated consumption data, the bill, all the CC’s advice, and the proof of any wrongdoings. However, as the GW is the only creator of Mainchain blocks, the Sidechain's miners could monitor those blocks thoroughly and report any misbehavior to the CC directly. Essentially, regulation of transmitting the proofs of any wrongdoings or misbehavior is based on the smart contract.

Finally, CC decrypts the acknowledged block’s aggregations based on auxiliary ciphertexts and obtains each dimensional-data’s aggregation by utilization of CRT’s properties. Moreover, after a predefined number of blocks, the GW calculates the bill ciphertext relying on the acknowledged data presented in the customer-sidechain’s blocks, and similarly, with the help of smart meters, the CC could decrypt the bill ciphertexts.

%system setup
\subsection{Phase 1: System Setup}
In this phase, KMC is the main responsible for system initialization and configuration of parameters. First, KMC determines the Chinese remainder theorem’s parameters to facilitate the smart meters in the transmission of multi-dimensional data. Then, relying on the utilization of the Elliptic curve and secret sharing technique, KMC generates each entity’s key pairs and computes the customer-side’s group key. 

Concretely, to make use of CRT for $L$-dimensional data transmission, the KMC selects $L$ relatively prime positive integers $m_1, m_2,\dots, m_L$. Subsequently, based on these $L$ integers, the following parameters will be defined:
\begin{IEEEeqnarray}{lll}
M=m_1\cdot m_2\cdot \dots \cdot m_L ~~\mbox{where}~ M<q ~,\\ \nonumber
M_{\ell}=M/m_{\ell} ~\mbox{and}~ y_{\ell}=M_{\ell}^{-1}\pmod{m_{\ell}} ~\mbox{for}~ \ell=1, 2, \dots , L .
\end{IEEEeqnarray}
Due to the employment of Elliptic curve $E(F_q)$ with equation $E: y^2=x^3+ax+b \pmod{q}$ wherein $a, b, x, y \in GF(q)$ and $ 4a^3+27b^2 \neq 0 \pmod {q}$, the KMC generates each entity’s key pairs for encryption and signature. Let us begin with CC’s key pairs. Relying on the notations and system parameters introduced above, the KMC chooses a random number $s_{CC}\leftarrow {\mathbb{Z}_q}^*$ as the CC’s private key $SK_{CC}=s_{CC}$. Subsequently, based on the chosen private key, it yields the public key by computing $PK_{CC}=\mathcal{S}_{CC}=s_{CC}\cdot G$. 

To utilize distributed decryption technique in our employed encryption scheme, the KMC deploys the secret sharing technique to create smart meters’ key pairs and customer-side’s group key (GK). At first, it chooses a random number $s_X$ from ${\mathbb{Z}_q}^*$. Then, it computes the customer-side’s group key as $GK=\mathcal{S}_{GK}=s_{GK}.G$ wherein $s_{GK}$ is the addition of CC’s private key $s_{CC}$ and the chosen $s_{X}$ \textit{i.e.}, $s_{GK}=s_{X}+s_{CC}$. At last, because at least $\mu$ number of smart meters would suffice to assist the CC in the decryption of aggregated data, the KMC produces smart meters’ key pairs by  the employment of secret sharing technique. Precisely, by defining an equation $f(t)$ with degree $\mu-1$ and setting the constant value to be $s_{X}$, each SM’s private key will be calculated as shown below:
\begin{IEEEeqnarray}{lll}
f(t)=s_{X}+a_1t^1+a_2t^2+\dots +a_{\mu-1}t^{\mu-1} ,\\
\mbox{where}~\mbox{$a_i$}\in {\mathbb{Z}_q}^* ~\mbox{for}~ i=1, 2,\dots,\mu-1 ,\nonumber\\
\mbox{and}~s_i:=f(i) ~\mbox{for}~ i=1, 2, \dots, n .
\end{IEEEeqnarray}
in which $SK_i=s_i$ for $i=1, 2,\dots, n$ are smart meters’ private key and their public key will be computed by the following multiplication: $PK_i=\mathcal{S}_{i}=s_i\cdot G$.

Similarly, to create GW’s key pairs, the KMC chooses a random number $s_{GW}\leftarrow \mathbb{Z}_q$ as the GW’s private key $SK_{GW}=s_{GW}$ and yields the corresponding public key by computing $PK_{GW}=\mathcal{S}_{GW}=s_{GW}\cdot G$.

Eventually, for the miner selection phase and the employed signature scheme, the KMC selects two secure cryptographic hash functions $H_1:\{0,1\}^*\rightarrow \mathbb{Z}_q^*$ and $H_2:\{0,1\}^*\rightarrow \mathbb{Z}_p$. In the end, KMC distributes each entity’s private key through a secure channel and publishes the public keys and all the parameters  in the network.

%Miner selection
\subsection{Phase 2: Miner Selection}
In the consortium blockchain employed in our proposed scheme, there are some powerful nodes ($SM'$s) who are capable of being the aggregator and the miner of Sidechain’s blocks. Since our proposed scheme’s blockchain operates based on proof of stake (PoS), miners must stake some amount of money in the network. This money which is more than a threshold could be linked to the blockchain via a bank account. The practical criteria with which we define a smart meter as powerful are: computational powers, strong hardware, storage space, its geographical location, and the money it stakes in the network. In the real world, these potent smart meters are typically large buildings, industrial or commercial buildings, and houses that could generate electricity alongside their consumption. 

Being a miner has many advantages which the least of them are low bill cost and regular maintenance of their equipment. It should be noted that each miner would be rewarded as much as it helped the network and the number of correct blocks it made. Nevertheless, although being a miner has many advantages, if a miner misbehaves or tries to deceive the network, it will be penalized based on the extent of its wrongdoing. What's more, the misbehaving miner's stakes will be burned and its ID will be removed from the miners' list. 

Generally, the miner of each round will be selected based on a random number  created by every $SM’$s' contribution. Concretely, due to $SM’$s' reputation in the network, the money they staked, and other criteria cited before, each $SM’$ has given a unique interval in the range of $0$ to $p$, in a way that all these intervals cover the whole range of $0$ to $p$. 

%Miners' intervals
\begin{table}[h]
\centering
\captionof{table}{Miners intervals}
\begin{tabular}{|l|c|c|c|c|c|r|}
\hline 0 & ~~~$SM'_1~~~$&$SM'_2$&$~~~~~~~SM'_3~~~~~~~$&$\dots$&$SM'_J$&p \\ \hline
\end{tabular}
\label{Tab}
\end{table}
Before each report generation and transmission phase, $SM’$s cooperate to achieve a random number in the range of $0$ to $p$. Then, relying on the $SM’$s' intervals and the created random number in the range of $0$ to $P$, the miner will be selected. As you can see in TABLE \ref{Tab}, while any normal smart meter could request to be on the miners’ list and may pass the minimum criteria, because of their low computational powers and the small amount of money they could stake, the probability of being chosen as a miner is pretty slim. 

To achieve this random number, we use a technique named, hash onion. Assume there is a predefined and publicly known constant integer $\theta$. In this technique each $SM’_j$ for $j=1, 2, \dots, J$ chooses a random number $\alpha_j$ and builds a sequence of hashes as follows:
\begin{IEEEeqnarray}{lll}
\alpha_j,~ h_{1j}=H_2(\alpha_j),  ~h_{2j}=H_2(h_{1j}),\nonumber\\
\dots, h_{(\theta-1)j}=H_2(h_{(\theta -2)j)}, ~h_{\theta j}=H_2(h_{((\theta-1)j}).
\end{IEEEeqnarray}
Then, each $SM’_j$ for $j=1, 2, \dots, J$ transmits the last number of the above sequence to the other $SM’$s in the miners' list as its random number. Note that only in the first round of communication the $SM’_j$ requires to sign his random number (the hash value) for participating in the miner selection phase; and after that, all the $SM’_j$ needs to do is to send his successive random numbers. The reason for which $SM’$s do not need to send the signature in every round besides the first round is based on the fact that their hash values are verifiable in the second round and the rounds after that. So, especially in the first round, $SM_j$ signs $h_{\theta j}$ with his private key as shown below\cite{Schnorr}:
\begin{enumerate}
\item He picks $v_j\leftarrow \mathbb{Z}_q$ randomly and computes $V_j=v_j\cdot G$.
\item Then, he sets $e_j=H_1(V_j, h_{\theta j}, TS)$.
\item $\sigma_j=[u_j=v_j+e_js_j\pmod{q}, V_j]$.
\end{enumerate}
And at last, he sends $h_{\theta j}, \sigma_j$ and $TS$ to the $SM’$s in the miners' list. After a predefined time-lapse, each $SM’_j$ verifies the signature of each random number. Particularly, to reduce the computational overhead, he uses the batch verification property of the employed signature scheme to verify the signatures:
\begin{enumerate}
\item he sets $b_1=1$ and chooses $b_k\leftarrow \mathbb{Z}_q: k=2, 3,\dots, J$ and $k\neq j$.
\item then he sets $e_k=H_1(V_k, h_{\theta k}, TS) : k=1, 2,\dots, J$ and $k\neq j$.
\item at last, he evaluates $\big(\sum_{k=1}^{J}b_ku_k\big)\cdot G=\bigoplus_{k=1}^{J} b_kV_k\oplus b_ke_k\mathcal{S}_k$.
\end{enumerate}
If the above equation holds, all the signatures are valid. Next, each $SM’_j$ transmits the antecedent member of the sequence $h_{(\theta-1)j}$ to each $SM’$ who has contributed their $h_{\theta j}$. Conversely, without any signing, each $SM’_j$ for $j=1, 2, \dots,J$ sends $h_{(\theta-1)j}$ to other $SM’$s. Similarly, each $SM’_j$ can verify the validity of random numbers by merely computing hashes of the previous random number. Subsequently, each $SM’_j$ aggregates the valid random numbers $\sum_{j=1}^{J}h_{(\theta-1)j}=h_{(\theta-1)}$ and hashes the result $H_2(h_{(\theta-1)})=\Theta_1$. Now, based on the created random number $\Theta_1\in \{0, P\}$ and each $SM’$’s interval, the miner of this round will be selected.

It is necessary to note that, this process will be performed before each report generation and transmission phase and it is only in the first round that each $SM’$ commits two random numbers. To be exact, except for the first block (\textit{i.e.}, genesis block) where each $SM’$ commits two random numbers and the miner’s ID gets broadcasted in the customer-side network, the miner of other blocks will be always identified in the previous block. Hence, by determination of miner’s ID in each Sidechain’s block, the system works more efficiently and the communication overhead will be reduced.

The number $\theta$ is defined relying on the number of blocks made in each week. To be precise, considering the weekly updates of the miners' list, the 15-minute period for each transmission phase, and the average number of repetition for crashing miners, we can set $\theta=15\times 4\times 24\times 7 \approx 1100$. And in each week, the CC publishes a valid miners' list, the number $\theta$, and the miners’ new intervals which are based on their credibility in the network and other criteria mentioned earlier. 

In essence, we peel one onion with $\theta$ layers each week. Let us turn to the validity and soundness of our hash onion technique\cite{Jahan}: 1)\textsc{Verifiability}: Based on $\theta$ subsequent hashes of each $SM’$’s random number, the validity of each $SM’$’s random number could be verified and the invalid random numbers would be discarded. While each $SM’$ computes the aggregated random number separately, they all reach to one similar number which is truly random. 2)\textsc{Unpredictability}: Due to the time-laps in the first round and the fact that miners who did not contribute in the first round cannot be miners for the rest of the week, and considering the one-way property of hashes, no $SM’$ even the one who sends the last random number could predict the next miner. 3)\textsc{Unbiased}: In the existence of malicious $SM’$s, the aggregated result of random numbers is still truly random and the malicious $SM’$s could not manipulate the result in their favor. 4)\textsc{Availability}: By the use of hash functions and the time-lapse in the first round, it can be guaranteed that we can reach a successful completion in each round and all valid $SM’$s reach one similar random number and subsequently a similar miner for each round. 5)\textsc{Scalability}: It is interesting to note that the number of $SM’$s does not affect the output distribution of our hash onion technique. Moreover, our hash onion technique achieves a truly random number by the means of limited computation and low communication overhead.

%User Report Gen
\subsection{Phase 3: User Report Generation}
At each time instants like every 15 or 30 minutes, each smart meter $SM_i$ for $i=1, 2, \dots, n$ encrypts his multi-dimensional data and signs all of his data, and transmits them to the miner. As described earlier, the miner of each round is identified in the previous block and for the first block’s miner, $SM’$s broadcast the identity of the miner in the network. 

Let us begin by introducing smart meters’ $L$-dimensional data $d_{i1}, d_{i2}, \dots, d_{iL}$ for $i=1, 2,\dots, n$. To encrypt and transmit data straightforwardly, we convert the $L$-dimensional data into one appropriate message via the utilization of the Chinese remainder theorem technique. Specifically, each $SM_i$ for $i=1, 2, \dots, n$ in the customer-side network performs the procedure shown below\cite{ECEnc}:
\begin{IEEEeqnarray}{ll}
D_i=d_{i1}(y_1M_1)+ d_{i2}(y_2M_2) + \dots +  d_{iL}(y_LM_L) \pmod{M}~.\nonumber
\end{IEEEeqnarray}
Then, relying on the employed Elliptic curve encryption scheme, each $SM_i$  encrypts the $D_i$ with the customer-side’s group key (GK) and obtains the ciphertexts as presented below:
\begin{IEEEeqnarray}{ll}
\mbox{Map Function:}~~ D’_i=D_i\cdot G~, \nonumber\\
C_i=[C_{1i}=r_i\cdot G, C_{2i}=r_i\cdot \mathcal{S}_{GK}+D’_i]~,
\end{IEEEeqnarray}
where $D’_i$ is a point on a curve and $r_i$ is a random number.

In each block beside the first block (genesis block), each SM should acknowledge or reject the last block he has seen. In order to accept or discard the last block of Sidechain, every smart meter scrutinizes the validity of each block by checking their own data and the general information incorporated in the block. Subsequently, if he wants to acknowledge or reject the last block, he will compute ACK=$H_1(+\mbox{last block’s hash})$ and NACK=$H_1(-\mbox{last block’s hash})$, respectively. 

Since we utilize the distributed decryption in our proposed scheme and the fact that the aggregated data cannot get decrypted without smart meters’ assistance, each SM should compute the auxiliary ciphertext to help the CC in the decryption of the aggregated consumption of the last block. To be precise, each $SM_i$ for $i=1, 2, \dots, n$ performs the following multiplication $\overline{C_{aux_i}}=s_i\cdot \overline{C_1}$ in which $\overline{C_1}$ is the aggregation of last block’s data $\overline{C_1}=\sum_{i=1}^{n}\overline{C_{1i}}$, and $s_i$ is $SM_i$’s private key. 

The next stage is to ensure the integrity and authentication of his data via a signature. Technically, each $SM_i$ performs the following procedure to sign his data with his private key:
\begin{enumerate}
\item He chooses $v_i\leftarrow \mathbb{Z}_q$ randomly and computes $V_i=v_i\cdot G$.
\item Then, sets $e_i=H_1(V_i, TS, C_{1i}, C_{2i}, \overline{C_{aux_i}}, ACK_i)$.
\item $\sigma_i=[u_i=v_i+e_is_i\pmod{q}, V_i]$.
\end{enumerate}
At the last stage, $SM_i$ transmits the ciphertexts ($C_{1i}, C_{2i}$), the last block’s verification $ACK_i$, the auxiliary ciphertext of the last block’s aggregated data $\overline{C_{aux_i}}$ and the signature of all these data $\sigma_i=(V_i, u_i)$ to the miner of this round.

%Sidechain
\subsection{Phase 4: Sidechain's Block Generation}
Once the miner receives all the smart meters’ data, he verifies the validity of each data, aggregates all the valid consumptions, builds and publishes the block in the customer-side’s network. 

Initially, the miner $SM’_j$ scrutinizes the TS and signatures. Precisely, to lower the computational overhead, the miner utilizes batch verification technique on all the received signatures:
\begin{enumerate}
\item He sets $b_1=1$ and chooses $b_k\leftarrow \mathbb{Z}_q: k=2, 3,\dots, n$.
\item Then, he sets $e_k=H_1(V_k, TS, C_{1i}, C_{2i}, \overline{C_{aux_i}}, ACK_i)$ for $k=1, 2,\dots, n$.
\item At last, evaluates the following equation:\\ $\big(\sum_{k=1}^{n}b_ku_k\big)\cdot G=\bigoplus_{k=1}^{n} b_kV_k\oplus b_ke_k\mathcal{S}_k$~,
\end{enumerate}
if the above equation holds, all the signatures used in batch verification equation are correct. Nevertheless, if some of the signatures were invalid and the batch verification equation could not hold, the miner would check the signatures one by one. After the verification procedure, the miner aggregates his data with all the verified data as follows:
\begin{IEEEeqnarray}{ll}
C_1=\sum_{i=1}^{n}C_{1i}~,~
C_2=\sum_{i=1}^{n}C_{2i}~.
\end{IEEEeqnarray}

Let us turn to auxiliary ciphertexts which are going to assist the CC in the decryption part. Concretely, relying on the secret-sharing technique, the Lagrange coefficients and the contribution of at least $\mu$ smart meters, the miner will be able to compute the final auxiliary ciphertext as presented beneath:
\begin{IEEEeqnarray}{ll}
\overline{C_{aux}}=\sum_{k=t_1}^{t_{\mu}}s_k\cdot \overline{C_1}\cdot \lambda_k=s_X\cdot \overline{C_1}~.
\end{IEEEeqnarray}

For acknowledgment of the last block’s legitimacy, the miner counts the $ACK/NACK$ of the users by counting $+1$ point for each $ACK$ and $-1$ point for every $NACK$. However, it is necessary to note that if a smart meter sends $NACK$ and claims that there is something wrong with his data or the general information of the block, based on the users’ smart contract he has to send the proof of that wrongdoing to the GW directly. To be exact, the wrongdoings in the Sidechain’s blocks typically are twofold: the faulty aggregation which could be modified by sending the correct aggregation; the omission of smart meter’s data in the aggregated consumption or the block itself, which could be rectified by sending the smart meter’s data to the GW directly.

Now, the miner builds and publishes the new block in the customer-side network as exhibited in TableII.

\begin{table}[h]
\centering
\captionof{table}{Sidechain's block structure}
\begin{tabular}{|c|c|c|}
\hline Previous block's hash & Hash Root&TS \\ 
\hline Block size and no.&$ID_{M}$&Miner's Proof\\\hline
\hline $C_1, C_2, \overline{C_{aux}}, \sigma_M$&$ID_{nextM}$&Next Miner's Proof\\
\hline $C_{11}, C_{21}$&$\overline{C_{aux1}}$&$ACK_1$\\
\hline $C_{12}, C_{22}$&$\overline{C_{aux2}}$&$ACK_2$\\
\hline \vdots&\vdots&\vdots \\
\hline $C_{1n}, C_{2n}$&$\overline{C_{auxn}}$&$ACK_n$\\\hline
\end{tabular}
\label{Tab2}
\end{table}

wherein $\sigma_M$ is the miner’s signature on the aggregated data which is obtained by the beneath computations:
\begin{enumerate}
\item The miner chooses $v_M\leftarrow \mathbb{Z}_q$ randomly and sets $V_M=v_M\cdot G$.
\item Then, he calculates the hash of the data which he wants to sign $e_M$:\\ $e_M= H_1(V_M, TS, C_1, C_2, \overline{C_{aux}}, ACKno)$.
\item Finally, he obtains the signature $\sigma_M=[u_M=v_M+e_Ms_M\pmod{q}, V_M]$.
\end{enumerate}
To specify the structure of Sidechain’s blocks exhibited in Table \ref{Tab2}, the block header contains the previous block hash, the Merkel root of all the block’s data, the TS, the block number and size, and the miner’s ID along with the random number that made him the miner of this round as a proof. Particularly, the block’s data consists of individual and aggregated consumption reports, $ACK/NACK$ of each SM, the number of acknowledgments $ACKno$, the auxiliary ciphertext of last block’s data, the miner's signature on these data and finally, the next miner’s ID with the random number that made him the next miner.

After receiving the block, each SM checks the validity of the block’s header and successively the data it contains. Concretely, each SM could verify his own data and the general information incorporated in the block such as the aggregated ciphertexts or miner’s signature. Therefore, if the SM approves the block’s legitimacy, he adds this block to his Sidechain. However, an SM could find some wrongdoings or problems in the block. Whether the block header or the miner’s legitimacy was invalid, the SM would discard the block. In the case of a problem with smart meters’ data like omission or alteration of his consumption report, his $ACK$, and the auxiliary ciphertext, the smart meter will send the correct data to the GW directly. Furthermore, if the aggregated consumption data included in the block are erroneous, the SM will transmit the real data to the GW directly as proof of this problem.

%Mainchain
\subsection{Phase 5: Mainchain's Block Generation}
In this phase, we concentrate on the Mainchain’s block structure and the data it contains. The GW is the only entity that constructs these blocks in parallel with the Sidechain’s block based on general information of the network such as aggregated consumptions, acknowledgments, and proofs of wrongdoings that were sent by the smart meters individually. Fundamentally, our proposed smart grid network operates based on Mainchain. In essence, Mainchain’s blocks are the only blocks to which all the network entities have access and store the general and necessary information of the network permanently. 

First, the GW examines the legitimacy of each Sidechain’s block he receives by validating all the information in the block. To be exact, the GW can check the validity of $C_1, C_2, \overline{C_{aux}}$, and $ACKno$ by computing them again based on the data incorporated in the block. Then, he can verify the miner’s signature by calculating the following evaluation:
\begin{IEEEeqnarray}{ll}
u_M\cdot G=V_M\oplus H_1(V_M, TS, C_1, C_2, \overline{C_{aux}}, ACKno)\cdot \mathcal{S}_M~.
\end{IEEEeqnarray}
If the above equation holds, the signature is valid. 

Second, to build the Mainchain’s block, the GW gathers all the information in the network, including: the aggregated consumption report, $\mu$ auxiliary ciphertexts of the Sidechain’s block, proof of any wrongdoings which were sent by some smart meters with which the GW could compute the newly aggregated data, the $ACKno$, the GW’s signature on these consumptions and the CC's warnings and notices which were sent to the GW. The Mainchain’s block structure is demonstrated in tableIII and the GW’s signature is created based on the following computations:
\begin{enumerate}
\item The GW selects $v_{GW}\leftarrow \mathbb{Z}_q$ randomly and computes $V_{GW}=v_{GW}\cdot G$.
\item Then, it computes the hash of the materials which it wants to sign $e_{GW}=H_1(V_{GW}, \\TS, C_1, C_2, \overline{C_{aux}}, ACKno)$ where $C_1, C_2$ could be the new aggregated data constructed based on the new correct data transmitted by some smart meters as proof of wrongdoings.
\item Finally, it calculates the signature $\sigma_{GW}=[u_{GW}=v_{GW}+e_{GW}s_{GW}\pmod{q}, V_{GW}]$.
\end{enumerate}

\begin{table}[h]
\centering
\captionof{table}{Mainchain's block structure}
\begin{tabular}{|c|c|c|}
\hline Previous block's hash & Hash Root&TS \\ 
\hline Block size and no.&-&-\\\hline
\hline $C_1, C_2, \overline{C_{aux}}, \sigma_{GW}$&no. of users: $n$&$ACKno$\\
\hline $ID_M$, Miner's Proof&$ID_{NextM}$, Next Miner's Proof&Bill\\
\hline Ciphertext Hashes& Wrongdoing Proofs& CC's Advice\\\hline
\end{tabular}
\label{Tab3}
\end{table}

Lastly, the GW publishes the Mainchain’s new block in the whole network. Since each entity in the network observes the Mainchain’s blocks, they can evaluate the authentication of blocks by verifying GW’s signature on the consumption reports. Moreover, all the smart meters and miners ($SM’$) can check the TS, hashes, aggregations, and acknowledgments in order to add this new block to their Mainchain. Since the GW is merely a relay between the customer-side area and the CC, it creates each block based on Sidechain’s block and CC’s communicated data. Although the GW individually does not profit from acting maliciously or changing the consumption data, the smart meters and miners ($SM’$) scrutinize each block he creates and reports the wrongdoings directly to the CC. For this reason, we can note that the customer-side network oversees GW’s activities.

%Report Reading
\subsection{Phase 6: Secure Report Reading}
Since the CC sees Mainchain’s blocks, akin to other entities in the network, the CC scrutinizes the correctness of each block and updates its Mainchain. In the same way, the CC checks the GW’s signature, TS, $ACKno$, and the general information incorporated in the block. Meanwhile, to authenticate the GW's signature, the CC carries out the beneath assessment:
\begin{IEEEeqnarray}{ll}
u_{GW}\cdot G=V_{GW}\oplus H_1(V_{GW}, TS, C_1, C_2,\nonumber\\ \overline{C_{aux}}, ACKno)\cdot \mathcal{S}_{GW}~.\nonumber
\end{IEEEeqnarray}
If the above equation holds, the GW's signature is valid and the GW is the true constructor of the block. After examination of the new block, it adds the block to its Mainchain. With the utilization of distributed decryption in our proposed scheme and by considering the finalized aggregated consumption data which have been acknowledged by most of the smart meters in the subsequent block, the following procedure will be performed by the CC:
\begin{enumerate}
\item Assume the aggregated consumption data $(C_1, C_2)$ are finalized which means the corresponding auxiliary ciphertext has been sent in the next block $C_{aux}=\sum_{k=t_1}^{t_{\mu}}s_k\cdot C_1\cdot \lambda_k=s_X\cdot C_1$.
\item The CC computes $s_{CC}\cdot C_1+s_X\cdot C_1=s_{GK}\cdot C_1$.
\item Afterwards, the CC decrypts the aggregated consumption by computing $C_2-s_{GK}\cdot C_1=\sum_{i=1}^{n}D'_i$.
\item And by the Pollard’s Lambda method \cite{PollardL}, we have: $\sum_{i=1}^{n}D'_i\xrightarrow{map} \sum_{i=1}^{n}D_i$.
\end{enumerate}
To obtain the aggregation of each dimensional-data, the CC employs CRT’s properties and performs as shown below:
\begin{IEEEeqnarray}{lll}
\sum_{i=1}^{n}D_i \pmod{m_1} = \sum_{i=1}^{n}d_{i1}~,\nonumber\\
\sum_{i=1}^{n}D_i \pmod{m_2} = \sum_{i=1}^{n}d_{i2}~,\nonumber\\
\vdots\nonumber\\
\sum_{i=1}^{n}D_i \pmod{m_L} = \sum_{i=1}^{n}d_{iL}~.
\label{Eq1}
\end{IEEEeqnarray}
To prove the soundness of equation \ref{Eq1}, we start by CRT’s properties:
\begin{IEEEeqnarray}{lll}
M_i\times y_i\equiv 1\pmod{m_i} ~,\nonumber\\
\mbox{and}~M_i\times y_i\equiv 0\pmod{m_j} ~\mbox{for}~ j\neq i~.
\end{IEEEeqnarray}
The correctness of equation \ref{Eq1} for the $\ell$-dimensional data where $\ell=1, 2, \dots, L$ is illustrated below:
\begin{IEEEeqnarray}{lll}
\sum_{i=1}^{n}D_i \pmod{m_\ell}=\sum_{i=1}^{n}[d_{i1}(M_1y_1)+d_{i2}(M_2y_2)+\nonumber\\
\dots+d_{iL}(M_Ly_L)]\pmod{m_\ell}=\sum_{i=1}^{n}d_{i\ell}
\end{IEEEeqnarray}
Now, the CC can evaluate and analyze each dimensional-data of every customer-side network and sends the necessary information and policies to the corresponding GWs. As you know, all the nodes (entities) in the network operate based on a smart contract which consists of codes that execute automatically when the conditions are triggered. The function of SMs' smart contract has been established during the scheme. And the summary of CC’s smart contract is depicted in TableIV. 
%Smart Contract
\begin{table}[h]
\centering
\captionof{table}{The CC's Smart Contract}
\begin{tabular}{l}
\hline \textbf{Algorithm:} ~The CC's Smart Contract\\\hline\hline
\textbf{Contract} CC's Feedback \{\\
\textbf{public} payed-in advance; delayed; seriously-delayed; price-policy;\\ predefined-threshold; proof-of-wrongdoing; impending-blackouts; \\
\textbf{function} ~payment()\{\\
\textbf{Case} bill-payment\\
\textbf{Case 1:} bill-payment$==$payed in advance\textbf{:} overall discount;\\
\textbf{Case 2:} bill-payment$==$delayed\textbf{:} send notification to the GW;\\
\textbf{Case 3:} bill-payment$=>$seriously delayed\textbf{:} force stop the SM; \\
\textbf{End case}\}\\
\textbf{function} ~price()\{\\
\textbf{if} price-policy$==$changed\textbf{:} transmit the new policy to the GW;\\
\textbf{else} output ``same price policy";\\
\textbf{End if}\}\\
\textbf{function} ~electricity-forecast()\{\\
\textbf{if} $\|$forecast-consumption$\|$$>=$predefined-threshold\textbf{:} \\check for faulty behavior or leakage;\\
\textbf{if} impending blackouts$==$1\textbf{:} notify the GW;\\
\textbf{End if}\}\\
\textbf{function} ~byzantine-behavior()\{\\
\textbf{if} proof of wrongdoing$==$valid\textbf{:} remove his ID from miners' list,\\ burn some of his stakes accordingly and reward the prover;\\
\textbf{elseif} proof of wrongdoing$==$invalid\textbf{:} penalize the prover by fine;\\
\textbf{elseif} proof of wrongdoing$==$after block finalization\textbf{:} discard the proof;\\
\textbf{if} proof of GW's wrongdoing$==$valid\textbf{:} suspend the GW and\\ introduce a new GW;\\
\textbf{End if}\}~\}\\

\\\hline
\end{tabular}
\label{Tab5}
\end{table}

%Bill
\subsection{Phase 7: Bill Calculation}
In spite of traditional schemes in which the CC calculates the bill every 30 days, in our proposed blockchain-based scheme, the GW calculates the bill after every 1100 blocks which are nearly one week. While other schemes typically calculate the bill based on plaintexts or with the transmission of the whole blockchain to the authorities, our proposed scheme acquires the bill through calculating temporal aggregation of ciphertexts. Moreover, despite other schemes that take one average price for bill calculation, our proposed SPDBlock scheme considers different price policies plus the fine for misbehaving smart meters. Specifically, the GW calculates the bill for each smart meter via their finalized consumption data corporated in the Sidechain blocks and publishes the resultant bill in one Mainchain block. Furthermore, the smart meters can verify the validity of bill block by computing each bill and especially their own bill based on the data stored in Sidechain. And similarly, smart meters assist the CC to decrypt these bills. 

Assume that the ciphertexts $(C_{1i\omega}, C_{2i\omega})$ for $\omega\in\{1100 ~\mbox{consecutive blocks}\}$ are SM$_i$’s data for $i=1, 2, \dots, n$. Then, relying on three different prices $P_{on}, P_{mid}$ and $P_{off}$ which are on-peak, mid-peak and off-peak prices, respectively, the GW computes the following procedure:
\begin{enumerate}
\item For on-peak instants $\omega_{on}$, computes $C_{1i\Omega}=(\sum_{\omega_{on}}C_{1i\omega})\times P_{on}, C_{2i\Omega}=(\sum_{\omega_{on}}C_{2i\omega})\times P_{on}$ which are the temporal aggregation of SM$_i$’s ciphertexts multiplied by on-peak price.
\item Similarly, for mid-peak and off-peak instants $\omega_{mid}, \omega_{off}$, the GW obtains $(C'_{1i\Omega}, C'_{2i\Omega})$ and $(C''_{1i\Omega},C''_{2i\Omega})$, respectively.
\item Then, it calculates the bill ciphertext as follows:\\ $C_{1i\beta}=C_{1i\Omega}+C'_{1i\Omega}+C''_{1i\Omega}, C_{2i\beta}=C_{2i\Omega}+C'_{2i\Omega}+C''_{2i\Omega}$~.
\item Finally, the GW publishes the bills in one Mainchain block. 
\end{enumerate}
After verification of this Mainchain block by smart meters, in order to assist in decryption of the bills, the SMs transmit the auxiliary ciphertexts to CC via GW. To be precise, assuming at least $\mu$ smart meters have sent $s_{t_k}\cdot C_{1i\beta}$ for decryption of SM$_i$’s bill wherein $k\in\{1, 2, \dots, \mu\}$, the GW publishes these auxiliary ciphertexts in a new Mainchain block. The auxiliary ciphertext for SM$_i$'s bill is created as presented below:
\begin{IEEEeqnarray}{lll}
C_{aux\beta_i}=\sum_{k=t_1}^{t_\mu}s_k\cdot C_{1i\beta}\cdot \lambda_k=s_X\cdot C_{1i\beta}~.
\end{IEEEeqnarray}
Finally, the CC carries out the following process to calculate each dimensional-data’s bill for each SM$_i, i=1, 2,\dots, n$:
\begin{enumerate}
\item Firstly, the CC computes $s_{CC}\cdot C_{1i\beta}+s_X\cdot C_{1i\beta}=s_{GK}\cdot C_{1i\beta}$.
\item Secondly, he computes $C_{2i\beta}-s_{GK}\cdot C_{1i\beta}=\mbox{Bill}'$ and by the utilization of Pollard’s Lambda method, we have: $\mbox{Bill}'\xrightarrow{map} \widehat{\mbox{Bill}}$, where ``$\widehat{\mbox{Bill}}$" is the composition of aggregated consumption and 3 types of prices.
\item Lastly, the CC calculates the bill for each dimensional-data by the utilization of CRT’s properties:
\begin{IEEEeqnarray}{lll}
\widehat{\mbox{Bill}} \pmod{m_1}=\mbox{bill}_1~,\nonumber\\
\widehat{\mbox{Bill}} \pmod{m_2}=\mbox{bill}_2~,\nonumber\\
\vdots\nonumber\\
\widehat{\mbox{Bill}} \pmod{m_L}=\mbox{bill}_L~,
\end{IEEEeqnarray}
\end{enumerate}
where bill$_\ell$ for $\ell=1, 2,\dots, L$ is the $\ell$-dimensional-data's bill relying on different prices. Later, the CC can add fines to these bills relying on the wrongdoing proofs incorporated in Mainchain’s block and transmit the bill and subsequent notices to each SM. It is interesting to note that in order to make our scheme more efficient and help the smart meters which have limited storage units, after the finalization of Mainchain’s bill-block, the SMs can erase their Sidechain’s blocks relevant to that Mainchain’s bill-block. To some extent, this is clear evidence that regardless of considering a blockchain-based network, the storage requirements of each SM are limited.

%security analysis
\section{Security Analysis}
Generally, we assess the security of our SPDBlock scheme based on three criteria: privacy preservation, integrity and authentication, and resistance against various attacks. Concretely, in each part, we presume there exists a strong adversary $\mathcal{A}$ who is capable of applying passive and active attacks with the purpose of disclosing privacy, invalidating integrity and authentication, or disrupting the smart grid network performance. 

More specifically, in the privacy preservation analysis, we aim to assess every potential attack on the privacy of the individual and aggregated consumption data since the SPDBlock is a privacy-preserving scheme. Moreover, we also consider other attacks on the privacy of each block or attacks based on collusion. In the next stage, we take every attack related to the integrity of transmitted data or each entity’s authentication into account. Thus, we evaluate every communication link between SG entities and all the multicast blocks. Finally, there are other attacks that are not especially categorized in the above classes. The attacks which are connected to the SG network are replay, modification, impersonation, MITM, \textit{etc.}; and the other attacks associated with the blockchain structure are fork, collusion, and $51\%$ attack. In the following subsections, we analyze the security of our proposed scheme in depth.

%privacy
\subsection{Privacy Preservation}
In this subsection, we analyze the main objective of this scheme which is the privacy preservation of individual and aggregated consumption data. Regarding the system model which is comprised of four components and two Consortium blockchains, in general, there exists multiple communication links between these components: SM-to-$SM’$, SM-to-GW, GW-to-CC, and $SM’$-to-CC. Based on these communication links and considering that each report is being transmitted through public channels, the strong adversary $\mathcal{A}$ could disclose and threaten the privacy of individual and aggregated consumption data in multiple phases. Furthermore, despite the fact that the blocks are being multicast instead of broadcast and we utilize Consortium blockchains in our proposed scheme wherein only the smart grid entities have access to them, the strong adversary $\mathcal{A}$ could somehow acquire access to them. Consequently, we should also scrutinize the privacy of users’ different types of data in every phase.

Technically, in each communication and interaction phase of the SPDBlock scheme, there exists no consumption data in the form of plaintext and every consumption data is encrypted via Elliptic curve-based constructions. In consequence, the only eavesdropped messages the adversary gets are in the form of ciphertexts and they could not obtain the real consumption data. For more precise evaluation, we assess every phase of our scheme more fundamentally.

Assume that the strong adversary $\mathcal{A}$ inhabits in the customer-side network and tries to eavesdrop on the individual consumption reports transmitted between SMs and the miners ($SM’$s). Since each $SM_i$ encrypts his data via the group-key $GK$ and the corresponding decryption key is built based on $\mu$ SMs and the CC’s secret key, the adversary cannot perform the decryption without the knowledge of all these secret keys. In addition, for obtaining the consumption data without keys, the adversary needs to break the ciphertexts and eventually face the computational Diffie-Hellman (CDH) hard problem \cite{CDH}. Hence, the adversary could not obtain the individual consumption data due to the hardness of solving the CDH problem and the fact that the secret keys are generated by the KMC and transmitted to each entity through a secure channel. However, owing to the high-frequent (every 15 or 30 minutes) of data transmission, the SM’s data are small values; and the adversary might try to apply the brute-force attacks. On the evidence of IND-CPA security of our employed encryption scheme \cite{IND-CPA}, it would be computationally infeasible for the adversary to obtain the messages $D_i$ or $D’_i$ from the ciphertext $C_i=(C_{1i}, C_{2i})$. Therefore, we can reliably deduce that our proposed scheme preserves the privacy of each consumption data in every interaction between the SMs and the $SM’$s.

Another phase that contains the individual consumptions is the Sidechain blocks. Although only the smart meters, $SM’$s, and the GW have access to the Sidechain blocks, envision that the adversary somehow acquires access to one of these blocks. In the same way, as mentioned above, this strong adversary $\mathcal{A}$ cannot obtain the individual’s consumption data because our employed encryption is IND-CPA secure, and the $\mathcal{A}$ does not have the relevant secret keys or needs to solve the CDH hard problem. Similarly, even if the curious or dishonest miner $SM’_j$ tries to break the received ciphertexts, he cannot perform any better than the adversary due to the aforementioned reasons.

The last part of our scheme where the adversaries could access or eavesdrop on the individual reports is when the SMs transmit their consumption data directly to the GW as proof of wrongdoing in blocks or for any other reason. In this scenario, the SM also encrypts his data via the group-key $GK$, and for the same reasons the eavesdropping adversary or the GW itself cannot reveal the privacy of this individual data. 

In aggregate, considering the above analysis, the privacy of users’ individual consumption data is preserved against the external adversaries, internal malicious entities, and in the form of block data.

%%%
The SPDBlock scheme not only preserves the privacy of each SM’s energy consumption, but also ensures the privacy of every spatial and temporal aggregated data. Due to the construction of our proposed scheme and the employment of an Elliptic curve-based homomorphic encryption, the calculation of various aggregations such as spatial and temporal aggregation is done in a privacy-preserving manner. Specifically, after the validation of signatures by the miner of that round $SM’_j$, the spatial aggregation of valid ciphertexts is obtained where $C_1=(\sum_{i=1}^{n}r_{i})\cdot G$ and $C_2=(\sum_{i=1}^{n}r_{i})\cdot S_{GK}+\sum_{i=1}^{n}D'_{i}$. Since the structure of this spatially aggregated data is exactly look like the individual data’s structure, for precisely the same reasons, no adversary could acquire the aggregated energy consumption data. It should be noted that the aggregation in the Sidechain and Mainchain blocks are almost always the same as each other. However, in the scenario where some SMs transmit their data to the GW directly, because of the homomorphic property of the employed encryption, the GW can calculate the new final aggregation by an addition operation. As a result, the privacy of the old and the new spatial aggregated consumption data is ensured. 

In other privacy-preserving schemes that employ homomorphic encryption in their scheme, when the CC behaves curiously or maliciously, it could get the individual ciphertexts and subsequently obtain the individual consumption data based on the decryption key it typically has. Even though the CC in the SPDBlock scheme does not have access to the Consortium Sidechain, imagine a situation wherein the CC somehow has access to the Sidechain blocks. To be precise, the CC could obtain the individual reports in the Sidechain blocks by the means of eavesdropping or collude with entities like SMs, $SM’$s, or GW. However, since the CC does not have at least $\mu$ smart meters’ secret keys, it cannot obtain the individual consumption data from the individual reports. As a result, in these scenarios, albeit the CC has part of the corresponding secret key, based on the utilization of distributed decryption in our scheme, the curious or disgruntled employees in the control center cannot acquire the individual energy consumption data. Besides, not only the CC cannot obtain the individual data, but also it cannot decrypt the aggregated data without the assistance of at least $\mu$ smart meters.

Relying upon the fact that the GW calculates the bills and executes the temporal aggregation based on ciphertexts, for similar justifications, the privacy of the temporal aggregated data and bills are guaranteed. Furthermore, even if one SM colludes with the malicious GW or CC to calculate the wrong temporal aggregation and subsequently the wrong bill, owing to the utilization of distributed-decryption in our proposed scheme, at least another $(\mu-1)$ SMs should corroborate with these wrong aggregation and bill so that the CC could decrypt them. Therefore, even when the CC, GW, and one SM are colluding, they could not achieve success in creating a wrong bill. In conclusion, the SPDBlock scheme can preserve the confidentiality of users’ individual and aggregated data even with strong collusion and various attacks.

%integrity
\subsection{Integrity and Authentication}
Besides privacy, other security criteria like integrity and authentication should be ensured for a scheme to works practically. To achieve data integrity and authentication, our proposed SPDBlock scheme uses a signature construction that works based on the discrete logarithm problem (DLP). For the block’s integrity and authentication, we utilize the same digital signature, hash functions, and the Merkel Tree-based structure. 

In general, excluding the miner selection phase wherein only the first round of communication contains signatures, the first step of every other phase is to validate the legitimacy of the message. Particularly, all the communication of individual and aggregated consumption data is done via the signature of the source along with the message. Fundamentally, our proposed scheme is comprised of 3 general communication links: SM-$SM’$, $SM’$-GW, and GW-CC. Accordingly, we assess the integrity of data and source authentication in every one of these links.
 
Technically, in the SPDBlock scheme, each SM signs its data via its secret key which is built by the KMC and transmitted to him through a secure channel. Therefore, relying on a secure cryptographic hash function, a specific secret key for every SM, and the Schnorr signature construction which is strongly unforgeable against adaptive chosen message attacks in the random oracle model (ROM) \cite{ROM}, no strong adversary $\mathcal{A}$ could discredit the integrity and authentication. Besides, due to the employed signature scheme, it is computationally infeasible for the strong adversary to forge a legitimate signature and invalidate the integrity of data or the authentication of each smart meter. Similarly, in the direct transmission of proofs to the GW or the CC, by the smart meters or the $SM’$s, respectively, due to the strong signature along with the message, no adversary could succeed in integrity or authentication related attacks.

Likewise, each miner and the GW who are the generator of Sidechain and Mainchain’s blocks respectively, create a signature on the aggregated data and include the signature in the block. Thus, for the same reasons, the integrity of the aggregated consumption data which are always included in the blocks is ensured. Moreover, since every block in the Sidechain or Mainchain contains the Merkel Root of all the block’s data, the adversary $\mathcal{A}$ could not alter any component of the blocks. Note that every component of each block is connected to the Merkel Root and each individual and aggregated consumption data, auxiliary ciphertext, acknowledgment, and the TS is signed via the block generator’s secret key. As a result, it could be considered as a secure construction wherein if the adversary changes the consumption data, he must be able to create a valid signature; and if he alters other components of the block beside the consumption data, he must change the Merkel Root; and subsequently due to the Merkel Root alteration, the block does not get finalized. Therefore, the SPDBlock scheme certainly ensures the integrity and authentication of blocks. 

Also, since the $SM’$s and the GW’s secret keys are distributed through a secure channel and considering the transmission of the signature in the blocks or along with the messages, the adversary $\mathcal{A}$ could not impersonate the miner or the GW. Additionally, due to the existence of miner selection proofs in the blocks, even the wrong miners could not generate any legitimate block. Hence, owing to the employed signature construction, Merkel Tree structure, and miners’ proofs, we can certainly infer that our proposed SPDBlock scheme ensures the integrity of each transmitted data or block. 

Furthermore, our employed Elliptic curve-based signature supports batch verification that makes our scheme much more practical. Since we utilize signature in each communication and typically it is required to simultaneously verify a bunch of signatures in the smart grid network, signatures with the batch verification capabilities are preferred. However, there exist some attacks on the signatures with batch verification wherein the attacker can succeed in the batch verification part even with invalid signatures.  Due to some randomization techniques, our employed signature scheme resists such attacks. Hence, even if the adversary colludes with some smart meters to impose bad data through an invalid signature in the batch verification parts, the SPDBlock scheme will detect and discard the invalid signatures and data.

We have demonstrated in every scenario that the adversary could impose his wrong data in the system, invalidate the integrity of data, and impersonate an SG entity such as SMs, $SM’$s, or the GW. Given the above assessment, it is pretty obvious that the SPDBlock provides data integrity, source authentication, and resistance against the imposition of invalid data in the SG network.

%attacks
\subsection{Resistance against other Attacks}
Aside from the main security criteria addressed above, our proposed scheme provides some additional features and resist various attacks like internal, external, and blockchain-based attacks. We demonstrate each attack thoroughly in the following paragraphs.

\textbf{General attacks and features:} Let us begin by the replay attack wherein an entity transmits the same message it has been transmitted in the previous rounds for the reason of changing his true consumption or stealing energy. Since the timestamp TS is included in every transmission and every block, no entity could succeed with the replay attack. 

Next, we can consider the impersonation and modification attacks which are correlated with invalidating the integrity and authentication. As mentioned in the previous section, and because of the transmission of signatures on every communication and every block, no external or internal entity or even a powerful adversary could yield a legitimate signature on their data. Therefore, impersonation of another entity or modification of the blocks or the communicated data is computationally infeasible and the detection of these wrongdoings is inevitable.

Comparatively, we consider the MITM attack which is related to integrity and authentication. Relying upon the fact that the verification of signatures is the first step of each phase, and regarding the unforgeability of our employed signature against adaptive chosen message attacks in the ROM, no adversary could carry out the MITM attack. More especially, let’s assume that somehow a malicious entity or an adversary succeeds in inhabiting the links between SM-to-$SM’$ or SM-to-GW and disturbs their communication or inject bad data or even build a new block. Although the signatures of SMs are not included in the Sidechain blocks, this newly built block could get accepted in the first round. However, as a result of the Mainchain blocks which works in parallel to the Sidechain blocks, these bad data or even this new Sidechain block does not get to be finalized. Because the entity whose data is altered can directly send the proof of this wrongdoing to the GW in the next round. Hence, even with the existence of a really powerful adversary in the SG network, the MITM attack is almost impossible.

Even though differential privacy has its specific definition and it is addressed by adding a specific noise to the aggregated data, our scheme intuitively resists against differential attacks $\textit{i.e.,}$ the malicious CC could collude with the miners to obtain the aggregation of two subsets of users: $n$ and $n-1$, and subsequently acquires that specific individual’s data. Concretely, by the employment of distributed decryption and secret sharing technique, the CC in our proposed scheme is not capable of decrypting any other aggregation besides the aggregation of all the participating users. Because he needs the assistance of at least $\mu$ number of SMs for the decryption of any data. Thus, the SPDBlock scheme resists differential attacks.

Transmitting multi-dimensional data is a necessary and useful property in the SG network. With respect to any typical data aggregation scheme, our proposed scheme not only achieves the spatial aggregation of multi-dimensional data, but also can acquire temporal aggregation of multi-dimensional data and calculates the bill for each dimensional data based on three kinds of prices (on-peak, mid-peak, and off-peak). It should be noted that all these temporal aggregations and bill calculations are performed based on ciphertexts, and thereby the privacy of these aggregations is preserved.

Because our system model is pragmatic and considers world-like situations, we take into account the real-world and practical properties such as fault-tolerance and dynamic users. It is fairly obvious that in real-world situations, SMs would leave the SG network, or conversely, some new SMs would be added to this network. As a result of employing an Elliptic curve-based homomorphic encryption, our proposed scheme could compute various aggregations of any number of users in a privacy-preserving manner. More precisely, in the case of adding a new SM to the customer-side network, our scheme only obliges the new SM to register itself to the KMC and obtains a valid pair of public-private keys, and the rest of our construction remains the same. If some smart meters leave the grid that happens for many reasons in the real world, the miner could still perform the aggregation and creates the new blocks without any key renewal. Furthermore, assume that some smart meters do not transmit their data whether it is the result of crashing or user manipulations, or as a consequence of adversary’s malicious behavior such as hacking the SMs. Similarly, because of the employed Elliptic curve-based homomorphic encryption, the miner of each round can still compute various aggregations on the rest of the received data; and there will be no disruption to the performance of our scheme. Moreover, CC can still compute the decryption of all the aggregations with the assistance of at least $\mu$ smart meters ($\mu$ is about half of the participating smart meters, $n/2$). Thus, the SPDBlock scheme provides the dynamic user and fault-tolerance property without any further costs.

\textbf{Blockchain-based attacks and features:}
Bearing in mind that the proposed SPDBlock scheme runs on two Consortium blockchains, there could exist some blockchain-based attacks on our scheme. Notwithstanding the fact that we have demonstrated the SM’s smart contract in the previous section and it clearly covers the resistance of our scheme against some blockchain-based attacks, we delineate some additional blockchain-based properties of our proposed scheme. 

As stated earlier, no entity besides the selected $SM’$ and the GW can generate a new block for the Sidechain and Mainchain, respectively. Therefore, the fork attack wherein an invalid miner generating a block is not feasible. Let us begin by attacks on the Sidechain blocks. Assume the miner tries to modify some smart meters’ data and succeeds in creating a new block with this data in it. Moreover, we can consider a harder situation wherein all the SMs besides $SM_i$ collude with each other to change the $SM_i$’s data. As a consequence of having two parallel Consortium blockchains, although $(n-1)$ users acknowledge this block as valid, this specific $SM_i$ could send the proof of the wrongdoing (which is the signature of his data, TS, \textit{etc.}) to the GW directly. Therefore, due to the independence of each entity in the SPDBlock scheme, this invalid block does not get finalized and the illegal data will be discarded. Due to the above analysis, the resistance against the $51\%$ attack is apparent. However, it is worth noting that if an SM sends $NACK$ in one round, he must submit his proof in the next round. Hence, if an SM tries to disturb the SG network by sending $NACK$s or tries to carry out a DoS attack, it will be detected in the next round and he will be punished and banned. 

Besides, typically in blockchain-based schemes, some users (SMs) could argue that they did not get the last block. As a result, the SM might claim that he did not know who the miner of the next round was. Since the result of not transmitting the consumption data in this scenario is the same as the result of not transmitting the consumption data for the reason of smart meter manipulation by the users, the users might succeed in disrupting the SM’s communication equipment. In our proposed scheme, owing to the existence of the parallel Mainchain blocks which every entity has access to, all the SMs could find the next miner even if they did not get the previous Sidechain block directly.

Finally, let us turn to the Mainchain blocks which are lightweight, and every entity in the SG network has an updated version of this chain in its memory. Given the fact that Mainchain blocks are parallel to the Sidechain blocks and GW is truly a relay and copies the final aggregation of Sidechain blocks in his blocks, the GW does not really perform any computations. In addition, the GW personally does not gain profit out of building invalid blocks or acting maliciously. Nevertheless, whether the GW builds an invalid block or changes the blocks’ content, the $SM’$ could send the proof of GW’s wrongdoings to the CC directly.

In the aggregate, based on the specific construction of two parallel Consortium blockchains and the existence of smart contract and PoS mechanism, we can safely infer that the SPDBlock scheme resists against various blockchain-based attacks such as fork, DoS, collusion, $51\%$ attack.

%performance evaluation
\section{PERFORMANCE EVALUATION}
The purpose of this section is to evaluate the practicality and the performance of our proposed SPDBlock scheme concerning its computational costs, communication overhead, and achievements. Precisely, we compare the performance of our proposed scheme with six blockchain-based schemes \cite{S2, S3, S4, S5, S6, S7} and three other non-blockchain-based schemes \cite{Cube-data, SMing, S8}. Note that the blockchain-based schemes are chosen on the grounds of their blockchain constructions and the similarity of their system models and goals with our proposed scheme. Although a blockchain-based scheme is supposed to be heavy in comparison to non-blockchain-based schemes, additionally, we have scrutinized the non-blockchain-based schemes in our comparison to indicate the efficiency of the SPDBlock scheme.

In the computation evaluation part, the timing of each operation is established, and relying on that we can illustrate the computational efficiency of the SPDBlock scheme in comparison with all these nine practical schemes. Subsequently, in the communication evaluation part, the length of each transmitted data and the network’s traffic are taken into account for the comparison of these schemes. Finally, after the above computation and communication evaluation, the achievement part is presented to show the differences between these schemes more clearly.

%Comp Cost
\subsection{Computation Costs}
In order to analyze the computational aspect of the SPDBlock scheme, we first highlight our criteria for comparison, establish the core computational costs, and then we can delineate the computational performance evaluation in detail.

Given that the system setup phase is typically a one-time expense in most privacy-preserving schemes and almost always one powerful entity or authority is in charge of this phase’s computation and configuration, like other schemes, we do not also consider this phase in our performance evaluation. In addition, we do not assess the control center’s computational burdens because the CC is in fact the most powerful entity in the smart grid network. Thereby, no matter how hard or heavy the computations are, it can get it done efficiently. 

For the sake of a fair comparison among different schemes which include blockchain-based and non-blockchain-based schemes that have miscellaneous system models, we concentrate the evaluation of the computational costs on the smart meters and aggregator’s computational responsibilities. Since each scheme has an aggregator in its network model, whether this aggregator is an SM or the GW, it could be a fair point of view for evaluating each scheme’s computational expenses. Hence, the best way to show the practicality and superiority of a privacy-preserving scheme in the smart grid network is to compare each SM’s computational burdens. 

As a result of utilizing various encryption and signature constructions in each scheme, for the sake of a reasonable comparison, we establish each scheme’s computational costs in the form of its core operations such as addition, multiplication, and exponentiation. Note that we base the computational evaluation on the costs described in \cite{S8}. Based on the MIRACL library \cite{MIRACL} running on an Intel Core i5-2430 2.4 GHz CPU with 2 GB memory and considering the standard SEPC160R1 for Elliptic curve cryptography, SHA1 for hash function, $|N^2|=2048$, and a $160-$bit group $\mathbb{G}$, we demonstrate the computational costs of the basic operation in TABLE \ref{TabV} below. In TABLE \ref{TabV}, the point-multiplication, exponentiation operation in $\mathbb{Z}_{N^2}$, exponentiation in $\mathbb{G}$, group-multiplication in $\mathbb{G}$, multiplication in $\mathbb{Z}_{N^2}$, and bilinear pairing are denoted by $C_{PM}$, $C_{ExpZ}$, $C_{ExpG}$, $C_{GM}$, $C_{ZM}$, and $C_{BP}$. It should be noted that operations such as computing hash functions, point-addition, modular inversion, simple integer addition and multiplication, $\textit{etc.}$, are considered negligible in relation to the operations depicted in TABLE \ref{TabV}. Therefore, we do not take these operations into account in the assessment of each scheme’s performance. 

%basic costs (Table X)
\begin{table}[h]
\centering
\captionof{table}{Basic Computational Costs.}
%\begin{adjustbox}{width=0.6\columnwidth,center}
\begin{tabular}{l|l}
Operation & Time (ms) \\ \hline\hline
$C_{PM}$ & 0.40  \\ \hline
$C_{ExpZ}$ & 9.78 \\ \hline
$C_{ExpG}$ & 9.80 \\ \hline
$C_{GM}$ & 1.18 \\ \hline
$C_{ZM}$ & 1.20 \\ \hline
$C_{BP}$ & 22.84 \\ \hline
\end{tabular}
\label{TabV}
%\end{adjustbox}
\end{table}

According to the computation of each scheme and the basic computational costs presented in TABLE \ref{TabV}, each SM’s cost is thoroughly illustrated in TABLE \ref{TabVI}.

%SM's costs (Table Y)
\begin{table}[h]
\centering
\captionof{table}{SM's Computational Costs.}
%\begin{adjustbox}{width=0.6\columnwidth,center}
\begin{tabular}{l|l|l}
Scheme & Computations &Time (ms) \\ \hline\hline
SPDBlock &$5C_{PM}$ or $4C_{PM}$ & 2 or 1.6 ms  \\ \hline
\cite{S2} &$\beta_1C_{ExpZ}+\beta_2C_{ExpZ}$ &$9.78\times (\beta_1+\beta_2)$ \\ \hline
\cite{S3}& $4C_{GM}$&$4.72$ ms \\ \hline
\cite{S4} &$2C_{ExpZ}+(n-1)C_{ZM}$&  \\ 
 &$+(n+3)C_{PM}$& $2.38n+21.9$ ms \\ \hline
\cite{S5} &$2C_{ExpZ}+C_{GM}$& $20.74$ ms \\ \hline
\cite{S6} & $3C_{ExpZ}$&$29.34$ \\ \hline
\cite{S7} &$(n+3)C_{PM}$& $0.4n+1.2$ ms \\ \hline
\cite{S8} &$(3\ell+2)C_{PM}$& $1.2\ell+0.8$ ms \\ \hline
\cite{Cube-data} &$2C_{ExpZ}+C_{GM}$& $20.74$ ms \\ \hline
\cite{SMing} &$2C_{ExpZ}$& $19.56$ ms \\ \hline
\end{tabular}
\label{TabVI}
%\end{adjustbox}
\end{table}

Now, based on TABLE \ref{TabVI}, we describe each scheme’s computational expenses comprehensively. At first, we evaluate the SPDBlock scheme. Our proposed scheme consists of 3 point-multiplication for encryption, one point-multiplication for computing the auxiliary ciphertext of the last block’s aggregated data, and one point-multiplication for creating a signature. Furthermore, it should be noted that although we considered the auxiliary ciphertext’s computation as an SM’s responsibility, only $\mu$ number of smart meters need to calculate the auxiliary ciphertext. Since the $\mu$ is about $n/2$, therefore nearly half of the smart meters in the SPDBlock scheme requires only 4 point-multiplications in each round ($4C_{PM}=1.6$ms).

Next, let us consider the scheme in \cite{S2} wherein each SM computes $\beta_1$ number of exponentiation in $\mathbb{Z}_{N^2}$ for signing its data and $\beta_2$ number of exponentiation in $\mathbb{Z}_{N^2}$ for validation of other SMs’ data. The $\beta_1$ is the number of each SM’s pseudonym and $\beta_2$ denoted all the pseudonyms in the network. Therefore, the scheme in \cite{S2} needs about $9.78(\beta_1+\beta_2)$ms per round.

The \cite{S3} scheme considers the amount of $4$ group-multiplications for the signcryption operations and the aggregated signature. Moreover, it should be noted that in order to have a fair comparison, we assumed that the number of receivers is one, otherwise, the \cite{S3} scheme requires $R_j\times 2C_{GM}+2C_{GM}$ computation for each round where $R_j$ is the number of receivers.

In the \cite{S4} scheme, each SM requires to compute the Paillier encryption, an Elliptic curve-based signature, Paillier homomorphic operation, and batch verification of all the received signatures which will lead to $2C_{ExpZ}+2C_{PM}+(n-1)C_{ZM}+(n+1)C_{PM}=2.38n+21.19$ ms.

Similarly, the \cite{S5} scheme also utilizes the Paillier encryption which costs 2 exponentiations in $\mathbb{Z}_{N^2}$ and one point-multiplication for creating a signature. Thus, it has a computational expense of around 20.47 ms.

The computational costs of each SM in the \cite{S6} scheme is comprised of 2 exponentiations for encrypting the data and one additional exponentiation operation for creating a signature on the encrypted data. Thereby, each SM requires at least $3C_{ExpZ}=29.34$ms. 

In the \cite{S7} scheme, besides the key generation that is computed at the beginning of the scheme, each SM needs to perform $(n+3)$ group-multiplications for encrypting the user power data and building a ring signature. As a result, the computational cost of each SM is at least $(n+3)C_{GM}=0.4\times(n+3)$ depending on the number of smart meters in the residential area ($n$).

The \cite{S8} scheme is able to transmit $\ell$-dimensional data and runs based on Elliptic curve encryption and signature construction. Hence, it costs at least $\ell$ point-multiplications for creating each dimensional data, $2\ell$ point-multiplications for encryption, and $2$ point-multiplications for producing a signature: $(3\ell+2)C_{PM}=0.4\times (3\ell+2)$.

In the \cite{Cube-data} scheme, each SM calculate some polynomial addition and multiplication to transmit $\ell$-dimensional data, 2 exponentiation operations for the Paillier encryption, and one group-multiplication for generating a signature. Thus, each SM requires no less than $2C_{ExpZ}+C_{GM}=20.74$ms at each round.

And lastly, the \cite{SMing} scheme employs the secret sharing technique and Paillier encryption. Consequently, besides computation of some additive secret sharing operation, each SM needs to calculate 2 exponentiation operations, and thereby it needs at least $2C_{ExpZ}=19.56$ms.

\begin{figure}
\includegraphics[width=15cm, height=10cm]{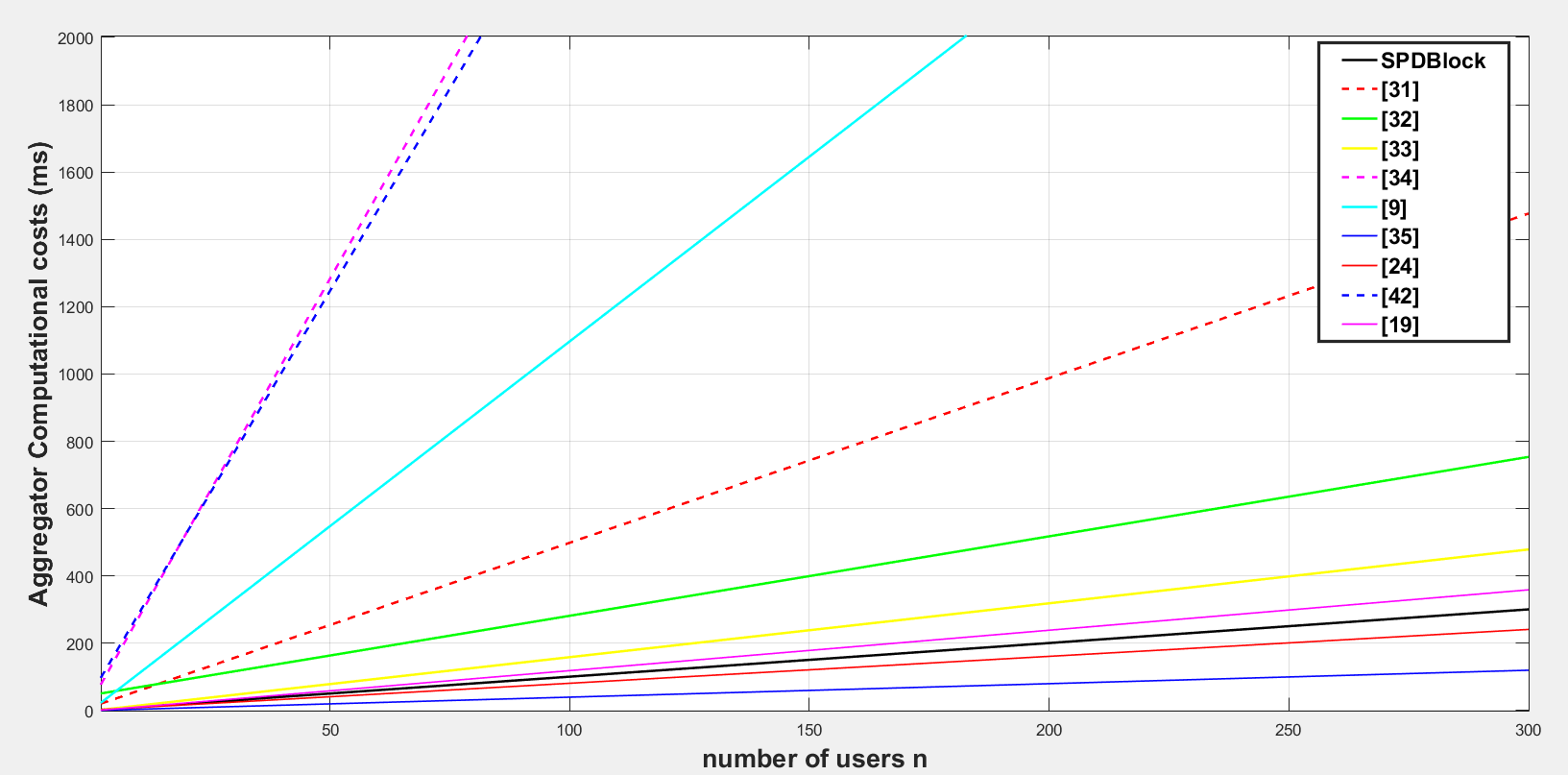}
\caption{Aggregator's Computational Costs}
\label{Fig-Aggr}
\end{figure}

In the next stage, for the aggregator’s computational responsibilities, we have represented the costs in TABLE \ref{TabVII}, depicted them in FIGURE \ref{Fig-Aggr} for better and clearer contrast, and described them in detail in the following paragraphs. The number of smart meters is signified by $n$.

%SM's costs (Table Y)
\begin{table}[h]
\centering
\captionof{table}{Aggregator's Computational Costs.}
%\begin{adjustbox}{width=0.6\columnwidth,center}
\begin{tabular}{l|l}
Scheme & Computations (ms) \\ \hline\hline
SPDBlock &$(2n+\mu+2)C_{PM}$  \\ \hline
\cite{S2} &$\beta_1C_{ExpZ}+\beta_2C_{ExpZ}$  \\ \hline
\cite{S3}& $2n\times C_{GM}+2C_{BP}$ \\ \hline
\cite{S4} &$(n-1)C_{ZM}+(n+3)C_{PM}$\\ \hline
\cite{S5} &$(n+1)C_{BP}+nC_{GM}+(n-1)C_{ZM}$\\ \hline
\cite{S6} & $nC_{ExpZ}+(n-1)C_{ZM}$\\ \hline
\cite{S7} &$nC_{PM}$ \\ \hline
\cite{S8} &$(2n+3)C_{PM}$ \\ \hline
\cite{Cube-data} &$(n+2)C_{BP}+C_{GM}+(n-1)C_{ZM}$ \\ \hline
\cite{SMing} &$(n-1)C_{ZM}$\\ \hline
\end{tabular}
\label{TabVII}
%\end{adjustbox}
\end{table}

Initially, we start by evaluating our SPDBlock scheme’s costs for the aggregation part. The aggregator in the SPDBlock scheme is one of the pre-determined powerful smart meters which are qualified for the aggregation computations. So, an $SM’$ in the SPDBlock scheme requires to compute $(2n+1)$ point-multiplications for batch verification, $\mu$ point-multiplications for the aggregation of auxiliary ciphertexts, and one point-multiplication for creating a signature on all these data. The result of which is $0.4\times (2n+\mu+2)$.

Next, we evaluate the \cite{S2} scheme wherein the PoW mechanism imposes that all the smart meters aggregate the consumption data and they all could be the miner of blocks. Accordingly, in this scheme, the aggregator’s cost is similar to each smart meter’s cost which is $9.78\times(\beta_1+\beta_2)$ms. 

In the \cite{S3} scheme, each pre-selected node aggregates the valid ciphertexts. Specifically, the pre-selected node firstly verifies the legitimacy of all the received ciphertexts and subsequently aggregates the consumption data. Hence, the aggregator’s computational cost is comprised of computing 2 bilinear pairing operations and $2n$ group-multiplications: $2.36n+45.68$ ms.

In the \cite{S4} scheme, each cluster gateway (CG) is responsible for the aggregation of data and creation of regional cluster’s blockchain blocks. So, due to the utilization of Paillier encryption, every CG needs to validate the signatures and then calculates Paillier homomorphic operation which is comprised of $(n-1)$ group-multiplications, $(2n-1)$ point-multiplications, and $(n-1)$ multiplications in $\mathbb{Z}_{N^2}$, respectively: $(3n-2)C_{PM}+(n-1)C_{ZM}$.

The DPPDA \cite{S5} scheme first chooses a miner and then it generates the keys and aggregates the valid receiving ciphertexts. More specifically, the mining node who is also the aggregator of each round, computes $n$ group-multiplications, $(n+1)$ bilinear pairing operations for the validation of signatures, and performs Paillier homomorphic operation on all these legitimate ciphertexts: $(n+1)C_{BP}+nC_{GM}+(n-1)C_{ZM}$.

For the DA-SADA scheme \cite{S6}, the aggregation node who is a strong SM in the user layer is in charge of aggregating the valid ciphertexts and creating a new block. Hence, it first validates the signatures and pseudonyms based on the bloom filter and $n$ exponentiation operations. Then, it computes the Paillier homomorphic operation. Therefore, the aggregation node’s cost is no less than $nC_{ExpZ}+(n-1)C_{ZM}$.

In the \cite{S7} scheme, the community area gateway (BG) is accountable for the batch verification of the ring signature and the generation of aggregated signcryption. Aside from the aggregation, each BG needs to compute at least $n$ point-multiplications for the batch verification. Thus, the computational expenses of the aggregator in the \cite{S7} scheme are around $0.4n$ms.

The aggregator (AGG) in the \cite{S8} scheme requires to perform batch verification on all the received ciphertexts and their signature on these data. Respectively, the AGG’s cost is $(2n+1)$ point-multiplications and 2 point-multiplications: $0.4\times(2n+3)$.

The \cite{Cube-data} scheme has two levels of aggregation. However, we only evaluate the first level of aggregation for the reason of a fair comparison. Accordingly, the residential area gateway (RAGW) at first divides the received ciphertexts into two groups so that it could perform a more attack-resistant batch verification. Successively, it computes the Paillier homomorphic operation on the valid ciphertexts and signs them with its private key. Therefore, its cost is $(n+2)C_{BP}+(n-1)C_{ZM}+C_{GM}$.

At last, in the \cite{SMing} scheme, each GW computes the spatial aggregation based on Paillier homomorphic operation. Thus, the computational cost of the aggregator (GW) is near $(n-1)C_{ZM}=1.20\times (n-1)$.

In conclusion, we have provided a comprehensive evaluation of each SM and the GW's computational expediency and from the TABLE \ref{TabVI}, TABLE \ref{TabVII}, and the FIGURE \ref{Fig-Aggr}, the efficiency of our proposed scheme is apparent .

%%%%%%%%%%%COMM
\subsection{Communication Overhead}
Another aspect of evaluation and comparison is the communication overhead. To determine the communication efficiency of our proposed scheme, after establishing the basic data sizes, we demonstrate the communication performance evaluation of each scheme.

Regarding the communication links between smart grid network entities and the limited power of each smart meter in the transmission of data regularly, it is fairly apparent that the most communication pressure is on the user (SMs).
As a result, we do not scrutinize the CC or the GW’s communication overheads, and we focus on our communication comparisons based on the SM’s responsibilities in each round of transmission. It is worth noting that due to the huge differences among the comparing schemes, especially between blockchain-based and non-blockchain-based schemes in respect to their communication model, it is quite impractical to compare them based on other communication overheads. Moreover, since these schemes aim to preserve the privacy of users’ consumption data and considering SM’s restricted communication capabilities, we specifically assess the network traffic of each scheme so that our communication performance evaluation would be completely fair.

By the fact that each scheme uses a different encryption and signature constructions or other techniques, we form each scheme’s communication overhead based on the length of their transmission elements.
More specifically, each element in $\mathbb{G}$, $\mathbb{G}_1$, $\mathbb{G}_2$, $\mathbb{Z^*}_q$, $\mathbb{Z}_N$, and $\mathbb{Z}_{N^2}$ has the length of 160 bits, 512 bits, 1024 bits, 160 bits, 1024 bits, and 2048 bits, respectively. All the comparing schemes’ overhead is comprised of the above terms for the same level of security. Furthermore, for the identity ID and the timestamp TS which are included in some schemes, 32 bits (separately) would be more than enough.

Relying on the aforementioned communication costs and the structure of each scheme, we have demonstrated the communication overhead of each SM in TABLE \ref{TabIV}. Regarding the TABLE \ref{TabIV}, besides the \cite{S3} scheme that requires to transmit only $694$ bits, the SPDBlock is also efficient and needs to transmit $992$ bits. Moreover, the communication overhead in the SPDBlock scheme does not depend on the number of smart meters or the $\ell$-dimensional data transmission. Therefore, considering the above analysis and the achievements of our proposed scheme, the SPDBlock is truly beneficial for practical uses.

%SM's Comm (Table Z)
\begin{table*}[t]
\centering
\captionof{table}{SM's Communication Overhead.}
%\begin{adjustbox}{width=0.6\columnwidth,center}
\begin{tabular}{l|l|l}
Scheme & Transmission &Overhead (bit) \\ \hline\hline
SPDBlock &$|C_i|+|TS|+|Ack_i|+|\bar{C_{aux_i}}|+|\sigma_i|$ & 992 bits  \\ \hline
\cite{S2} &$(n-1)\times(|m_i|+|TS|+|EK_{Sk_i}|+|Pk_i|)$ &$(n-1)\times 2208$ bits \\ \hline
\cite{S3}& $|\omega|+|ID_i|+|ID_r|+|TS|+|D|+|\sigma_i|$&694 bits \\ \hline
\cite{S4} &$|E(u_i)|+|\mbox{Address}|+|\mbox{Version~no.}|+|Sig|$& 2592 bits \\ \hline
\cite{S5} &$|C_i|+|TS|+|Pk_i|+|\sigma_i|$& 2754 bits \\ \hline
\cite{S6} & $(n-1)\times(|C_i|+|TS|+|\sigma_i|+|\mbox{Pseudo}|)$&$(n-1)\times 5152 $bits \\ \hline
\cite{S7} &$|C_i|+|f|+|R|+|A_i|+|U|+$& $(n+4)\times 160 $ \\ 
&$|U_1|+...+|U_n|+|TS|$&$+32$ bits\\\hline
\cite{S8} &$|C_i|+|ID|+|TS|+|Sig|$& $864+320\times \ell$ bits \\ \hline
\cite{Cube-data} &$|C_i|+|ID_i|+|ID_r|+|TS|+|\sigma_i|$& 2652 bits \\ \hline
\cite{SMing} &$(k-1)|\mbox{Plaintext}|+|C_i|+2|ID|$& 2240 bits \\ \hline
\end{tabular}
\label{TabIV}
%\end{adjustbox}
\end{table*}

%%%Achievements
\subsection{Achievements}
In brief, this subsection presents each scheme’s achievements. Particularly, owing to the structure of each scheme and its goals, we illustrate and compare the achievements of each scheme based on their system model, security criteria, and smart grid features. Additionally, we compare the miner selection and the block creation of each blockchain-based scheme in this subsection. Considering the communication and computational evaluation presented above, employing this achievement comparison we can completely and clearly depict the efficiency and practicality of each scheme.

%%%%Sys Model
\subsubsection{\textbf{System models}}
As you know, the more the smart grid’s network model is akin to the real-world situations, the harder and costlier it gets to preserve the privacy of users’ consumption data and accomplish other security criteria.

In the SPDBlock scheme, besides the TA who is customarily an external entity that is responsible for generating the keys at the beginning of the protocol and has no access to the consumption data, no other smart grid entity is assumed to be honest. Moreover, we also consider various attacks related to all the internal entities and collusion among them. Hence, it is pretty obvious that our network model has the most similarity with the real-world situations among all the proposed schemes for the smart grid.

However, in \cite{S2} scheme, with the exception of the key management center (KMC) who creates the keys and builds the bloom filter, the billing center, and smart meters are considered to be trusted and the CC is assumed to be honest-but-curios.
For the \cite{S3} scheme, the CC is responsible for key management, and thereby it is like smart meters presumed as completely trusted. Besides, other receivers like the grid operator and the supplier have also access to the keys that CC has, therefore they are presumed as trusted.
In \cite{S4} scheme, both the CC and the CG are assumed to be honest-but-curious and smart meters are considered to act trustworthy.
The DPPDA scheme \cite{S5} states that although it has no TTP or CA in its network model, the smart meters are honest-but-curious and the CC is trusted.
The network model of the DA-SADA scheme \cite{S6} is comprised of honest-but-curious fog and cloud nodes along with a completely trusted TA and several trustworthy smart meters. Particularly, malicious nodes are assumed to be less than $1/3$ of all the nodes in the network so that the blockchain-based mechanism works correctly.
The \cite{S7} scheme considers a trusted key generation center (KGC), an honest-but-curious CC along with a BG who can act maliciously. However, it assumes that no more than $1/3$ of the nodes (BG) could be byzantine nodes or have a failure for their mechanism to operate properly. 
The \cite{S8} scheme has a typical network model wherein both the CC and the GW are honest-but-curious and the TTP who configures the system is fully trusted.

In the \cite{Cube-data} and \cite{SMing} schemes, the CC who is in charge of system initialization is assumed to be trustworthy and both of the GWs in the residential and district area for \cite{Cube-data} and the GW in \cite{SMing} is presumed to be honest-but-curious. Additionally, the SMs are naturally considered to act reliably. 

By and large, regarding the above inspection of each scheme’s network model, we can safely draw the conclusion that the SPDBlock scheme has the most world-like network model.

%%%Security Criteria
\subsubsection{\textbf{Security Criteria}}
To conduct a fair comparison among different schemes, besides the communication and computational assessments, we represent each scheme’s security achievements in TABLE \ref{TabV}. For instance, a scheme can have lower communication, but does not provide integrity or authentication. Thus, knowing every aspect of a scheme makes a more reasonable and effective contrast among these schemes. Based on the above analysis, the SPDBlock accomplishes to have all the required security criteria.

%Security Table (Table A)
\begin{table*}[t]
\centering
\captionof{table}{Security Criteria Achievements.}
%\begin{adjustbox}{width=0.6\columnwidth,center}
\begin{tabular}{l|l|l|l|l|l|l|l|l|l|l}
Scheme & Priv.&Int.&Auth. &Rep.&Modif.&Imper.&MITM&DoS&F.-T.&Col. \\ \hline\hline
SPDBlock &Yes&Yes&Yes& Yes&Yes&Yes&Yes&Yes&Yes&Yes \\ \hline
\cite{S2} &Yes&Yes&Yes&Yes &Yes&Yes&Yes&No &Yes&No\\ \hline
\cite{S3}& Yes&Yes&Yes&Yes&Yes&Yes&Yes&No&Yes&No\\ \hline
\cite{S4} &Yes&No&Yes&Yes&Yes&No&Yes&No&Yes&No\\ \hline
\cite{S5} &Yes&Yes&Yes&Yes&Yes&Yes&Yes&No&Yes&No\\ \hline
\cite{S6} & Yes&Yes&Yes&Yes&Yes&Yes&Yes&No&Yes&No\\ \hline
\cite{S7} &Yes&Yes&Yes&Yes &Yes&Yes&Yes&No&Yes&No\\ \hline
\cite{S8} &Yes&Yes&Yes&Yes&Yes&Yes&Yes&No&Yes&No\\ \hline
\cite{Cube-data} &Yes&Yes&Yes&Yes&Yes&Yes&Yes&No&Yes&No\\ \hline
\cite{SMing} &Yes&No&No&No&No&No&1No&No&Yes&No\\ \hline\hline
\end{tabular}
\label{TabV}
%\end{adjustbox}
\end{table*}

In the above table, the columns' names are abbreviated for a more clear demonstration. In TABLE \ref{TabV}, Priv., Int., Auth., Rep., Modif., Imper., F.-T., and Col. denote Privacy, Integrity, Authentication, Replay, Modification, Impersonation, Fault-Tolerance, and Collusion, respectively.

\subsubsection{\textbf{Features of Interest in the SG Network}}
Apart from security criteria, some features should be provided in a practical scheme for the smart grid network. Especially, in this subsection, we present temporal aggregation, the transmission of multi-dimensional data, and the calculation of bills in a privacy-preserving way. 

Since the smart meter’s data is truly comprised of diverse components, the transmission of multi-dimensional data would be beneficial in the smart grid network. To transmit multi-dimensional data in a privacy-preserving manner, some schemes employ various techniques such as super-increasing sequence, Horner’s rule, or a special algorithm for packing. Among these comparing schemes, only the \cite{S8} and \cite{Cube-data} schemes consider the transmission of multi-dimensional data. 

The \cite{S8} scheme expands its employed Elliptic curve-based encryption and creates $\ell$ ciphertexts. Consequently, the transmission of multi-dimensional data in a privacy-preserving way costs $2(\ell -1)$ ms additional computations and $320(\ell -1)$ bits extra communication. Therefore, it is apparent that the transmission of multi-dimensional data in the \cite{S8} scheme is really heavy concerning the communication overhead and computation costs.

In the \cite{Cube-data} scheme, the smart meters utilize Horner’s rule to create a single polynomial containing the multi-dimensional data. For the creation of this polynomial, each smart meter must perform numerous polynomial addition and multiplication even if we consider precomputations for Horner’s rule parameters. Besides, at the final phase, the CC must execute a recursive algorithm to reconstruct each dimensional data and it cannot be done separately for each data. Therefore, although the \cite{Cube-data} scheme’s multi-dimensional technique does not impose any communication overhead, it has numerous polynomial computations.

And lastly, the SPDBlock proposes a Chinese remainder theory-based technique that adds no communication or heavy computational costs. More specifically, this proposed CRT-based technique is independent of the employed encryption scheme. Therefore, in relation to other techniques, it could be used with almost all the encryption constructions and adds no additional burdens on the encryption. Moreover, since this technique works on plaintexts, there is no extra communication overhead. For the computational costs, in the SPDBlock scheme that employs Elliptic curve-based encryption, it only requires $\ell -1$ additional integer multiplications which is negligible comparing to any encryption operations. Thus, we can conclude that the transmission of multi-dimensional data in the SPDBlock scheme accomplishes the lowest communication and computational costs.

Besides the spatial aggregation which is obtained for privacy reasons in the smart grid network, the computation of temporal aggregation is necessary for better analysis of the network and the calculation of bills. Although most schemes focus on privacy and do not consider the temporal aggregation, among the comparing schemes, the SPDBlock and \cite{SMing} schemes compute temporal aggregation of users’ data in each billing period. However, the \cite{SMing} scheme requires each smart meter to send all its data in a billing period to the GW so that the GW can compute the temporal aggregation. Hence, for the computation of temporal aggregation, the assistance of smart meters is obliged, and thereby it burdens the smart meters. But, in the SPDBlock, due to the blockchain-based construction of our scheme, the GW can compute the temporal aggregation relying on the data in Sidechain blocks. Furthermore, because of our proposed CRT-based technique for transmission of multi-dimensional data, we can also accomplish in acquiring the multi-dimensional temporal aggregation of users’ consumption data. In conclusion, the SPDBlock achieves a more useful and lightweight temporal aggregation in the SG network.

Also, the computation of temporal aggregation is necessary for the calculation of the bills in a privacy-preserving manner. In the comparing schemes, the bill calculation is generally accomplished via transmitting the whole blockchain to the billing center, transmitting data directly for bill calculation purposes, or the computation of temporal aggregation. However, allowing the billing centers or authorities to access the blockchain does not preserve privacy or requires a trusted powerful entity in the SG network. And for the same reasons, sending the users’ consumption data to the CC or the billing center for bill calculation is not secure. Finally, the computation of temporal aggregation is the heavy but effective way to calculate the bills without revealing any critical information. Whereas, due to the employed homomorphic encryption and the lightweight computation of temporal aggregation in SPDBlock, the GW can calculate the bills based on three kinds of prices in a privacy-preserving style. Additionally, because of the employed distribution decryption technique, the CC cannot decrypt the bills without the approval of at least $\mu$ number of SMs. Consequently, the SPDBlock scheme not only calculates the bill based on ciphertexts with lightweight communication and computations, but also can preserve the privacy of the bills and allows for only the approved-bill decryption.

\subsubsection{\textbf{Miner Selection}}
Since our proposed scheme runs based on blockchains, one especial aspect of comparison is in fact the structure of the deployed or the proposed blockchain. As opposed to previous parts, we only can compare the efficiency of our proposed SPDBlock scheme with 6 other blockchain-based schemes \cite{S2, S3, S4, S5, S6, S7}.

Initially, relying on the consensus mechanism and the miner selection phase, we delineate the efficiency of each blockchain-based scheme. The \cite{S2} scheme proposes a novel PoW mechanism wherein the miner of each block is selected based on every smart meter’s contribution. Specifically, each SM transmits its data to every other SM in the network and after the calculation of the average consumption, the SM who has the closest consumption data to the average result is the miner of that round. Some drawbacks of this mechanism are as follows: every SM in the network calculates the average consumption, all SMs heavily communicate with each other, and there might be more than one miner for each round. Hence, it is apparent that the \cite{S2} scheme’s proposed mechanism is not quite practical from the view of computation and communication costs.

In the \cite{S3} scheme, the delegated PoS mechanism is used in a Consortium blockchain wherein $101$ pre-selected powerful SMs have been chosen to be the miner of each round sequentially. As a result of publishing pre-selected nodes (SMs) in the SG network, it would be considerably easier for an adversary to focus on the pre-determined miner and conduct some passive and active attacks such as collusion, malware, or DoS attacks. 

Both \cite{S4} and \cite{S7} schemes employ PBFT mechanisms in their blockchain construction. As presented in their schemes, although this mechanism supports $f$ number of failure where $n>3f+1$, it requires lots of computations and multiple rounds of communication among all nodes. Consequently, given the great number of nodes in the SG network and the restriction on their communication and computational capabilities, it would be inefficient to utilize such a mechanism in a privacy-preserving scheme for the SG network.

The \cite{S5} DPPDA scheme presents a miner selection mechanism that works based on $3$ states and an election. Particularly, each SM could play the role of a follower, candidate, and miner. While this election needs no special computations, it necessitates multiple rounds of communication to select the mining node. Therefore, it is also not suitable for a system with a great number of nodes like the SG network. 

In the DA-SADA scheme \cite{S6}, only the aggregation node of each subarea creates the new block for each round based on the received and the aggregated consumption data. And relying on the $2n/3+1$ confirmation messages, the block is considered to be legitimate. According to the fact that this scheme assumes less than $1/3$ of the nodes are malicious, this mechanism works properly. On the evidence of at least $2n/3+1$ confirmation messages among all the nodes, we can safely conclude that this mechanism is heavy concerning the network traffic.

Finally, the SPDBlock scheme which operates on the PoS mechanism employs a novel and specialized hash onion technique to select the miner of each round. Technically, before each round of transmission, a few pre-approved and strong smart meters ($SM’$) contribute to select the miner of the next round. The criteria for being on the list of $SM’$s is quite reasonable, and the only computation they need to perform is a simple hash function. Thereby, it won’t be computationally heavy or burdensome. Furthermore, due to the small number of $SM’$s, the lightweight communication overhead of the messages (only one hash function), and the fact that they need to communicate only one time, we can reliably infer that our proposed miner selection is extremely efficient in comparison with other scheme’s construction. 

Notwithstanding the communication and computational efficiency of our proposed mechanism, smart meters do not need to transmit the block in the network for validation. More precisely, because of our two parallel Consortium blockchain construction, the SMs are only obliged to transmit one $Ack/Nack$ for the previous block along with their consumption data in the next round. 
Considering each scheme’s block size, our scheme achieves the lowest communication and computational costs related to blockchain features such as miner selection and block validation.
Additionally, our proposed miner selection technique and blockchain-based construction accomplish the 5 criteria cited in the miner selection phase and resist against all the potential BC-based attacks such as collusion, DoS, fork, or $51\%$ attack.

From the perspective of blocks, our SPDBlock scheme also prospers greatly. Particularly, the SG network in the SPDBlock scheme runs on Mainchain’s blocks and the aggregated ciphertext and GW’s signature are the only regular data in the Mainchain’s blocks. Therefore, the block size of our scheme is considerably efficient in comparison with other schemes that are comprised of all the user’s consumption data. 

One further significant problem of blockchain-based scheme is blockchain bloat \textit{i.e.,} the blockchain ledger becomes drastically larger over time. It should be noted that blockchain bloat is in fact a serious concern in SG network due to the limited storage size of each SM. The existence of two parallel blockchains in the SPDBlock scheme helps to address this problem accurately. Concretely, after one week (the billing period which is about $1100$ blocks), all the Sidechain’s blocks get deleted and only the lightweight Mainchain’s blocks remain. It is worth noting that all the required and necessary information of the SG network is included in the Mainchain’s blocks, therefore the removal of Sidechain’s blocks does not affect the network in any way.

%conclusion
\section{CONCLUSION AND FUTURE WORK}
With the advent of blockchain and regarding the privacy problems of the Smart grid network, in this chapter, we have proposed a secure privacy-preserving data aggregation scheme for blockchain-based smart grid (SPDBlock). Based on proposing a novel blockchain-based structure, we could manage to preserve the privacy of users’ individual and aggregated  data in a network model akin to real-world situations. Furthermore, the SPDBlock scheme not only preserves the privacy of users’ data, but also ensures data integrity, source authentication, and it is resistant to numerous attacks. Besides the security criterion, in the performance evaluation, we have shown other achievements such as fault-tolerance, multi-dimensional data transmission, temporal aggregation, and bill calculation. Precisely, relying on the Chinese remainder technique and employed Elliptic curve encryption, the SPDBlock scheme can transmit multi-dimensional data in a way that CC could acquire spatial and temporal aggregation each dimensional-data in a privacy-preserving manner. Furthermore, SPDBlock allows for the calculation of the bills based on ciphertexts; and therefore the privacy of the bills is guaranteed. Finally, in the performance evaluation, we have shown that our proposed scheme is quite practical and accomplishes the lowest communication and computational costs with multiple achievements.

%-----------------------------------------------------------------------------------------------------------
% Conclusion
%--------------------------------------------------------------------------------------------------------
\chapter{Conclusion and Future Work}
\lhead{\emph{Conclusion}} % Set the left side page header to "Conclusion"
\section{Conclusion}
This master’s thesis defines the smart grid, establishes its most important challenges, and then focuses on addressing these challenges. Specifically, in this thesis, we introduce two novel privacy-preserving data aggregation schemes based on two state-of-the-art approaches, utilizing lattice-based cryptography and the architecture of blockchain technology. 
Besides privacy which is the pivotal criterion for the smart grid implementation, we also try to guarantee other security criteria and accomplish some additional smart grid properties. The first proposed scheme, LPM2DA, is constructed based on lattice-based cryptography to achieve a higher level of security that is resistant against post-quantum attacks. Likewise, the second scheme, SPDBlock, is designed relying on blockchain technology to achieve a higher level of security against insider attackers and resolve centralization problems.

\begin{enumerate}
\item LPM2DA: Due to the fact that one of the most important criteria in smart grid networks is preserving the privacy of users’ data, in the first part of this thesis, we have introduced a secure lattice-based multi-functional and multi-dimensional data aggregation scheme (LPM2DA). In situations where networks are in imminent danger of quantum attacks, our LPM2DA scheme not only preserves the privacy of users’ consumption data, but also ensures data integrity and source authentication. Moreover, our LPM2DA scheme maintains fault-tolerant, allows dynamic users, and is resistant against various attacks like impersonation, modification, MITM, and replay attack. Based on a homomorphic encryption scheme and Chinese remainder theorem, the control center could acquire temporal and spatial aggregations of users’ multi-dimensional data efficiently; and it is capable of calculating various statistical functions like mean, variance, and skewness. Finally, we have demonstrated the computational and communication efficiency of our proposed scheme via comparing its performance with other data aggregation and lattice-based schemes.

\item SPDBlock: With the advent of blockchain and considering the privacy problems of the Smart grid network, in this paper, we have introduced a secure privacy-preserving data aggregation scheme for blockchain-based smart grid (SPDBlock). Based on proposing a blockchain-based construction, we could manage to preserve the privacy of users’ individual and aggregated consumption data in a network model akin to real-world situations. Moreover, the SPDBlock scheme not only preserves the privacy of users’ data, but also ensures other security criteria such as data integrity, source authentication, and resistance against numerous passive and active attacks. Besides the security strength of our proposed scheme, in the performance evaluation, we have also presented other achievements such as fault-tolerance, multi-dimensional data transmission, temporal aggregation, and bill calculation. To be precise, relying on the Chinese remainder technique and employed Elliptic curve encryption, we can transmit multi-dimensional data in a way that CC could acquire spatial and temporal aggregation of each dimensional-data in a privacy-preserving manner. We have shown that the transmission of multi-dimensional data in our scheme is more efficient than other schemes in SG. Furthermore, as opposed to other schemes, the SPDBlock allows for the calculation of the bills based on ciphertexts; and thereby the privacy of the bills is guaranteed. Finally, in the performance evaluation, we have demonstrated that the SPDBlock is quite practical and accomplishes the lowest communication and computational costs with multiple achievements. 
\end{enumerate}

In summary, we have proposed two novel privacy-preserving data aggregation schemes based on state-of-the-art approaches wherein the privacy of users’ individual and aggregated data is preserved. In both schemes, the security criteria like integrity and authentication are guaranteed and each of them resists various attacks such as impersonation, modification, replay, and MITM attacks. Furthermore, we have proposed a practical Chinese remainder theorem-based technique that is independent of the employed encryption and could be utilized along with several encryption schemes and different structures. More precisely, this CRT-based technique only requires to be adapted for the plaintext space of the employed encryption; and that is achieved through choosing the appropriate construction of the CRT. For example, to use number-theoretic encryptions with integer plaintext or lattice-based encryptions with polynomial plaintext, providing an appropriate selection of CRT’s parameters, we could utilize an integer CRT or polynomial CRT, respectively. Finally, regarding the exhaustive analysis of each scheme’s security and the evaluation of their performance, we can safely infer that both of these schemes are suitable for practical use in the smart grid network.

\section{Future Work}
It is apparent that there exist some additional security measures and extra properties that could be beneficial for the smart grid network which we did not consider in our proposed schemes. 

Concerning the LPM2DA scheme, in future work, we aim to achieve differential privacy and detection of any collusion among different entities. To accomplish differential privacy, the basic approach is to add special noise to the users’ data. However, we tend to dispense the burden of noise addition in the whole network for better computation and communication efficiency. Furthermore, bearing in mind that lattice-based cryptography contains heavy communication overhead, utilizing a different structure such as forecasting-demand instead of data aggregation would be critical in increasing the communication efficiency. Finally, due to the utilization of a lattice-based secret sharing technique, we can also consider multiple servers for the gateway or the control center itself to preserve the privacy of users’ data in an efficient manner. 

With regard to the SPDBlock scheme, for the ensuing work, we are currently working on a forecasting-demand structure in a blockchain-based smart grid wherein we employ payment channel networks. The only communication in forecasting-demand structure is when a consumer changes its electrical demand. Moreover, in payment channel networks, instead of writing every data on the blockchain, only the final result of electrical changes goes on the blockchain. Due to the low communication and computational costs of Elliptic curve cryptography, it is the perfect fit for the system model of this scheme. Consequently, we are concentrating on achieving the lowest communication overhead while preserving consumers’ privacy and data security.

Finally, it is worth noting that both schemes could be viewed as a foundation of a novel idea. Exactly, for future work, a brilliant innovation is to design a post-quantum secure privacy-preserving scheme for the blockchain-based smart grid. Technically, to the best of our knowledge, the combination of lattice-based cryptography and blockchain technology has never been implemented in the smart grid or any other related application. Hence, the proposed schemes in this thesis would be the groundwork of this novel combination.

%\input{Chapters/Chapter3}
%\input{Chapters/Chapter4} 
%\input{Chapters/Chapter5} 
%\input{Chapters/Chapter6} 
%\input{Chapters/Chapter7} 

%----------------------------------------------------------------------------------------
%	THESIS CONTENT - APPENDICES
%----------------------------------------------------------------------------------------

%%\addtocontents{toc}{\vspace{2em}} % Add a gap in the Contents, for aesthetics

%%\appendix % Cue to tell LaTeX that the following 'chapters' are Appendices

% Include the appendices of the thesis as separate files from the Appendices folder
% Uncomment the lines as you write the Appendices

%%\input{Appendices/AppendixA}
%\input{Appendices/AppendixB}
%\input{Appendices/AppendixC}

%%\addtocontents{toc}{\vspace{2em}} % Add a gap in the Contents, for aesthetics

%%\backmatter

%----------------------------------------------------------------------------------------
%	BIBLIOGRAPHY
%----------------------------------------------------------------------------------------
\nocite{*}
\label{Bibliography}

\bibliographystyle{apalike} % Use the "custom" BibTeX style for formatting the Bibliography

%bib

%%\bibliography{Bibliography} % The references (bibliography) information are stored in the file named "Bibliography.bib"

\end{document}